\documentclass[onecolumn]{aastex7}
\usepackage{color}
\usepackage{graphicx}
\usepackage[normalem]{ulem}
\usepackage{rotating}
\let\tablenum\relax
\journalinfo{}
\usepackage{siunitx}

\usepackage{braket}
\DeclareFontShape{OT1}{cmr}{m}{scit}{<->ssub * cmr/m/it}{}

\shorttitle{Exploration of the \textit{Chandra} Source Catalog with Self-Organizing Maps}
\shortauthors{Scaudo et al.}

\date{\today}

\begin{document}

\title{Self-Organizing Maps for the Exploration and Classification of X-Ray Sources\\ in the \textit{Chandra} Source Catalog}

\author[0009-0002-9618-0212]{Andrea Scaudo}
\affil{INAF -- Osservatorio Astronomico di Brera-Merate, via E. Bianchi 46, I-23807 Merate (LC), Italy}
\affil{ICSC --  National Research Centre for High-Performance Computing, Big Data and Quantum Computing, Spoke 2: Fundamental Research \& Space Economy, Italy}
\email{andrea.scaudo@inaf.it}
\makeatletter
\begingroup
\long\def\hyper@@anchor#1#2{#2}%
\correspondingauthor{Andrea Scaudo, \href{mailto:andrea.scaudo@inaf.it}{andrea.scaudo@inaf.it}}
\endgroup
\makeatother

\author[0000-0003-3073-0605]{Raffaele D'Abrusco}
\affil{Center for Astrophysics \textbar\ Harvard \& Smithsonian, 60 Garden Street, Cambridge, MA 02138, USA}
\email{rdabrusc@cfa.harvard.edu}

\author[0009-0000-5483-2652]{V. Samuel Pérez-Díaz}
\affil{Courant Institute, New York University, 60 Fifth Avenue, New York, NY 10011, USA}
\email{samuelperez.di@gmail.com}

\author[0000-0002-5069-0324]{Juan Rafael Martínez-Galarza}
\affil{Center for Astrophysics \textbar\ Harvard \& Smithsonian, 60 Garden Street, Cambridge, MA 02138, USA}
\email{juan.martinez.galarza@cfa.harvard.edu}

\author[0000-0001-5570-5081]{Marco Landoni}
\affil{INAF -- Osservatorio Astronomico di Brera-Merate, via E. Bianchi 46, I-23807 Merate (LC), Italy}
\email{marco.landoni@inaf.it}

\begin{abstract}
In this work, we apply Self-Organizing Maps (SOMs) to the \textit{Chandra} Source Catalog (CSC) to investigate the distribution of observation-level detections in the \textit{feature} space defined by their spectral and variability properties. SOMs are neural-network-based models that project points in high-dimensional space onto a two-dimensional latent space while preserving their local topological structure. From CSC 2.1.1 we extract a sample of 36,904 ACIS-only, observation-level, high statistical significance detections, belonging 
to 17,927 unique sources. Among these detections, 25,192 (10,078 sources) are associated with a label from 
SIMBAD via a $1\arcsec$ cross-match that provides information about their astrophysical nature. Training a $40\times40$ hexagonal SOM reveals structured regions in the U-Matrix and activation response map, and \textit{feature}-projection maps trace coherent patterns in spectral shape and variability. Overlaying SIMBAD classes shows that labeled populations occupy distinct, partially adjacent regions of the map and that generic categories can be interpreted via their SOM locations. We then retrain the SOM for classification using a source-grouped train/validation/test split and introduce \textit{SOMClassifier}. When applied to 11,712 detections without SIMBAD labels, it assigns valid (non-abstained) labels to 6,225 detections (coverage $\sim\!53\%$). Furthermore, we release the trained models and visualization/classification workflows through a public web application built on the most up-to-date CSC release.
\end{abstract}

\section{Introduction}
\label{sec:introduction}
Since its launch in 1999, the NASA \textit{Chandra} X-ray Observatory has significantly advanced the field of X-ray astronomy, thanks to its exceptional spatial resolution. The observatory operates with two primary instruments: the Advanced CCD Imaging Spectrometer (ACIS) and the High-Resolution Camera (HRC). These instruments provide high-resolution imaging crucial for studying a broad range of astrophysical phenomena. In its more than 25 years of operation, \textit{Chandra} has contributed to numerous groundbreaking discoveries, solidifying its role as a cornerstone in high-energy astrophysical research~\citep{2019cxro.book.....W}.

A structured public archive has been maintained to systematically organize, store, and distribute the vast amounts of observational data gathered by \textit{Chandra}. The resource derived from this archive is the \textit{Chandra} Source Catalog (CSC), which compiles a comprehensive inventory of X-ray sources detected in publicly available \textit{Chandra} observations. This catalog offers extensive details on each source, including astrometric measurements, photometric and spectroscopic properties, model-based flux estimations, variability assessments, and information regarding source morphology and extension~\citep{2024ApJS..274...22E}.

The CSC, with its latest updates culminating in version 2.1.1 (October 2024), provides a vast repository of X-ray sources and detections from \textit{Chandra} observations. Within this framework, detections represent photon clusters recorded in detector images, while sources denote distinct astrophysical X-ray emitters. As many of these sources lack secure multi-wavelength counterparts, their classification remains uncertain, underscoring the need for advanced analytical methods to improve their identification and characterization.

As astronomical surveys continue to grow in scale and complexity, the use of advanced statistical and machine learning methodologies has become crucial. In particular, unsupervised learning techniques provide valuable insights into extensive datasets by identifying underlying patterns without requiring predefined labels. Among these, Self-Organizing Maps (SOMs) serve as a powerful tool for dimensionality reduction, offering a structured and interpretable representation of high-dimensional data. By leveraging competitive learning, SOMs project intricate datasets onto a two-dimensional plane while maintaining the essential topological relationships of the original \textit{feature} space.

SOMs~\citep{2001som..book.....K} have been extensively applied in astrophysics to identify and visualize elusive patterns in a diverse landscape of usually complex astronomical datasets. They have been used, over time, as a tool to investigate the color distribution of galaxies in large photometric surveys in order to optimize the reconstruction of their photometric redshift probability density 
functions (PDFs)~\citep{2014MNRAS.438.3409C,2025arXiv250802779R}, and to improve the photometric redshift calibration by identifying underrepresented regions of the training set parameter space~\citep{2015ApJ...813...53M}. In exoplanet research, they have been employed to distinguish genuine planetary transits from false positives, significantly improving candidate selection in the Kepler mission~\citep{2017MNRAS.465.2634A}.

More recently, SOMs have been instrumental in determining the clustering of different classes of galaxies in their observational space and mapping the clusters to the physical properties of the galaxies such as stellar mass, color, and 
morphology~\citep{2022MNRAS.513.1972H}. SOMs have also been used to estimate the physical parameters of galaxies from multi-band, deep photometric data of statistically significant samples of galaxies~\citep{2025arXiv250604138A}, and the bolometric luminosity of quasars from multi-wavelength, large-band
spectral energy distributions~\citep{2025ApJ...988..204C}. Several aspects of
the discovery of elusive Active Galactic Nuclei (AGNs) have also been successfully tackled 
through this model:~\citet{2025ApJ...992..138S} used SOMs to assess the efficiency and observational biases affecting the 
selection of low-luminosity AGNs hosted by dwarf galaxies from multi-wavelength datasets,
while~\citet{2025ApJ...986...36L} demonstrated that SOMs are effective at recognizing 
massive black holes that are accreting at sub-Eddington rates.

The versatility of SOMs has also been demonstrated by their application to time-series analysis in different 
wavelength regimes. SOMs have been successfully employed to classify AGNs based on their optical variability~\citep{2025A&A...699A.303M} and have been used to uncover patterns in X-ray light curves of 
sources in the \textit{XMM-Newton} catalog, aiding in the identification of flaring and eclipsing sources~\citep{2022A&A...659A..66K}. 

While the SOM ability to determine clustering in high-dimensional \textit{feature} spaces is often used to 
infer similarities in the properties of large populations of astronomical sources, SOMs can also be used to detect anomalies, which has proven invaluable for accelerating the discovery of novel or rare astrophysical behaviors. Recent studies have employed a variety of machine learning techniques to address the classification of generic, unknown X-ray sources in the CSC~\citep{2022ApJ...941..104Y,2023MNRAS.520.5065K,2024MNRAS.528.4852P} using unsupervised learning 
to classify observation-level detections with no
prior spatial constraints, or to refine the classification of specific X-ray sources located in the vicinity of unidentified \textit{Fermi} $\gamma$-ray sources~\citep{2024ApJ...971..180Y} or specific areas of the sky where confusion between stellar and extragalactic sources is significant~\citep[e.g.,][for the Orion Nebula Cluster]{2023ApJ...949...12H}. 

In this study, we utilize SOMs to analyze the CSC, focusing on uncovering hidden structures within the 2.1.1 release. By reducing the dimensionality of the dataset while preserving essential relationships, our approach enables a systematic classification of X-ray sources based on shared characteristics. Through this method, 
we aim to reveal new insights into both known and previously unidentified sources. To support the broader research community, 
we have developed an interactive web-based tool~\footnote{\href{https://somcsc.streamlit.app/}{somcsc.streamlit.app}} that 
allows users to explore the trained SOM and interact with the most recent CSC dataset, facilitating pattern discovery in 
X-ray astronomy. The source code of the web application, including the SOM training, the visualization tools used to produce the SOM maps, and the \textit{SOMClassifier} implementation used to
produce the classifications presented in this paper, is released on GitHub~\footnote{\href{https://github.com/andreascaudo/SOM-CSC}{github.com/andreascaudo/SOM-CSC}}.

\section{Catalog Overview}
Since its launch, the \textit{Chandra} X-ray Observatory has collected 15,533 observations, which have been processed to 
extract the CSC, the largest available catalog of X-ray sources detected by \textit{Chandra}. The latest 
release, CSC 2.1.1\footnote{Access to data from the CSC is described at~\href{https://doi.org/10.25574/csc2}{doi:10.25574/csc2}~\citep{csc2_doi}} (October 2024), is the most comprehensive to date, containing 407,806 unique X-ray sources and 
1,304,376 individual detections from publicly available \textit{Chandra} observations covering $\sim\!730$ deg$^2$ of the sky. The catalog organizes these detections into energy bands based on X-ray photon 
energies. For the ACIS, these bands include broad (\numrange[range-phrase=--]{0.5}{7.0}\,\unit{\kilo\electronvolt}), hard (\numrange[range-phrase=--]{2.0}{7.0}\,\unit{\kilo\electronvolt}), 
medium (\numrange[range-phrase=--]{1.2}{2.0}\,\unit{\kilo\electronvolt}), 
soft (\numrange[range-phrase=--]{0.5}{1.2}\,\unit{\kilo\electronvolt}), and 
ultra-soft (\numrange[range-phrase=--]{0.2}{0.5}\,\unit{\kilo\electronvolt}). Meanwhile, HRC detections cover the whole energy interval of the detector and their observational properties are only estimated on the entire energy band (spectral \textit{features} are not available for HRC detections).

In the catalog, X-ray sources are initially detected within individual observations~\citep{2024ApJS..274...22E}. When 
multiple observations of the same region overlap and their aimpoints are all within $60\arcsec$ from each other, the CSC 
merges these data into 'stacks'. 'Stack-level' detections are also determined independently from the single 
observation-level detections found in each separate observation included in the stack. 
The properties of the spatially matching, stack-level detections are used to characterize master sources. 
These master sources provide the most accurate characterization of the actual astrophysical entity 
responsible for the observed X-ray emission. Their properties are derived from the best groups of observations 
('blocks'). 
For aperture photometry-derived quantities, the best estimates are typically determined from the flux-ordered Bayesian block with the highest exposure time across multiple observations. While the characterization of the observed properties of 
master sources and stack-level detections is an interesting endeavor, in this paper we will focus on 
observation-level detections because they provide the most granular representation of the changing 
behavior of variable sources.

The CSC 2.1.1 contains a large number of master sources and observation-level detections, which support a variety of data-driven investigations by allowing the exploration of an extensive range of source- and detection-level behaviors. The characterization of these X-ray sources can be performed by studying the distribution of several \textit{features}, including estimates of variability 
between and within observations, spectral characteristics, positional localization, source morphology, statistical significance, multi-band flux measurements, hardness ratios, and variability indices. Additionally, the catalog provides various data products, such as full-field images, linked directly to 
archival observations and existing literature.

\subsection{Feature Selection and Data Cleaning}
\label{subsec:dataset}

For our analysis, we focused on \textit{features} primarily related to spectral properties and variability, as these are fundamental in characterizing the nature of X-ray sources. Since spectral parameters are derived from single-observation detections using ACIS, our study is restricted to ACIS-based observation-level data, excluding HRC observations and stack-level detections. The following are the selected input \textit{features}:

\begin{itemize}
    \item \textbf{Hardness Ratios} (\texttt{hard\_*}): Hard-Medium (\texttt{hm}), Hard-Soft (\texttt{hs}), Medium-Soft (\texttt{ms})
    \item \textbf{Spectral Properties:} Blackbody temperature (\texttt{bb\_kt}), Power-law index (\texttt{powlaw\_gamma})
    \item \textbf{Variability Metrics:}  
        \begin{itemize}
            \item Probabilities (\texttt{var\_prob\_*}): Broad (\texttt{b}), Hard (\texttt{h}), Soft (\texttt{s})
            \item Ratios (\texttt{var\_ratio\_*}): Broad (\texttt{b}), Hard (\texttt{h}), Soft (\texttt{s})
        \end{itemize}
\end{itemize}

The detailed descriptions of these \textit{features} are presented in Table~\ref{tab:som_features}. The selection prioritizes distance-independent parameters, as opposed to flux values, which depend on the distance of the source, unknown for the overwhelming majority of sources in our dataset.

To ensure data quality, filtering criteria were applied during preprocessing. Only observation-level detections with a \texttt{flux\_significance\_broad} exceeding 5 and non-null values for all selected \textit{features} were retained. After applying these constraints, the final dataset consisted of 36,904 valid detections, corresponding to 17,927 distinct sources. This preprocessing step guarantees that the extracted statistical properties remain reliable and meaningful for subsequent analyses. The mean fractional error on all \textit{features} (computed as the ratio of the reported uncertainty to the measurement value, expressed as a percentage) 
is smaller than 10\% for $\sim\!95\%$ of the detections 
in the sample used to train the model. For this reason, we expect the effect of the measurement
errors on the final trained SOM model
to be small. Even assuming large shifts from the expected \textit{feature} values for the remaining detections, the mapping of the original dataset to 
the SOM map would not deviate significantly from the results presented in this paper on the scale of the whole SOM map, and lead to different mapping only locally.
All experiments with the SOM model described in subsequent sections use this same dataset, which we refer to as the ACIS-SV-CSC dataset. This dataset includes all CSC sources meeting the above criteria, focusing on spectral and variability measures (SV).

\begin{deluxetable}{lccc}
    \tablecaption{Summary of selected \textit{features}. Hardness ratios: * = hm (Hard-Medium), hs (Hard-Soft), ms (Medium-Soft). Energy bands: ** = b (Broad), h (Hard), s (Soft). Descriptions are from the CSC documentation.}
    \label{tab:som_features}
    \tablehead{
        \colhead{Property Name} & 
        \colhead{Description} & 
        \colhead{Log Trans.\tablenotemark{a}} & 
        \colhead{Norm.\tablenotemark{b}}
    }
    \startdata
    \rule{0pt}{4em}\texttt{hard\_*} & \parbox{3cm}{Hardness ratio: HR\textsubscript{xy} = \(\frac{F(x) - F(y)}{F(x) + F(y)}\), where \(F(x)\) and \(F(y)\) are aperture fluxes for two energy bands.} & -- & -- \\\relax
    \rule{0pt}{3.5em}\texttt{bb\_kt} & \parbox{3cm}{Temperature (kT) of the best-fitting absorbed black-body model spectrum.} & Yes & Yes \\\relax
    \rule{0pt}{3.5em}\texttt{powlaw\_gamma} & \parbox{3cm}{Photon index of the best-fitting absorbed power-law model spectrum.} & -- & Yes \\\relax
    \rule{0pt}{3.5em}\texttt{var\_prob\_**} & \parbox{3cm}{Probability that the count rate lightcurve varies across time bins.} & -- & -- \\\relax
    \rule{0pt}{3.5em}\texttt{var\_ratio\_**} & \parbox{3cm}{Ratio of flux variability standard deviation (\texttt{var\_sigma}) to its mean (\texttt{var\_mean}).} & Yes & Yes \\\relax
    \enddata
    \tablecomments{
        (a): Indicates whether a logarithmic transformation is applied (Yes or --);
        (b): Indicates whether normalization is applied (Yes or --).
    }
\end{deluxetable}

\subsection{SIMBAD cross-match}
\label{subsec:simbad}

The nature of X-ray sources detected by \textit{Chandra} in the CSC cannot always be exclusively determined based on
their X-ray properties. 
For this reason, we performed a cross-match with the SIMBAD database~\citep{2000A&AS..143....9W} to identify likely
multi-wavelength counterparts to CSC sources and take advantage of this information to interpret 
the results (Section~\ref{sec:results}) of the SOM analysis presented in this paper. 
SIMBAD is one of the most extensive 
repositories of classified astronomical objects: it compiles multi-wavelength classifications and general observational 
characteristics, aggregating data from published astrophysical literature for community access.

We performed a positional cross-match between our dataset and SIMBAD using a $1\arcsec$ matching radius. 
Each independent observation-level detection was assigned the sky coordinates of its corresponding master source, ensuring 
that all detections of the same astrophysical object were consistently linked to a single counterpart. To categorize the 
matched sources, we utilized the \texttt{main\_type} field in SIMBAD, which follows a hierarchical classification system. 
This system prioritizes the intrinsic physical nature of objects rather than their emissions in specific spectral regions 
(e.g., Radio, IR, Red, Blue, UV, X-ray, or gamma).

\begin{figure*}[t]
    \includegraphics[]{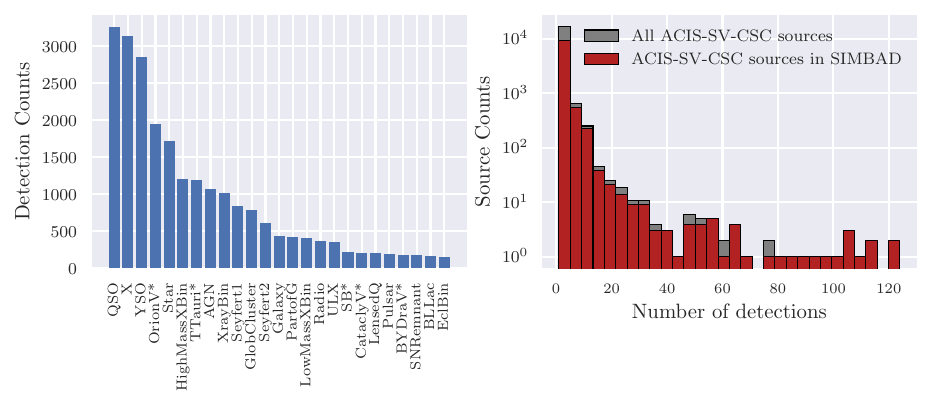}
    \caption{Left: Histogram of the 25 most frequently cross-matched object classes from SIMBAD in the ACIS-SV-CSC dataset, ranked by detection count. Right: Histograms of the number of detections of all sources 
    in the ACIS-SV-CSC dataset (gray) and the ACIS-SV-CSC sources cross-matched with SIMBAD (red).}
    \label{fig:simbad_combined}
\end{figure*}

We successfully matched 25,192 out of 36,904 detections ($\sim\!68.2\%$) and 10,078 out of 17,927 sources ($\sim\!56.2\%$) 
in the ACIS-SV-CSC dataset with SIMBAD entries. The distribution of the 25 most frequently matched source types 
is visualized in the left plot of Figure~\ref{fig:simbad_combined}, while the histogram of the number of detections for all sources in the ACIS-SV-CSC dataset (gray) and the subset of sources matched with a SIMBAD record (red) is shown in the right plot.

The \texttt{main\_type} categories span a variety of different classes of astrophysical sources, 
and they are not all at the same granularity level: some classes are broad (like \texttt{QSO}, \texttt{Star}), 
while other classes are more specific (\texttt{YSO}, \texttt{HMXB}), and overlaps may occur. 
The two labels \texttt{PartOfG} (Part of a Galaxy) and \texttt{X} are vague and offer little detail about the nature of the sources 
they are associated with: the \texttt{PartOfG} \texttt{main\_type} is attributed to unknown sources that are located in the 
vicinity of a known galaxy, while \texttt{X} labels are attached to sources whose only known characteristic is to be detected in the X-rays. These uninformative labels reflect the inherent inhomogeneous and incomplete nature of the classifications collected by SIMBAD from the scientific literature.

Below, we provide basic information on the twenty-one most common \texttt{main\_type} classes
sorted according to decreasing number of detections. These classes span a broad range of astrophysical 
objects and, together, they account for $90\%$ of the total detections included in the dataset, 
ensuring a robust analysis of the trained SOM model.

\begin{itemize}
    \item \textbf{\texttt{QSO} (Quasar)}: Extremely bright point-like X-ray sources, a subset of the active galactic 
    nuclei, usually located in the center of galaxies, fueled by supermassive black holes (3,259 detections).
    \item \textbf{\texttt{X} (X-ray Source)}: Generic (i.e., unclassified) objects with an X-ray counterpart (3,141 detections).
    \item \textbf{\texttt{YSO} (Young Stellar Object)}:  Early-stage stars still evolving (2,856 detections).
    \item \textbf{\texttt{OrionV*} (Orion variable)}: Young stars that exhibit irregular and eruptive
    variations in their luminosity, usually associated with diffuse nebulae and named after the Orion constellation where
    most of these objects have been discovered (1,955 detections).
    \item \textbf{\texttt{Star}}: Self-luminous bodies driven by core nuclear fusion (1,723 detections).
    \item \textbf{\texttt{HMXB} (High-Mass X-ray Binary)}: Binary stellar systems where a usually 
    massive star is coupled with a compact object (e.g., neutron star or black hole), emitting X-rays 
    from accreting material (1,213 detections).
    \item \textbf{\texttt{TTauri*} (T Tauri stars)}: Orion variable stars that are also characterized
    by fluorescent violet emission lines in their spectra (1,197 detections).
    \item \textbf{\texttt{AGN} (Active Galactic Nucleus)}: Central region of a galaxy characterized
    by bright multi-wavelength emission from a supermassive black hole accreting gas (1,077 detections).
    \item \textbf{\texttt{XRayBin} (X-ray Binary Stars)}: Binary stellar systems where a 
    star whose mass is not known is coupled with a compact object (e.g., neutron star or black hole), emitting X-rays 
    from accreting material (1,023 detections).
    \item \textbf{\texttt{Seyfert1} (Seyfert 1 Galaxies)}: Active galaxies with a very bright, compact 
    nucleus associated with an accreting supermassive black hole. Their optical spectra, containing bright broad 
    and narrow emission lines, suggest an unobstructed view of the central AGN and face-on host galaxy morphology 
    (840 detections).
    \item \textbf{\texttt{GlobClu} (Globular Clusters)}: Compact, dense stellar systems composed by $\sim\!10^4$
    to $\sim\!10^6$ stars that are bound together by gravity (788 detections).
    \item \textbf{\texttt{Seyfert2} (Seyfert 2 Galaxies)}: Active galaxies with a very bright, compact 
    nuclear region whose emission is caused by an accreting supermassive black hole. Their 
    optical spectra show only strong, narrow emission lines (unlike Seyfert 1 galaxies), indicative of a partially
    obscured view of the central AGN and edge-on host galaxy morphology (611 detections).    
    \item \textbf{\texttt{Galaxy} (Galaxy)}: Generic (i.e., unclassified) objects spatially matching the position 
    of a galaxy (431 detections).
    \item \textbf{\texttt{PartOfG} (Part of a Galaxy)}: Generic labels of objects spatially associated with a 
    galaxy but lacking a more specific classification (427 detections).
    \item \textbf{\texttt{LMXB} (Low-Mass X-ray Binary)}: Binary stellar systems where a low-mass 
    star is coupled with a compact object (e.g., neutron star or black hole), emitting X-rays 
    from accreting material (407 detections).
    \item \textbf{\texttt{Radio} (Radio Source)}: Generic (i.e., unclassified) objects with a radio 
    counterpart (371 detections).
    \item \textbf{\texttt{ULX} (Ultra-Luminous X-ray Source)}: Non-nuclear X-ray sources whose luminosity
    can exceed $10^{39}$ erg/s, likely associated with compact objects accreting matter to super-critical levels (352 detections).    
    \item \textbf{\texttt{SB*} (Spectroscopic Binary)}: Binary stellar systems with very small angular separation
    between the two stars, only confirmed via the detection of periodic Doppler shifting of spectral lines caused by relative
    velocity of the stars (215 detections).   
    \item \textbf{\texttt{CataclyV*} (Cataclysmic Binary)}: Binary stellar systems undergoing irregular, 
    significant changes in their brightness, caused by instability in the accretion disk from the donor to the 
    accretor star (213 detections).  
    \item \textbf{\texttt{LensedQ} (Gravitationally Lensed Image of a Galaxy)}: Images of a background
    galaxy (lensed objects) produced by an intervening lens (another galaxy or a cluster of galaxies) via gravitational 
    lensing (206 detections).
    \item \textbf{\texttt{Pulsar} (Pulsar)}: Rapidly rotating highly magnetized neutron stars that emit
    electromagnetic radiation from its magnetic poles in highly regular, periodic patterns (190 detections).       
\end{itemize}

As evident from the definitions reported above, the SIMBAD \texttt{main\_type} field mixes broad 
and fine-grained classes, as well as classes that do not provide information about the nature 
of the sources but only characterize them based on one specific, heterogeneous property (i.e., 
emitting in the X-rays or radio band, and/or being located close to a known galaxy). 
This heterogeneous classification, caused by the serendipitous nature of both 
the CSC catalog and the mechanism employed by SIMBAD to automatically harvest classes for sources 
reported in astrophysical literature, is not ideal for the detection-level classification we present later 
(Section~\ref{subsec:classification}). To ensure that the labels used to identify
the type of source that typically populate different regions of the SOM map are comparable in hierarchical specificity 
and sufficiently populated to support stable neighborhood pooling on the SOM, we aggregate fifteen 
of the original \texttt{main\_type} labels (with the exclusion of the generic \texttt{PartOfG} and 
\texttt{X}) into four target classes defined at a similar generality 
level as follows:

\begin{itemize}
  \item \textbf{\texttt{neoYSOs}}: \texttt{YSO} + \texttt{OrionV*} + \texttt{TTauri*} (6,008 detections)
  \item \textbf{\texttt{Binaries}}: \texttt{HMXB} + \texttt{XRayBin} + \texttt{LMXB} + \texttt{CataclyV*} + 
  \texttt{SB*} + \texttt{BYDrav*} (3,250 detections)
  \item \textbf{\texttt{Stars}}: \texttt{Star} (1,723 detections)
  \item \textbf{\texttt{neoAGNs}}: \texttt{AGN} + \texttt{QSO} + \texttt{Seyfert1} + \texttt{Seyfert2} + \texttt{BLLAc} (5,957 detections)
\end{itemize}

This consolidation achieves more similar hierarchical levels and balanced sizes for each aggregated label, 
while merging \texttt{main\_type} of similar physical systems. Although~\texttt{neoYSOs} 
and~\texttt{Binaries} are specific stellar subclasses and should be incorporated in the more general
class~\texttt{Stars}, they contain sufficiently specific and widely represented X-ray sources that
cannot be merged with the generic sample of stellar source labeled as~\texttt{Stars}. Some overlap 
is expected between \texttt{Stars} and \texttt{neoYSOs}, which we consider when interpreting the results. 
The~\texttt{Binaries} aggregated class, in particular, is intrinsically heterogeneous as it 
spans multiple types of binary stellar systems characterized by a variety of emission mechanisms. This heterogeneity translates to a wide range of values of the \textit{features} used to 
investigate the parameter space distribution of the ACIS-SV-CSC dataset.

\subsection{Pre-processing}
\label{subsec:preprocessing}

To ensure consistency across numerical properties with significantly different value ranges, the ACIS-SV-CSC dataset 
underwent pre-processing steps that included logarithmic transformations and normalization. As summarized in 
Table~\ref{tab:som_features}, the \textit{features} \texttt{bb\_kt} and \texttt{var\_ratio\_*} were both log-transformed 
and normalized. Conversely, \texttt{powlaw\_gamma} was solely normalized, while~\texttt{hard\_ms},~\texttt{hard\_hm},~\texttt{hard\_hs}, 
and~\texttt{var\_prob\_*} remained in their original form without any transformations or normalization.

\subsubsection{Logarithmic Transformation and Normalization}

\begin{itemize}
    \item The \textbf{logarithmic transformation} was used to reduce asymmetry in the distributions of 
    specific properties, namely \texttt{bb\_kt} and \texttt{var\_ratio\_*} (where * represents energy bands: b, h, s). 
    This transformation helps mitigate the influence of extreme values by converting the data to a logarithmic scale. 
    The natural logarithm function from \texttt{numpy}~\citep{harris2020array} was applied as follows:

        \[
            x_{\text{log}} = \log(x)
        \]

where:
\begin{itemize}
    \item \( x_{\text{log}} \) is the logarithmically transformed value.
    \item \( x \) represents the original \textit{feature} values.
\end{itemize}

To handle zero values, we replaced them with $x_0$, defined as one-tenth of the smallest non-zero value in the respective column:

\[
x_{0} = \frac{1}{10}\, \min\bigl\{\, x_i : x_i > 0 \,\bigr\}
\]

\item We applied \textbf{normalization}, using the \texttt{MinMaxScaler} function from the \texttt{sklearn} library~\citep{2011JMLR...12.2825P}. This scaling maps the data to a fixed range $[0,\,1]$, making comparisons between \textit{features} more straightforward. The normalization formula is:

\[
x_{\text{norm}} = \frac{x - x_{\text{min}}}{x_{\text{max}} - x_{\text{min}}}
\]

where:
\begin{itemize}
    \item \( x_{\text{norm}} \) is the normalized value.
    \item \( x \) is the logarithmically transformed \textit{feature} value.
    \item \( x_{\text{min}} \) and \( x_{\text{max}} \) are the minimum and maximum values of the feature, respectively.
\end{itemize}

Normalization was applied to \texttt{bb\_kt} and \texttt{var\_ratio\_*} after logarithmic transformation, and directly (without log transformation) to \texttt{powlaw\_gamma}.
\end{itemize}

\subsection{Validating the dimensionality of the \textit{feature} space}
\begin{figure}[t]
    \begin{center}
    \includegraphics[width=\linewidth]{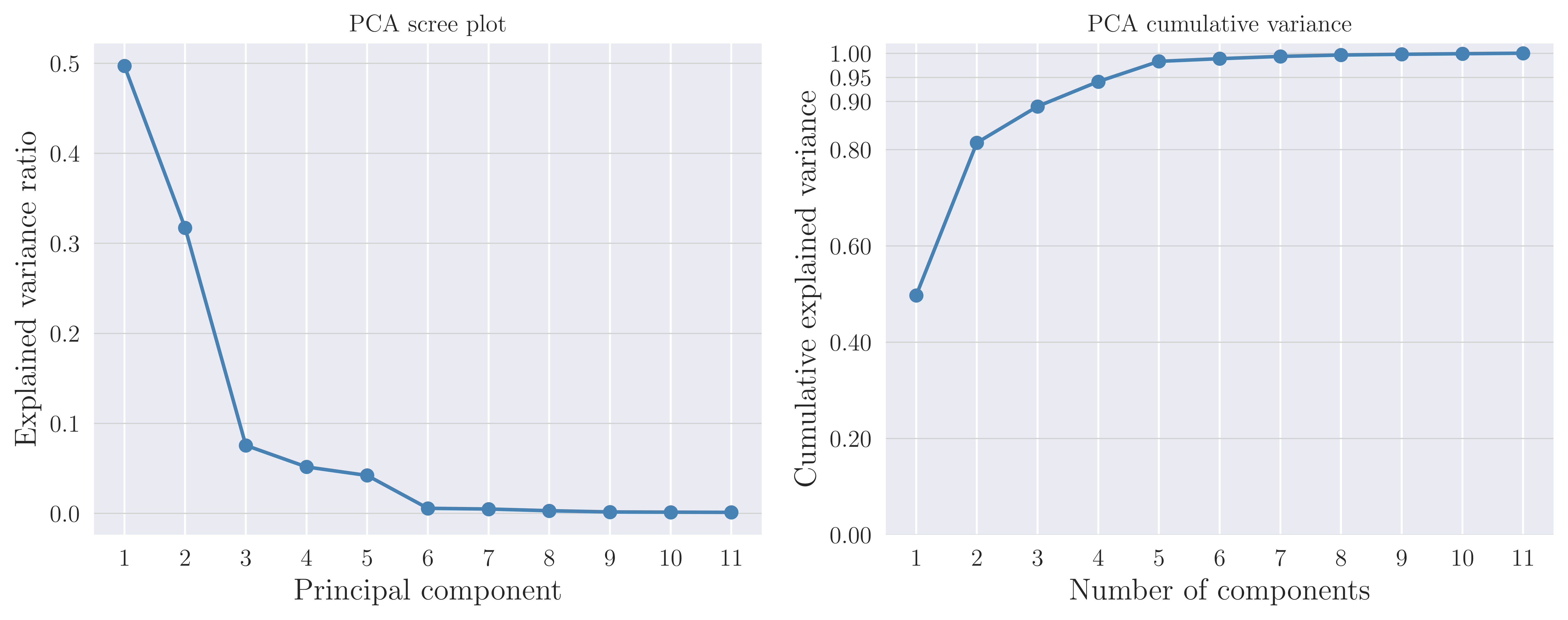}
    \caption{Principal component analysis of the 11-dimensional SOM input space.
     \textit{Left:} explained variance ratio per principal component. \textit{Right:}
    cumulative explained variance.}
    {\label{fig:step1_pca_scree_cumulative}}
    \end{center} 
\end{figure}

The eleven input \textit{features} that generate the original parameter space and the geometric properties of the input dataset in such space determine how faithfully a two-dimensional projection such as the SOM can preserve the original topology. To quantify this, we performed a
principal component analysis (PCA) of the 11-dimensional pre-processed \textit{features}. Figure~\ref{fig:step1_pca_scree_cumulative} shows the
resulting scree and cumulative variance curves (left and right figures, 
respectively).

The first two principal components account for $81.4\%$ of the total variance.
The linear dimensionality of the input space is therefore
larger than two, and a two-dimensional SOM cannot preserve the full topology of the 11-dimensional distribution. The variance is,
however, dominated by two physically meaningful axes, whose linear combinations
of the original \textit{features} are listed in Table~\ref{tab:pca_loadings}:
PC1 ($49.7\%$) is an overall spectral-hardness axis, driven by the three
hardness ratios, which together contribute $97.4\%$ of its variance; PC2
($31.7\%$) is an overall variability axis, driven by the three variability
probabilities, which contribute $97.2\%$ of its variance. These loadings indicate correlated structure within these feature groups, as expected from their related definitions.

\begin{deluxetable}{lcc}
    \tablecaption{Loadings of the eleven \textit{features} on the
    two dominant principal components of the SOM input space, rounded to three decimal places.}
    \label{tab:pca_loadings}
    \tablehead{
        \colhead{\textit{Feature}} &
        \colhead{PC1} &
        \colhead{PC2}
    }
    \startdata
    \texttt{hard\_hm}              & $\mathbf{+0.417}$ & $-0.032$ \\
    \texttt{hard\_hs}              & $\mathbf{+0.753}$ & $+0.023$ \\
    \texttt{hard\_ms}              & $\mathbf{+0.481}$ & $+0.061$ \\
    \texttt{var\_prob\_b}          & $-0.017$ & $\mathbf{+0.669}$ \\
    \texttt{var\_prob\_s}          & $+0.063$ & $\mathbf{+0.416}$ \\
    \texttt{var\_prob\_h}          & $-0.074$ & $\mathbf{+0.592}$ \\
    \texttt{bb\_kt}     & $+0.092$ & $-0.008$ \\
    \texttt{powlaw\_gamma} & $-0.085$ & $+0.023$ \\
    \texttt{var\_ratio\_b} & $-0.007$ & $+0.111$ \\
    \texttt{var\_ratio\_h} & $-0.025$ & $+0.083$ \\
    \texttt{var\_ratio\_s} & $+0.019$ & $+0.060$
    \enddata
\end{deluxetable}

To test whether the spatial structure of the input data can nonetheless be represented on two axes, we used the Uniform Manifold Approximation and Projection (UMAP)~\citep{2018arXiv180203426M}
model. UMAP is a non-linear manifold-learning method that performs dimensionality reduction by approximating the manifold underlying the distribution of detections in the \textit{feature} space with a fuzzy topological structure. Similarly to SOM, the low-D embedding is produced by UMAP through a low dimensional projection 
of the data with fuzzy topological structure as close as possible to that of the dataset in the
input space. In order to assess how well a two-dimensional representation preserves the topological structure of our dataset, we apply UMAP to the first 
three ($k=3$, accounting for $89.0\%$ of
the total variance) and the first four ($k=4$, accounting for $94.1\%$ of the total variance) PC scores, 
and evaluate the resulting two-dimensional embeddings with the trustworthiness and continuity metrics.
For all configurations considered (both $k$ values and neighborhood sizes of 15
and 30), both metrics remain above approximately $0.95$ at every evaluation scale, indicating
that the two-dimensional nonlinear embeddings preserve the local structure of
the \textit{feature} space with high fidelity. This supports the practical use
of a two-dimensional SOM as an accurate and interpretable low-dimensional representation 
of the general topological structure of the input distribution used in this paper in its \textit{feature} space.

\section{Self-Organizing Map Approach}
\label{sec:method}

This section outlines the unsupervised machine learning technique applied in our study. We utilize SOMs to analyze and extract patterns from a subset of the X-ray detections in the CSC 2.1.1 catalog. SOMs are a type of 
artificial neural network~\citep{2001som..book.....K} designed for dimensionality reduction, mapping high-dimensional datasets 
onto a structured two-dimensional grid while maintaining the local spatial relationships between data points. This characteristic makes them particularly useful for visualization and interpretation. For implementation, we used the MiniSom Python 
library~\citep{vettigliminisom}, which provides a lightweight yet effective framework for training and visualizing SOMs. 
The strengths of SOMs that make them ideal for our analysis include:

\begin{itemize}
    \item \textbf{Unsupervised Learning:} SOMs categorize similar input vectors into nearby neurons without the need for labeled data. This structure provides a similarity-based mapping that can be used to group similar vectors based on their \textit{features} and determine clusters in a lower-dimensional space.
    \item \textbf{Classification:} Once trained, a SOM can map new, unseen data points to their nearest neuron, enabling label transfer from known samples.
    \item \textbf{Dimensionality reduction:} By preserving the local topological relationships of the high-dimensional \textit{feature} space, SOMs enable intuitive visual representation, making it easier to explore and interpret the data.
\end{itemize}

\subsection{How SOMs Work}

The training process consists of iterative steps that gradually refine the map to reflect the structure of the 
input data. Initially, input vectors are presented to the SOM, where each neuron is associated with a weight 
vector of the same dimensionality. Throughout this paper, we use the terms neuron and cell interchangeably to refer to the individual units of the SOM grid. The algorithm identifies the Best-Matching Unit (BMU), the neuron whose 
weight vector is closest to the input vector, using a similarity metric, typically Euclidean distance.

Once the BMU is identified, the weight vectors of the BMU and its neighboring neurons are updated by the following 
equation to move closer to the input vector:

\[
W(t + 1) = W(t) + \theta(t) \cdot \alpha(t) \cdot (V - W(t))
\]

where:
\begin{itemize}
    \item \(W(t)\) represents the weight vector at iteration \(t\),
    \item \(\theta(t)\) is the neighborhood function, which controls how much nearby neurons adjust,
    \item \(\alpha(t)\) is the learning rate, which decreases over time to refine the adjustments,
    \item \(V\) is the input vector.
\end{itemize}
This process produces a structured 2D representation in which similar data points tend to be placed closer together.

\section{Model Training}
\label{sec:training}

The initial phase of training involved optimizing the SOM hyperparameters by testing various 
configurations to minimize both the quantization error, a metric of data representation accuracy, and the topographic 
error, a measure of topology preservation.  
The \textbf{quantization error} (\( E_q \)) quantifies the average distance between each input vector and its corresponding BMU. A lower \( E_q \) indicates that the SOM effectively represents the input data in a reduced-dimensional space. It is defined as:
\[
E_q = \frac{1}{N} \sum_{i=1}^{N} \| \mathbf{x}_i - \mathbf{w}_{\text{BMU}(\mathbf{x}_i)} \|
\]

where:
\begin{itemize}
    \item \( N \) is the total number of input vectors,
    \item \( \mathbf{x}_i \) is the \(i\)-th input vector,
    \item \( \mathbf{w}_{\text{BMU}(\mathbf{x}_i)} \) is the weight vector of the Best-Matching Unit (BMU) corresponding to the input vector \( \mathbf{x}_i \),
    \item \( \| \cdot \| \) represents the Euclidean distance between the input vector and the BMU.
\end{itemize}

The \textbf{topographic error} (\( E_t \)) evaluates how well the SOM preserves the topological relationships 
among input vectors, calculated as the proportion of vectors whose first and second BMUs are not adjacent on 
the SOM grid. A smaller \( E_t \) reflects better topology preservation and is defined as:

\[
E_t = \frac{1}{N} \sum_{i=1}^{N} e_t(\mathbf{x}_i)
\]

where:

\[
e_t(\mathbf{x}_i) = 
\left\{
\begin{array}{ll} 
1, & \text{if BMU, 2nd-BMU non-adjacent for } \mathbf{x}_i \\ 
0, & \text{if BMU, 2nd-BMU adjacent for } \mathbf{x}_i 
\end{array}
\right.
\]
The most effective SOM training results for the ACIS-SV-CSC dataset were achieved by training the model with the following parameters, reaching a quantization error of 0.164 and a topographic error of 0.037:
\begin{itemize}
    \item \textbf{Map Dimension}: $40\times40$. Although the heuristic method suggested by~\citet{vesanto2000som} would recommend a $31\times31$ grid ($\sim\!960$ neurons) for a dataset of 36,904 entries, we opted for a slightly larger $40\times40$ grid (1,600 neurons) based on the results of a grid search evaluation. This increase improved the ability of the model to capture intricate structures in the data while maintaining topological coherence.
    \item \textbf{Neighborhood Function}: Gaussian. A smooth Gaussian kernel (initial $\sigma = 4$) was chosen to balance global ordering and local refinement.
    \item \textbf{Learning Rate}: 0.04 was selected to effectively balance both quantization and topographic errors.
    \item \textbf{Topology}: Hexagonal. The hexagonal structure was chosen over a rectangular grid due to its enhanced ability to preserve topological relationships. Each neuron in a hexagonal grid is surrounded by six neighbors, facilitating a more natural representation of the input space.
    \item \textbf{Iterations}: 1,000,000 ($\sim\!27$ epochs). To evaluate convergence, we monitored the quantization and topographic errors continuously throughout the training process. While the topographic error stabilized rapidly during the initial iterations, the quantization error exhibited a steady but decelerating decrease. Training was halted at 1,000,000 iterations because the error curve plateaued. At this stage, the reduction in error became negligible, indicating that the map had fully stabilized.
\end{itemize}

The SOM model was trained using our interactive web application, which also provides all the visualization 
tools used to generate the results and figures presented in the following sections (see Appendix~\ref{sec:webapp} for details).

\section{Results}
\label{sec:results}

\begin{figure}[htbp]
    \begin{center}
    \includegraphics[width=0.48\textwidth]{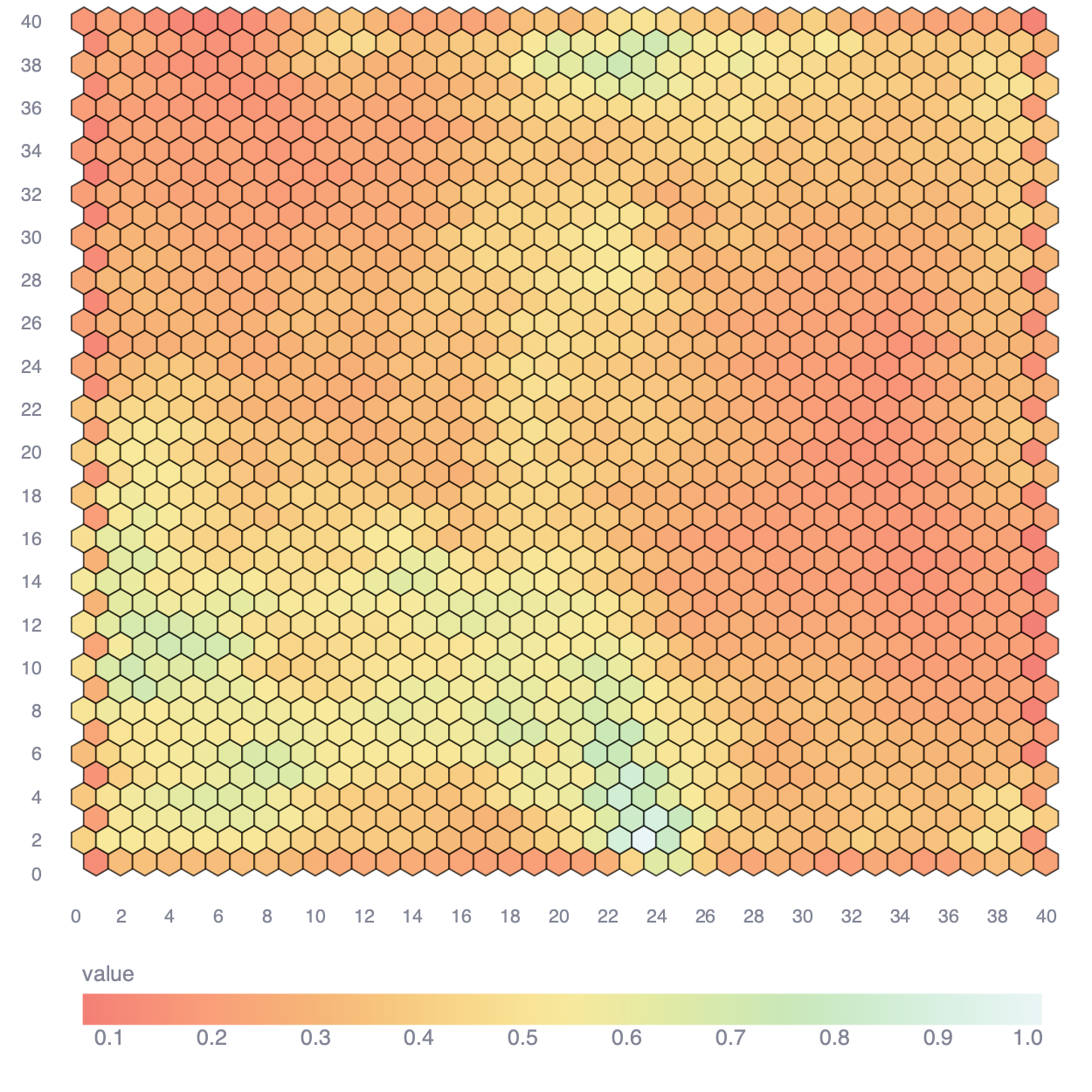}
    \includegraphics[width=0.48\textwidth]{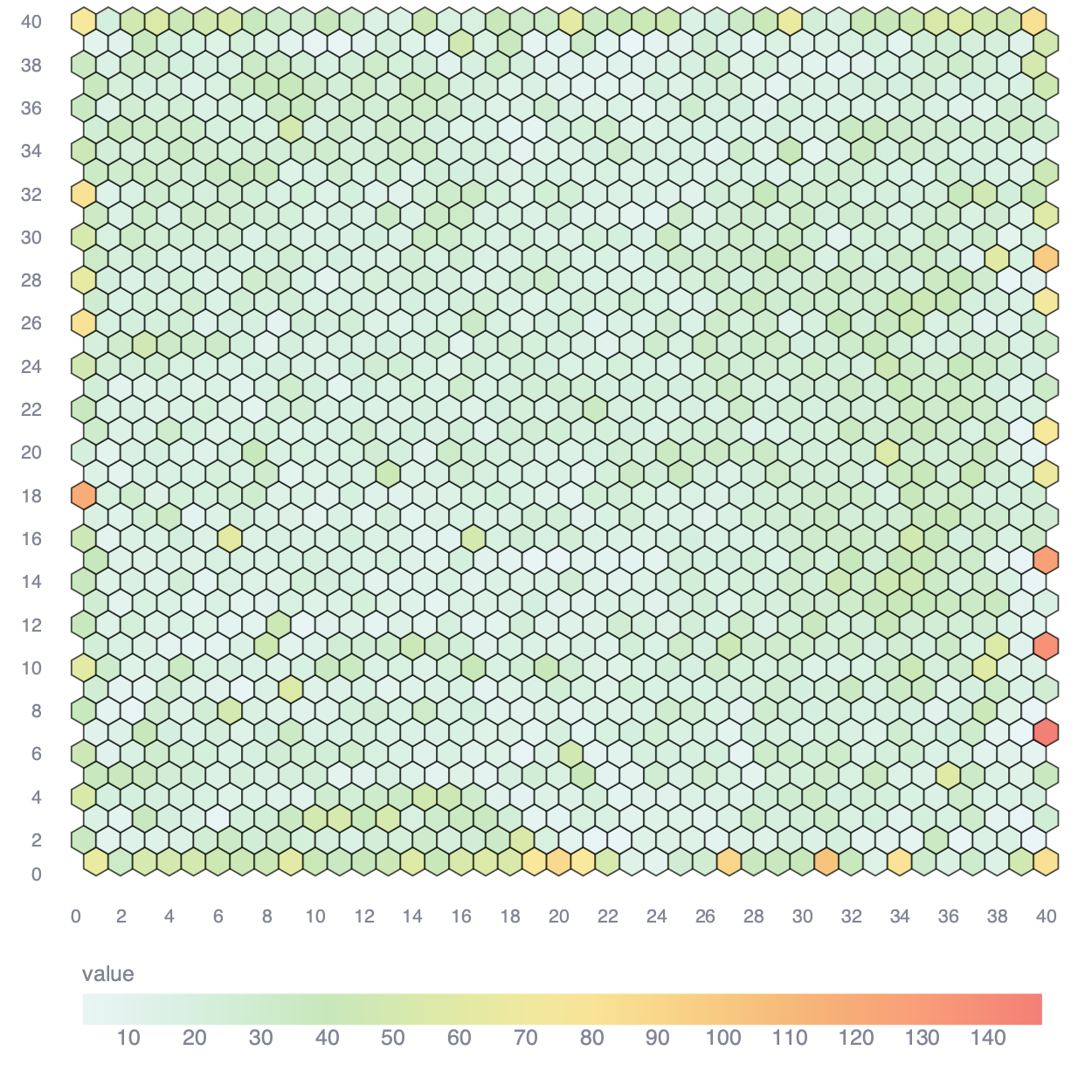}
    \caption{
        (Left) U-Matrix of the trained SOM, showing the average distance of each neuron from all neighboring neurons with a color gradient (red to green). 
        (Right) Activation Response, indicating the number of data elements (in this case, single CSC detections used to train the
        SOM) mapped to each neuron, with higher values highlighting more frequently activated neurons.
    }    
    \label{fig:som_analysis}
    \end{center}
\end{figure}

The SOM provides a powerful visualization of the underlying structure of the \textit{feature} space 
distribution of CSC detections. Figure~\ref{fig:som_analysis} shows two key visualizations of the result of training of 
the SOM model (Section~\ref{sec:training}) on the ACIS-SV-CSC dataset (Section~\ref{subsec:dataset}). The left panel of 
Figure~\ref{fig:som_analysis} displays the U-Matrix, i.e., the map of the average normalized distance between each neuron and all 
its adjacent neurons, that reveals the inner structures of the high-dimensionality \textit{feature} space mapped into 
the two-dimensional 
SOM grid. The right panel of Figure~\ref{fig:som_analysis} 
shows the activation response map, i.e., the number of data samples mapped to each neuron. These figures together illustrate both 
the emergent topological structure and the density of the data projections on the trained SOM grid. This specific U-Matrix 
is characterized by an extended region of large average distance between cells located 
in the lower-left quadrant of the map, with additional, smaller areas of increased inter-neuron distance 
in the top and along the right margin of the SOM grid. The activation response shows that neuron occupancy varies
between 10 and 30 with an average of 21.2 detections per cell when neurons on the margins are excluded. In the margins, 
the average 
occupancy grows to $\sim\!41$ detections/neuron, hinting at potential boundary effects, where edge neurons
have fewer neighbors, receive asymmetric neighborhood updates, and therefore come to represent a larger area that 
attracts more samples.

This section presents a series of astrophysical applications of the SOM model trained on the ACIS-SV-CSC dataset. While different in nature, these applications all highlight the ability of this model to reveal patterns and properties of X-ray sources. In particular, we first examine the \textit{feature} space structure captured by the trained SOM (Section~\ref{subsec:feature_space_mapping}), then analyze how available classes are distributed on the SOM grid (Section~\ref{subsec:labels}). Next, we perform detection-level classification with a SOM retrained on the same \textit{features} but restricted to the training and unlabeled subsets, holding out the validation and test splits to avoid data leakage (Section~\ref{subsec:classification}).

\begin{figure}[t]
    \begin{center}
    \includegraphics[width=0.73\linewidth]{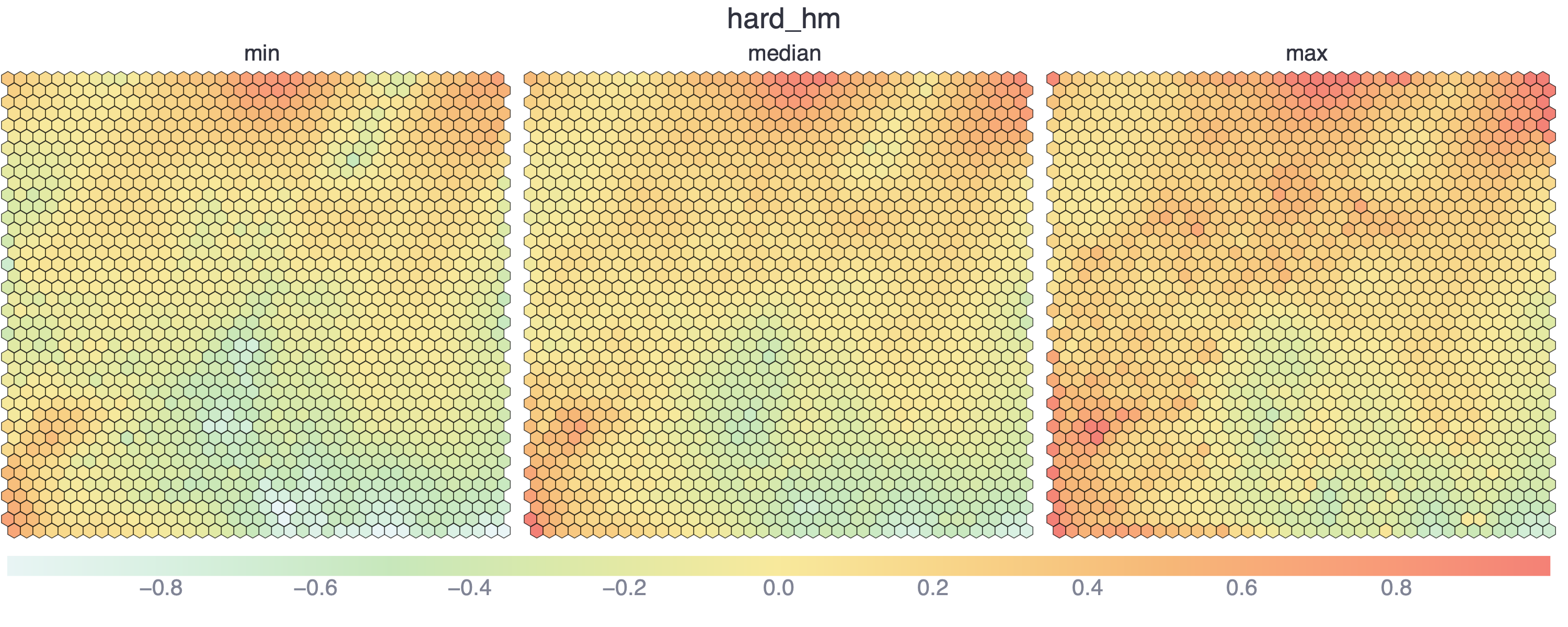}
    \includegraphics[width=0.73\linewidth]{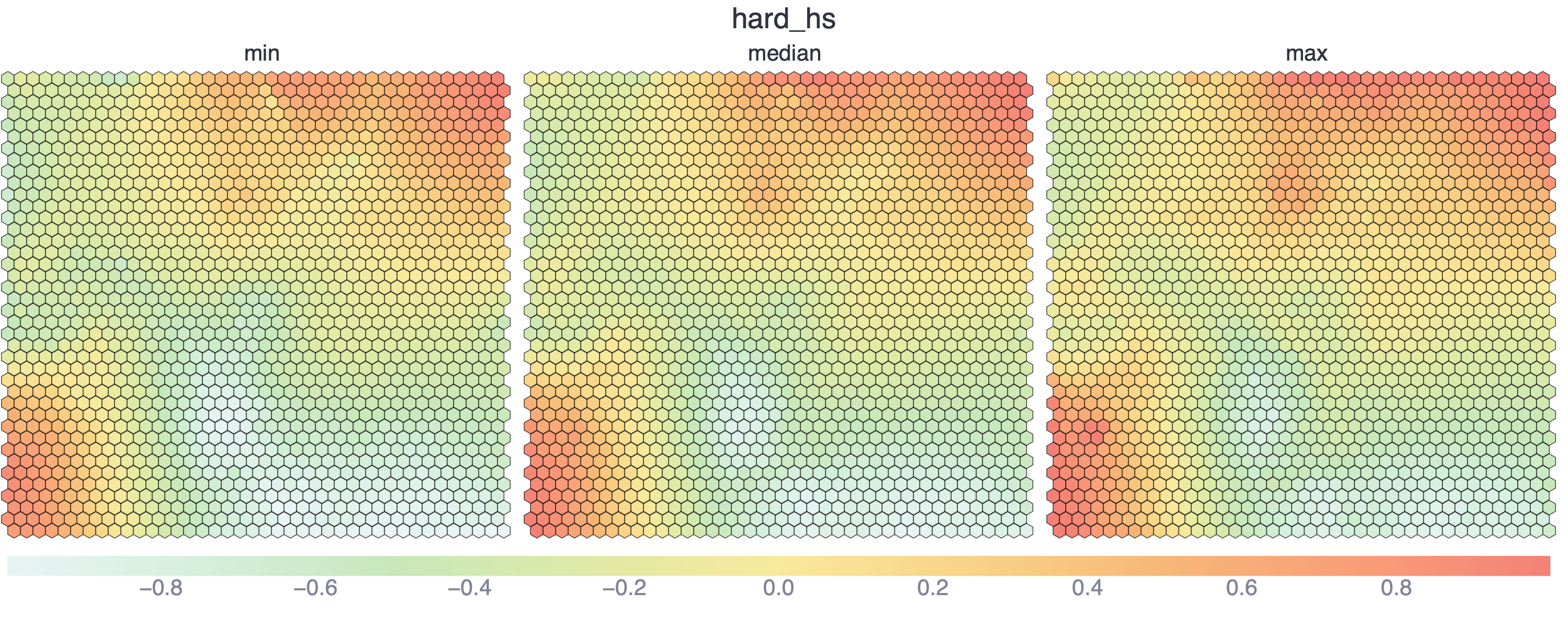}
    \includegraphics[width=0.73\linewidth]{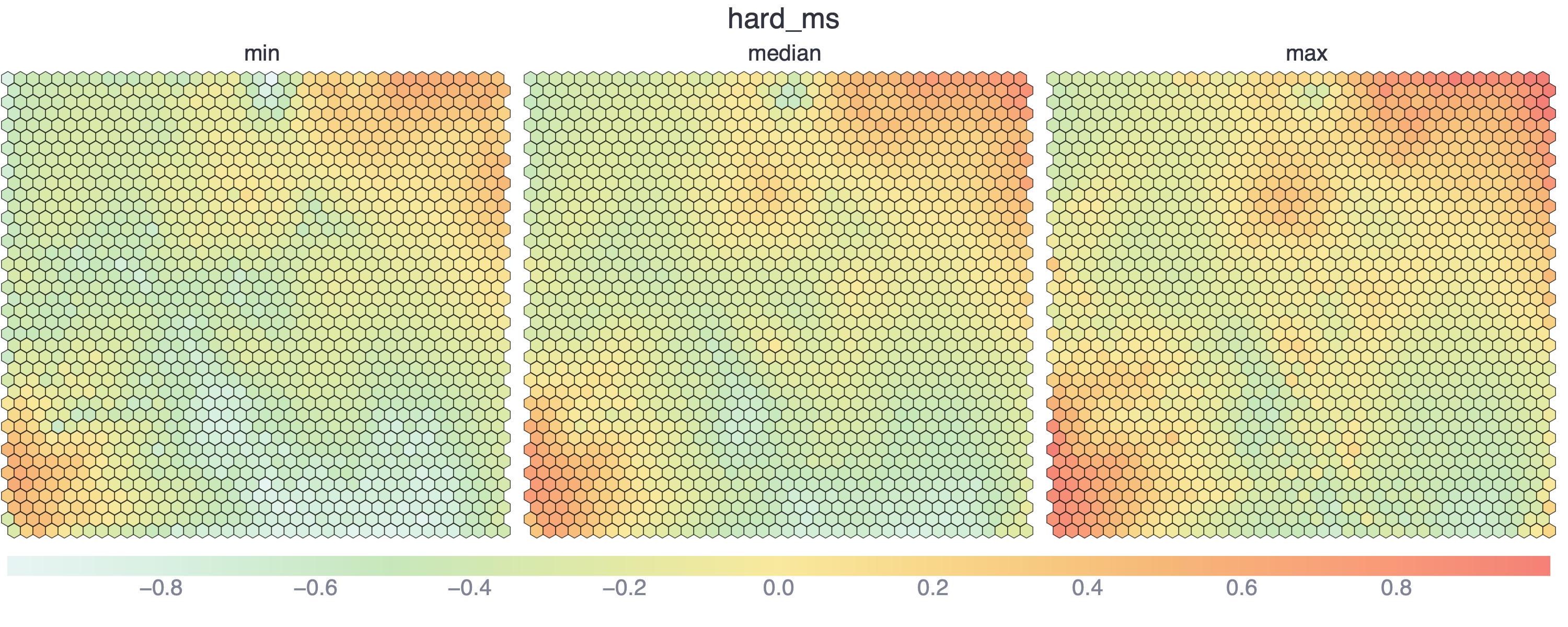}
    \includegraphics[width=0.73\linewidth]{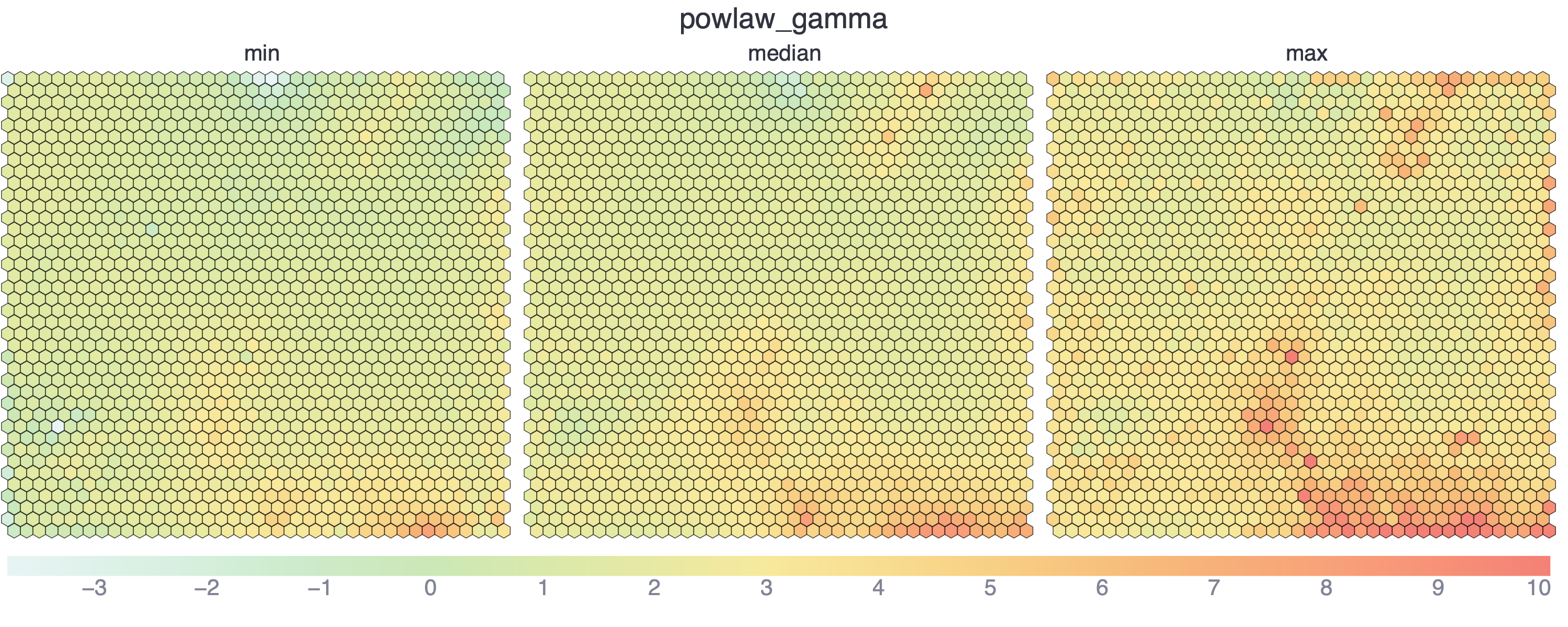}
    \caption{From top to bottom, maps of the minimum, median, and maximum values for the following \textit{features}: \texttt{hard\_hm}, \texttt{hard\_hs}, \texttt{hard\_ms}, \texttt{powlaw\_gamma}. Each row corresponds to a feature, with the three statistics shown side by side.}
    \label{fig:features_values_maps_1}
    \end{center}
\end{figure}

\begin{figure}[t]
    \begin{center}
    \includegraphics[width=0.73\linewidth]{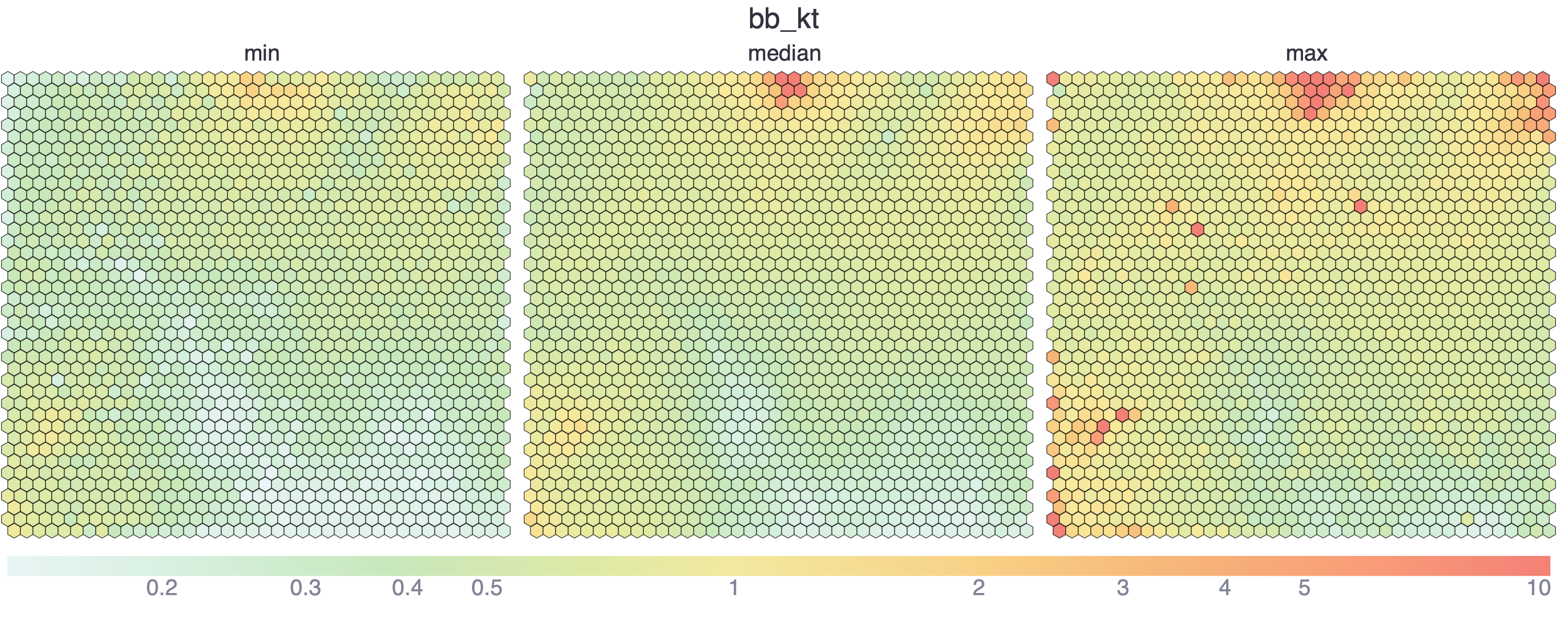}
    \includegraphics[width=0.73\linewidth]{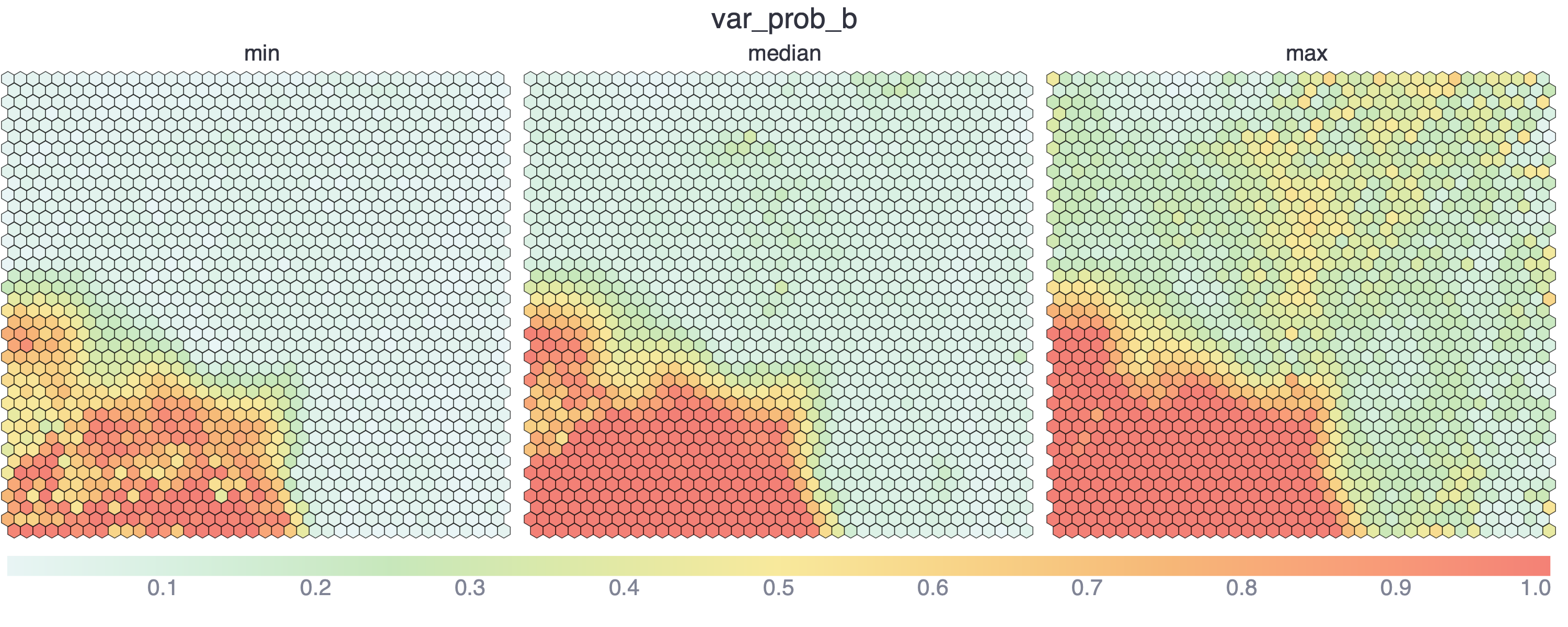}
    \includegraphics[width=0.73\linewidth]{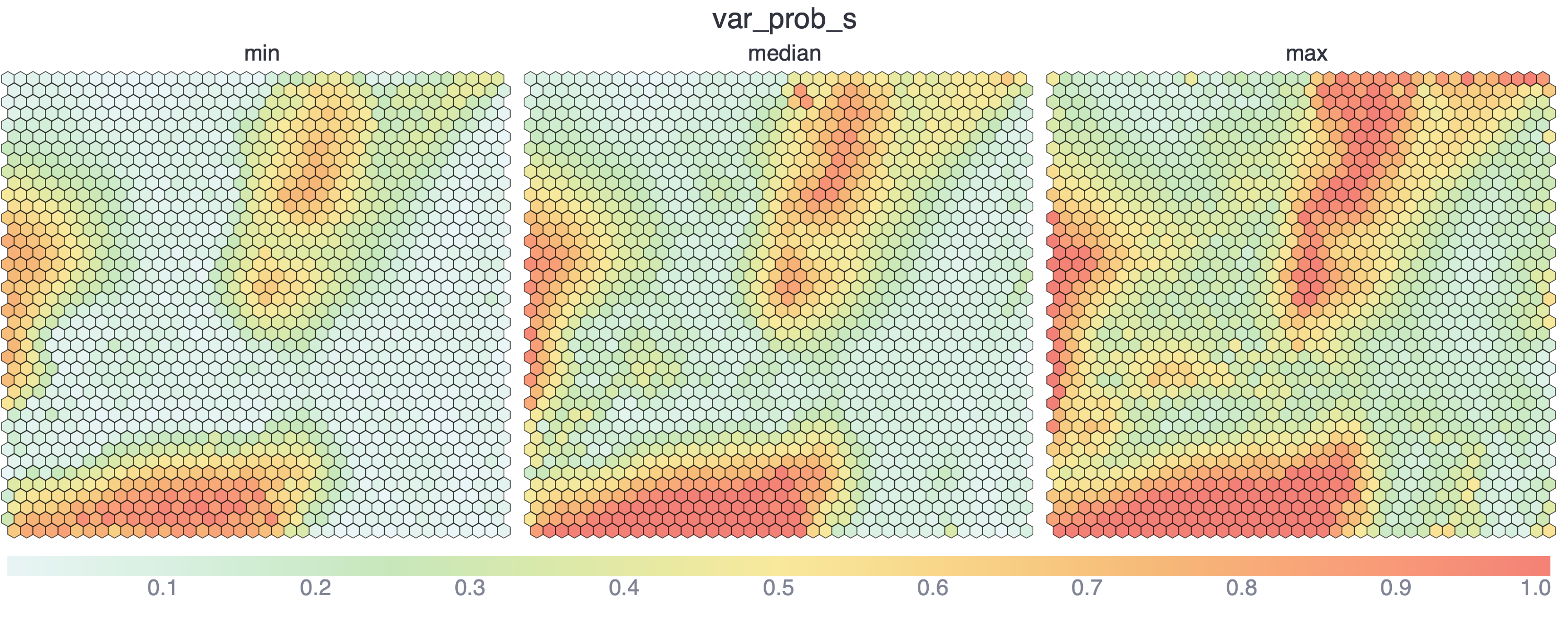}
    \includegraphics[width=0.73\linewidth]{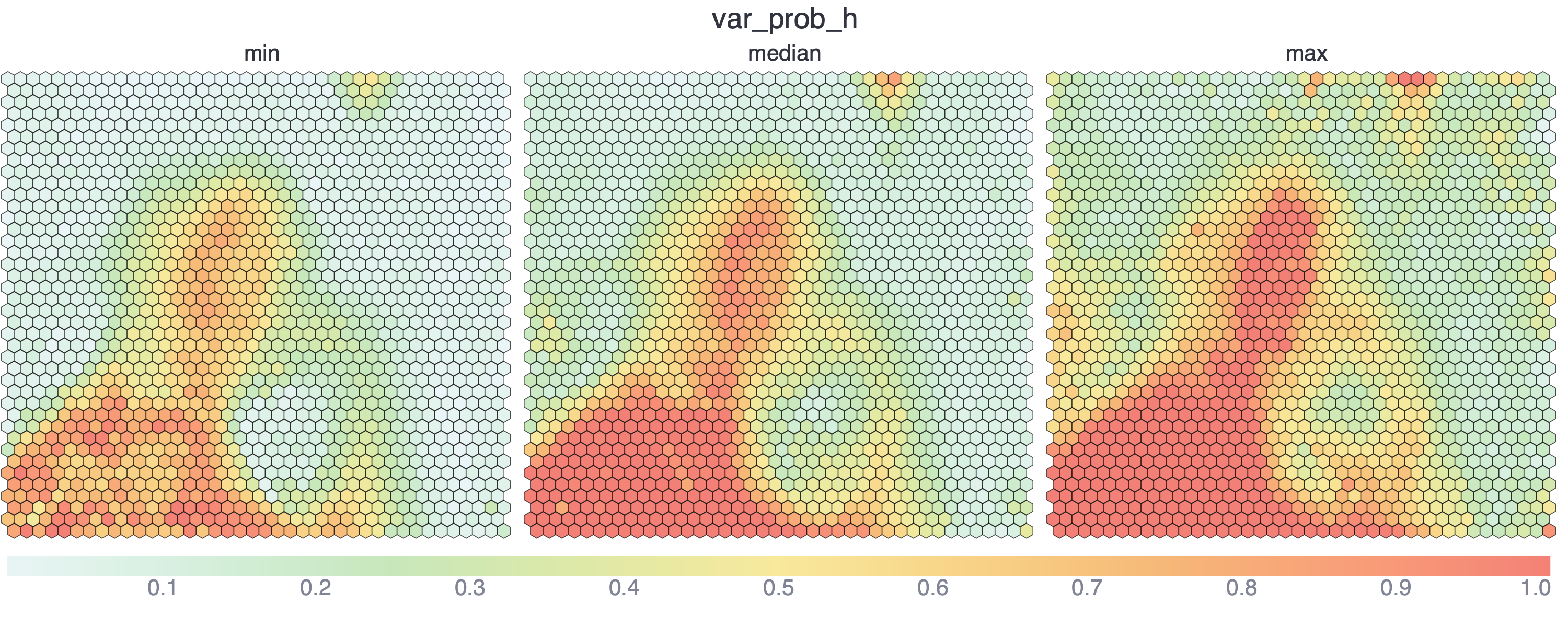}
    \caption{From top to bottom, maps of the minimum, median, and maximum values for the following \textit{features}: \texttt{bb\_kt}, \texttt{var\_prob\_b}, \texttt{var\_prob\_s}, \texttt{var\_prob\_h}. Each row corresponds to a feature, with the three statistics shown side by side.}
    \label{fig:features_values_maps_2}
    \end{center}
\end{figure}

\begin{figure}[t]
    \begin{center}
    \includegraphics[width=0.73\linewidth]{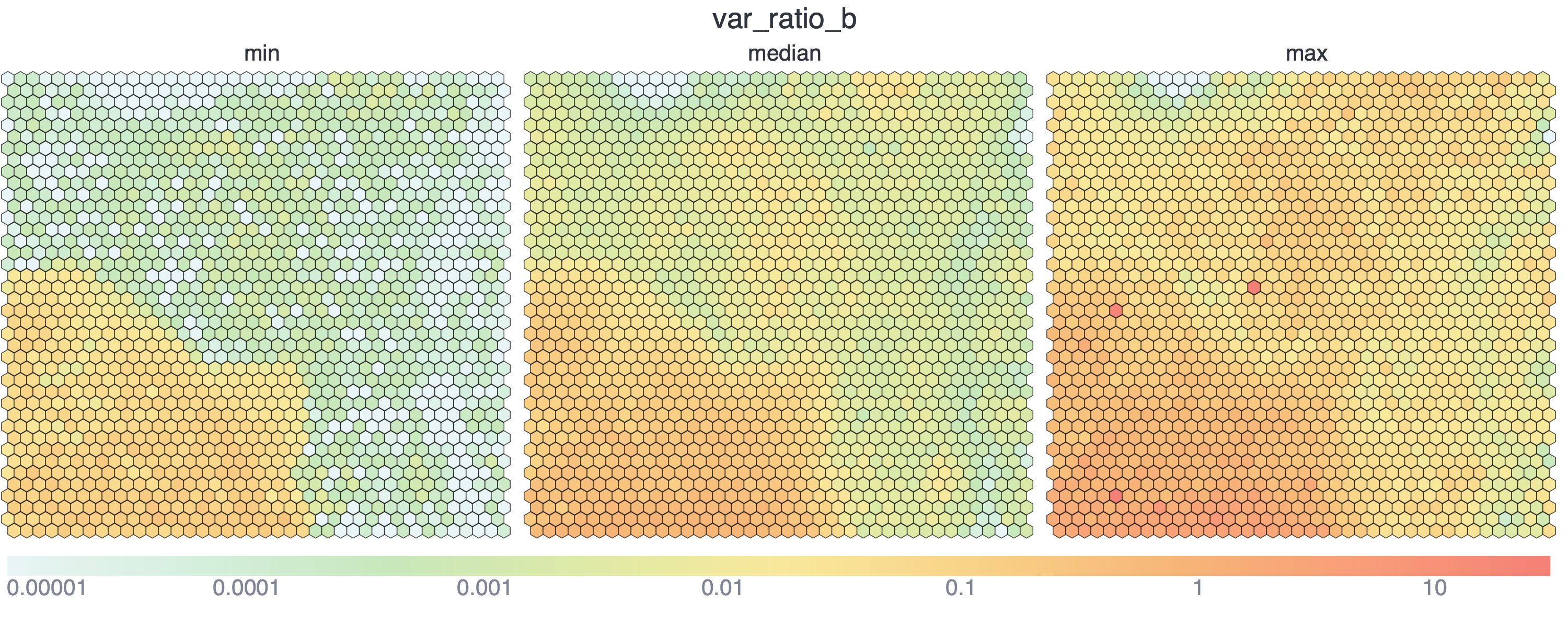}
    \includegraphics[width=0.73\linewidth]{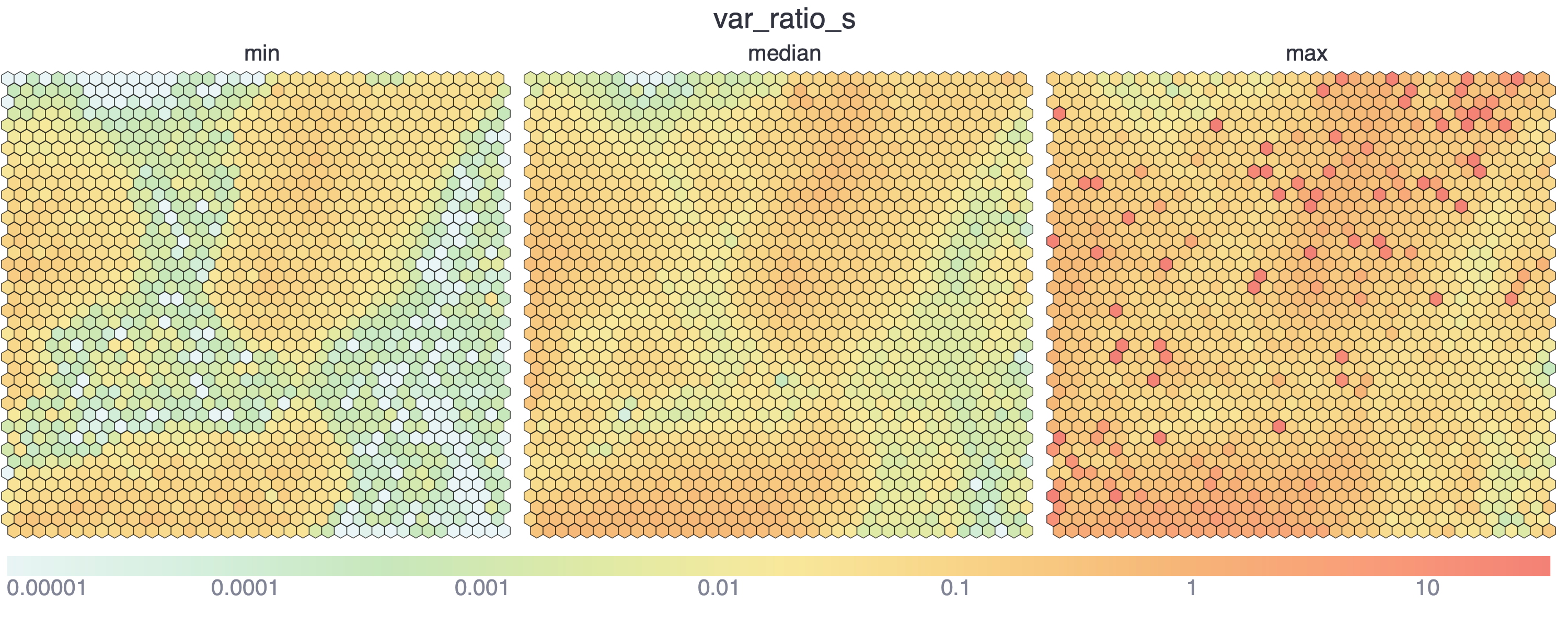}
    \includegraphics[width=0.73\linewidth]{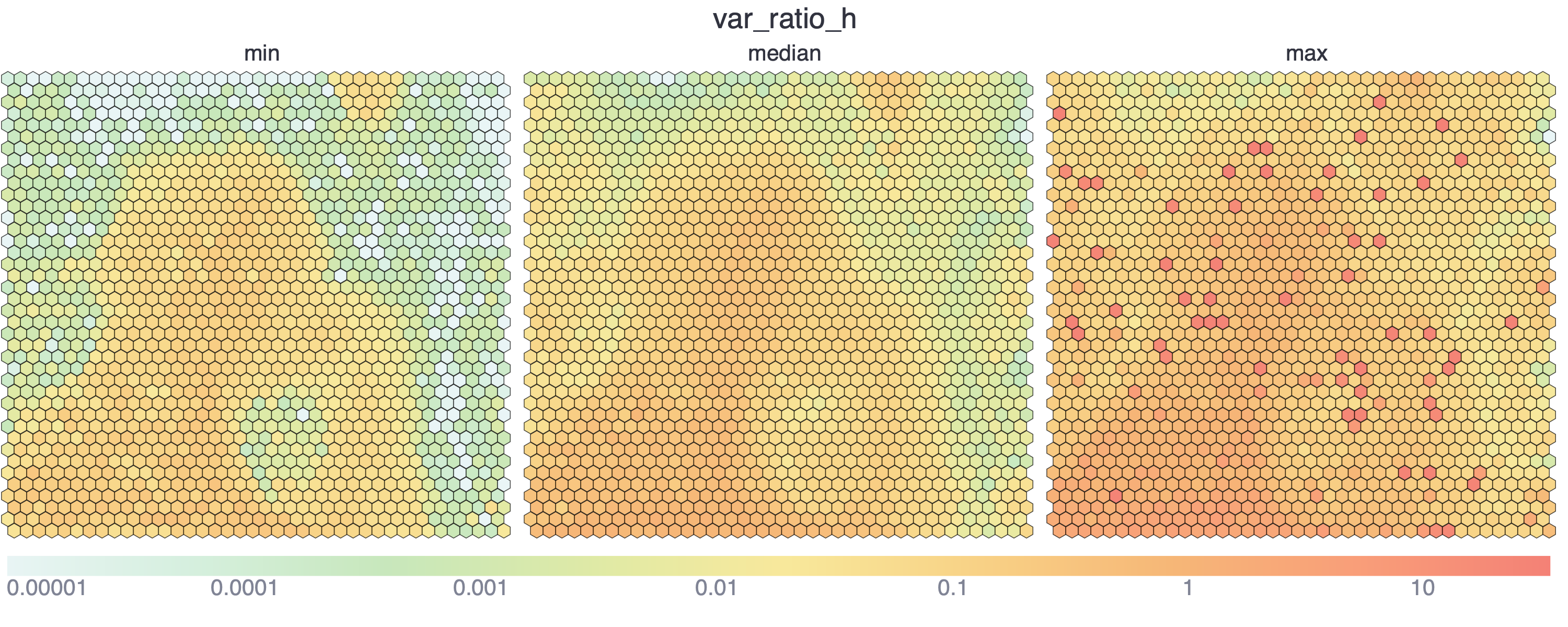}
    \caption{From top to bottom, maps of the minimum, median, and maximum values for the following \textit{features}: \texttt{var\_ratio\_b},
    \texttt{var\_ratio\_s},
    \texttt{var\_ratio\_h}. Each row corresponds to a feature, with the three statistics shown side by side.}
    \label{fig:features_values_maps_3}
    \end{center}
\end{figure}

\subsection{SOM \textit{feature} space mapping}
\label{subsec:feature_space_mapping}

Classical clustering methods explicitly define groups, usually after the number of clusters represented in the dataset has 
been determined (e.g., using validation criteria), and assign each data vector to a single cluster or calculate the probability 
of it belonging to multiple clusters.

Following a different approach, a SOM arranges the input data in 
the two-dimensional projected space according to similarity in the \textit{feature} space without directly 
assigning each of the usually large number of neurons composing the SOM map
to pre-specified clusters. We can explore the factors that drive how the 
SOM model separates detections in the original \textit{feature} space across the neurons of the map 
by investigating the distribution of the \textit{features} values over the SOM map and their interesting patterns. 
Figures~\ref{fig:features_values_maps_1},~\ref{fig:features_values_maps_2} and~\ref{fig:features_values_maps_3} 
show the SOM map of the minimum, median, and maximum values 
of all \textit{features} listed in Table~\ref{tab:som_features}, obtained from the values of all 
detections associated with each neuron. The inspection of these maps reveals the following general trends:

\begin{itemize}
    \item The distributions of median values of the three Hardness Ratios (HRs) \texttt{hard\_hm},~\texttt{hard\_hs} and 
    \texttt{hard\_ms} (three upper plots in Figure~\ref{fig:features_values_maps_1}) follow 
    similar patterns: neutral or moderately 
    positive HR values indicating flat or hard X-ray emission over the energy interval 
    spanned by the CSC bands are observed across a majority of the SOM map, with the maximum 
    values located in the bottom left and the top right corners of the SOM grid. Negative HRs 
    values, typical of softer
    X-ray spectra, populate the lower areas of the grid and decrease to their minimum values along 
    the lower
    margins of the SOM grid. The distributions of minimum and maximum HR values in each neuron follow 
    the same general trend. A similar pattern is visible in the bottom panels of 
    Figure~\ref{fig:features_values_maps_1}, where the minimum, median and maximum values of the 
    index $\nu$ of the 
    best-fitting power-law model $F_{E}\!\propto\!E^{-\nu}$ of the spectrum 
    of each detection (\texttt{powlaw\_gamma}) are shown. Large, positive values of $\nu$, indicating 
    a soft X-ray spectrum, are observed along the lower
    margin and the bottom left corner of the SOM grid, where negative values of the HRs are also observed.
    \item The map of the median temperature (\texttt{bb\_kt}) of the best-fitting absorbed black-body 
    model of the spectrum of each detection~(mid panel in the top row of Figure~\ref{fig:features_values_maps_2})
    shows that neurons with temperatures in the 1 to 5 keV interval are mostly located 
    in the bottom left corner and top right area of the SOM map, while lower temperatures 
    (kT$\in[0.5, 1]$ keV) are found in the bottom right corner of the SOM. A small region 
    dominated by very large temperatures is located in the middle of the top margin of the SOM map.  
    \item The distributions of the minimum, median and maximum values of the variability probabilities 
    \texttt{var\_prob\_b}, \texttt{var\_prob\_s}, \texttt{var\_prob\_h}
    measured in the Broad, Soft and Hard energy bands~(lower three plots in 
    Figure~\ref{fig:features_values_maps_2}) display different trends. In the Broad
    band (upper diagram), the detections that are almost certainly variables 
    (\texttt{var\_prob\_b} ${\sim}\,1$) are almost entirely found in the lower left 
    corner of the SOM map, with a slight excess of detections with max probability of being
    variable $\leq\!0.5$ spanning the upper right quadrant of the SOM 
    (second right plot from the top in Figure~\ref{fig:features_values_maps_2}). The
    variability in the Soft band \texttt{var\_prob\_s} indicates that time-variable detections 
    spans three clearly distinct areas, located along the lower and left margins and 
    in the upper half of the grid, respectively. Conversely, the distribution of 
    \texttt{var\_prob\_h} (lower diagram in Figure~\ref{fig:features_values_maps_2}) shows
    that the bulk of the variable detections in the Hard band are located in a region 
    spanning from the lower left corner
    of the SOM grid to its center, partially overlapping with the area characterized by 
    high-probability variability in the Soft band. The differences among the distributions of the 
    three variability probabilities in the three energy bands can be explained by the fact 
    that the Hard and Soft bands are entirely included in the Broad band: sources that are characterized
    as variable only in either the S or H bands may not be flagged as variable in the Broad band, 
    because variability in those bands may be diluted over the entire energy interval observed by 
    \textit{Chandra}. 
    \item Unsurprisingly, the distributions of the variability ratios 
    \texttt{var$\_$ratio$\_\{$b,h,s$\}$} in the Broad, Hard and Soft energy bands 
    (Figure~\ref{fig:features_values_maps_3}) trace the same general trends visible in the distributions 
    of variability probabilities in Figure~\ref{fig:features_values_maps_2}. In particular, the largest 
    values of the median value of \texttt{var$\_$ratio$\_$b}, 
    indicating larger variations in the properties of the detections compared to their average values, 
    are again located in the lower left quadrant of the SOM map, where the largest values of variability
    probability can also be found. Similarly, the variability ratios in the Hard
    and Soft bands (lower two diagrams in Figure~\ref{fig:features_values_maps_3}) display similar
    patterns to those visible in the \texttt{var$\_$prob$\_$h} and \texttt{var$\_$prob$\_$s} SOM diagrams.
    The isolated cells with very large maximum values, especially in the \texttt{var$\_$ratio$\_$h} and 
    \texttt{var$\_$ratio$\_$s} maps, can be explained by the presence of single outliers in the input 
    catalog. Since \texttt{var$\_$ratio$\_$*} is defined as the ratio between \texttt{var$\_$sigma$\_$*} 
    and \texttt{var$\_$mean$\_$*}, extreme values arise when a source has an anomalously 
    large \texttt{var$\_$sigma$\_$*} and, at the same time, a comparatively small \texttt{var$\_$mean$\_$*}.
\end{itemize}

The comparison of the SOM maps of the median values of the spectral and 
variability \textit{features} highlight some interesting patterns: A) the bulk of the most likely 
variable detections characterized by 
large-amplitude variability in the B band (lower left corner of the SOM grid in the second 
from top panel of Figure~\ref{fig:features_values_maps_2} and top 
panel of Figure~\ref{fig:features_values_maps_3}, respectively), corresponds to generally large
values of HRs (top three panels in Figure~\ref{fig:features_values_maps_1}); B) the large area 
occupied by detections with hard spectra
(upper right area of the SOM in the three upper panels of Figure~\ref{fig:features_values_maps_1}) 
matches the positions of detections with 
very small variability probability in the B band (second panel from top in 
Figure~\ref{fig:features_values_maps_2}), 
and relatively small variability probability in the H and S bands (third panel from top and lower panel 
in Figure~\ref{fig:features_values_maps_2}); C) the median distributions of 
\texttt{var$\_$ratio$\_\{$b,h,s$\}$} parameters on the
SOM map (Figure~\ref{fig:features_values_maps_3}) indicate that harder X-ray spectra 
located in the upper right regions of the SOM grid correspond to variability amplitudes similar 
or larger than those observed for detections in the lower-left corner (this trend is more clearly 
visible in the S and H bands, lower plots of Figure~\ref{fig:features_values_maps_3}). 

These results seem to contradict the correlation between variability behavior and spectral shapes
of X-ray emission known for some classes of sources. For example, X-ray stellar binaries are known 
to undergo coordinated variations of their luminosity and spectral shape, usually described by the X-ray 
hardness-intensity diagram~\citep{2004MNRAS.355.1105F}. Their variability properties also change with 
their spectral behavior and luminosity: hard, brighter states are usually characterized by more 
frequent, larger amplitude short time-scale variability than soft states (where longer timescales, 
less significant flux variations have been observed). This behavior is driven by the interplay of
different emission 
processes that produce X-rays in the inner, optically thin corona (usually caused
by Inverse Compton (IC) scattering and responsible for the hard component) where significant fluctuations can occur more 
frequently and quickly than in the more stable, optically thick accretion disks (dominated by black-body emission) responsible for the bulk of the X-ray 
emission during soft states. A given observation would more likely reveal variability of a source 
if taken during its hard state than its soft state. 

\begin{figure}[t]
    \begin{center}
    \includegraphics[width=0.45\linewidth]{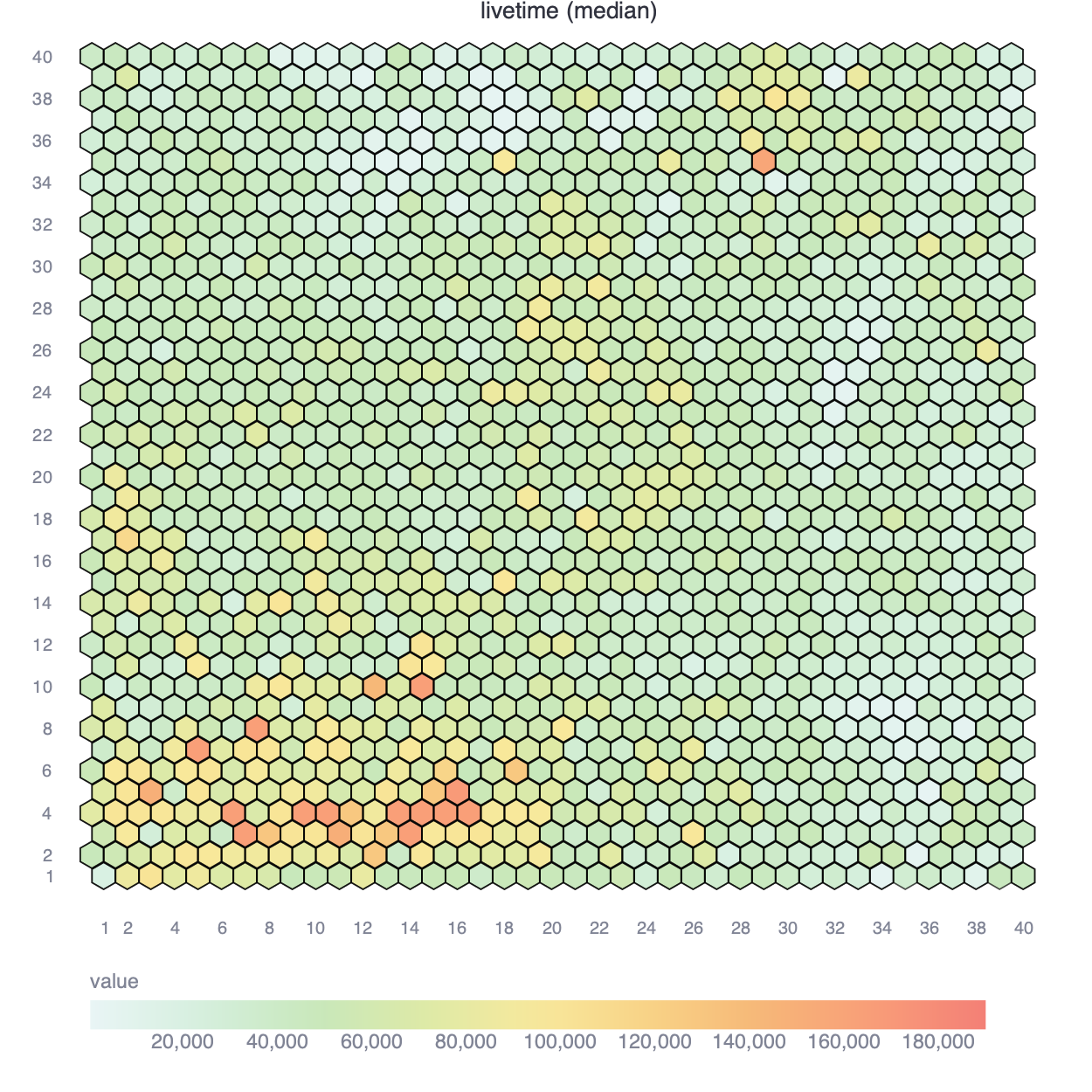}
    \includegraphics[width=0.45\linewidth]{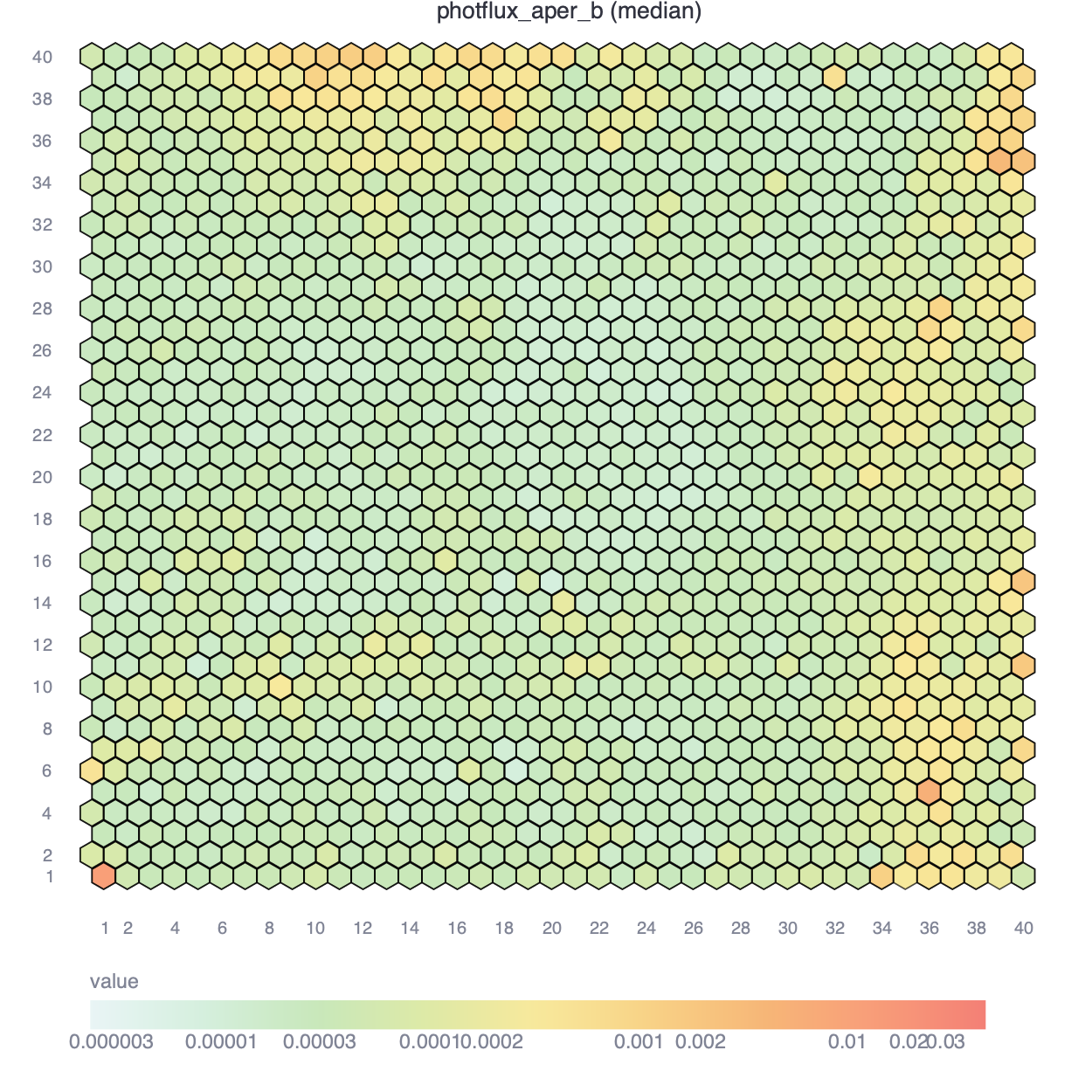}
    \caption{SOM maps of the median \texttt{livetime} (ks) of the 
    observations used to extract each detection (left), and median photon 
    flux in the Broad band (photons/s/cm$^{2}$) on the right.}
    \label{fig:exptime_photflux_maps}
    \end{center}
\end{figure}

\begin{figure}[t]
    \begin{center}
    \includegraphics[width=0.45\linewidth]{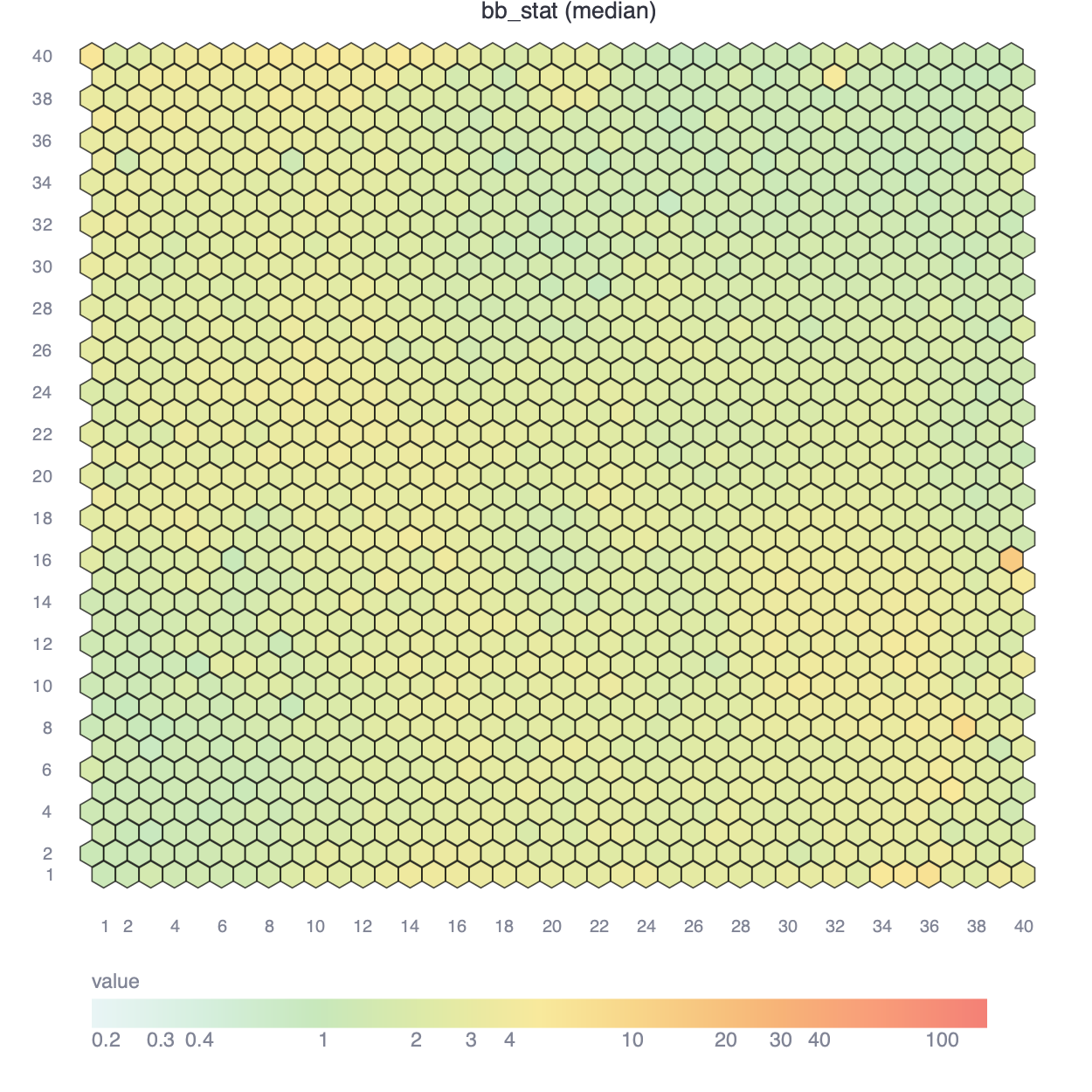}
    \includegraphics[width=0.45\linewidth]{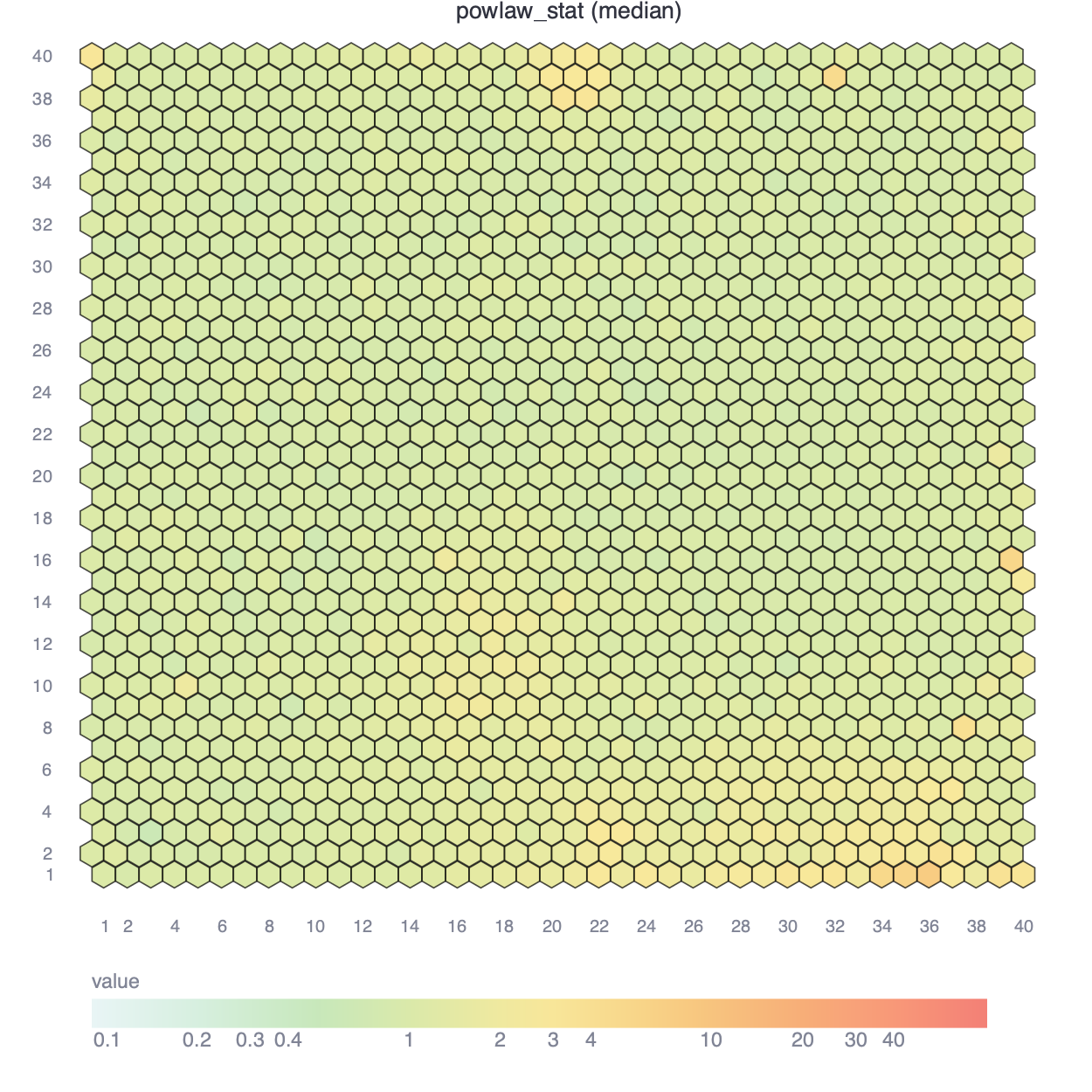}
    \caption{Median SOM \textit{feature} maps of the reduced $\chi_\nu^2$ statistic for the best-fitting absorbed spectral models to the source-region PI spectrum: black-body (\texttt{bb\_stat}, left) and power-law (\texttt{powlaw\_stat}, right).}
    \label{fig:bb_pwl_stat_maps}
    \end{center}
\end{figure}

It is important to stress that the serendipitous nature of the CSC, 
containing X-ray sources
observed in very different regions of the sky with different observational strategies that are
optimized to achieve the scientific goals of each specific program and not to achieve homogeneous 
sampling of the general population of X-ray sources of the sky, complicates the interpretation of 
our results. For example, 
both variability diagnostics used for the training of the SOM 
(\texttt{var$\_$prob$\_\{$b,h,s$\}$} and \texttt{var$\_$ratio$\_\{$b,h,s$\}$}) can only 
assess flux variability on timescales equal or shorter than the duration of the observation 
during which each detection takes place. While intra-observation variability is known to correlate with 
inter-observation variability for sources observed multiple times by \textit{Chandra} and associated 
with multiple single detections, the duration of the single observation affects what type of 
variability can be observed and measured. The SOM distributions of median \texttt{livetime}, 
obtained from the exposure time of the \textit{Chandra} observations where each detection 
is observed after applying the good time intervals and the deadtime correction factor (left 
plot in Figure~\ref{fig:exptime_photflux_maps}), shows that the longest median \texttt{livetime} values are located in the lower-left corner of the 
SOM, a region also populated by detections characterized by large median values of 
\texttt{var\_prob\_b}, \texttt{var\_prob\_h} and \texttt{var\_ratio\_s} with relatively hard
X-ray spectra. The similarities observed between the patterns followed by the distribution 
of median \texttt{livetime} and spectral/variability \textit{features} used to train the SOM
model suggest that these parameters are correlated. 

Similarly, the fluxes of the X-ray detections
also correlate with the spectral and variability properties used to 
train the SOM model. The plot on the right of Figure~\ref{fig:exptime_photflux_maps} 
displays the median values of the photon fluxes of all detections in the Broad band, which anti-correlate with the \texttt{livetime} of the observations, as
the larger values can be found in regions of the SOM map characterized by 
shorter observations. Brighter X-ray detections also are associated with sources that are harder (see the \texttt{hard\_hm} and \texttt{hard\_hs} maps in Figure~\ref{fig:features_values_maps_1})
and less variable in the B band (Figure~\ref{fig:features_values_maps_2}). 
The two properties~\texttt{powlaw\_gamma} and~\texttt{bb\_kt} used in this work as \textit{features} 
in the space where the SOM model is trained are used regardless of the likelihood that the 
corresponding power-law and black-body models approximate well the observed shape of the 
spectrum of each detection. We made
this choice to be able to train the SOM model on the same set of parameters for 
all X-ray detections in the CSC, independently of the best-fitting spectral model. 
Figure~\ref{fig:bb_pwl_stat_maps} displays the per-cell median values of the 
reduced $\chi_{\nu}^{2}$ statistics for~\texttt{bb\_kt} (left) and~\texttt{powlaw\_gamma} (right). 
Good fits have $\chi_{\nu}^{2}\!\approx\!1$, $\chi_{\nu}^{2}\!\gg\!1$ usually indicate that
the fitted model does not correctly model the spectrum, while $\chi_{\nu}^{2}\!\ll\!1$ is 
indicative of over-fitting. The map of the $\chi_{\nu}^{2}$ median values 
for the~\texttt{bb\_kt} parameter in Figure~\ref{fig:bb_pwl_stat_maps} suggests that the black-body might not be the best model to fit the spectra of detections located in a large region of
the SOM map stretching from the lower-right corner to the upper margin. This area
spans all observed values of the three hardness ratios used to train the SOM model 
(three upper plots in Figure~\ref{fig:features_values_maps_1}) and is characterized
by relatively large $\gamma$ parameter values (lower plot in Figure~\ref{fig:features_values_maps_1}), in particular in the bottom-right corner, and 
large values of the~\texttt{bb\_kt} parameter especially along the upper margin of the map 
(upper plot in Figure~\ref{fig:features_values_maps_2}). Similarly, large median values 
of the $\chi_{\nu}^{2}$ for the~\texttt{powlaw\_gamma} \textit{feature}, indicating that the 
power-law is a poor model, are observed in the lower-right area of the SOM map and across 
the upper margin of the map. A potential interpretation of these findings is that these 
two simple spectral models fail to capture the spectral properties of higher S/N detections
whose spectra show a variety of different behaviors (as highlighted by the large scatter
of the hardness ratios for these detections), while they work reasonably well 
for fainter detections (cp. with the~\texttt{photflux\_aper\_b} distribution in the right 
plot in Figure~\ref{fig:exptime_photflux_maps}) whose spectra are well-fitted by simplified
models.

\subsection{Distribution of classes}
\label{subsec:labels}

The relations between the distributions of typical values of the spectral 
and variability \textit{features} used to train the SOM are complicated by the heterogeneity of the 
population of objects available in the CSC, which is composed of different classes of astronomical 
sources whose X-ray emission is determined by different physical processes.
We can in part reduce this complexity by investigating the distribution on the SOM map
of the \textit{bona fide} classes available for some of the X-ray sources in the training sample. 
The available source classification was incorporated as labels of a subset of detections in
the ACIS-SV-CSC dataset~(see Section~\ref{subsec:simbad} for details on how this data was collected and 
cross-matched against the CSC dataset used in this paper).

\begin{figure} [t]
    \begin{center}
        \includegraphics[width=0.48\linewidth]{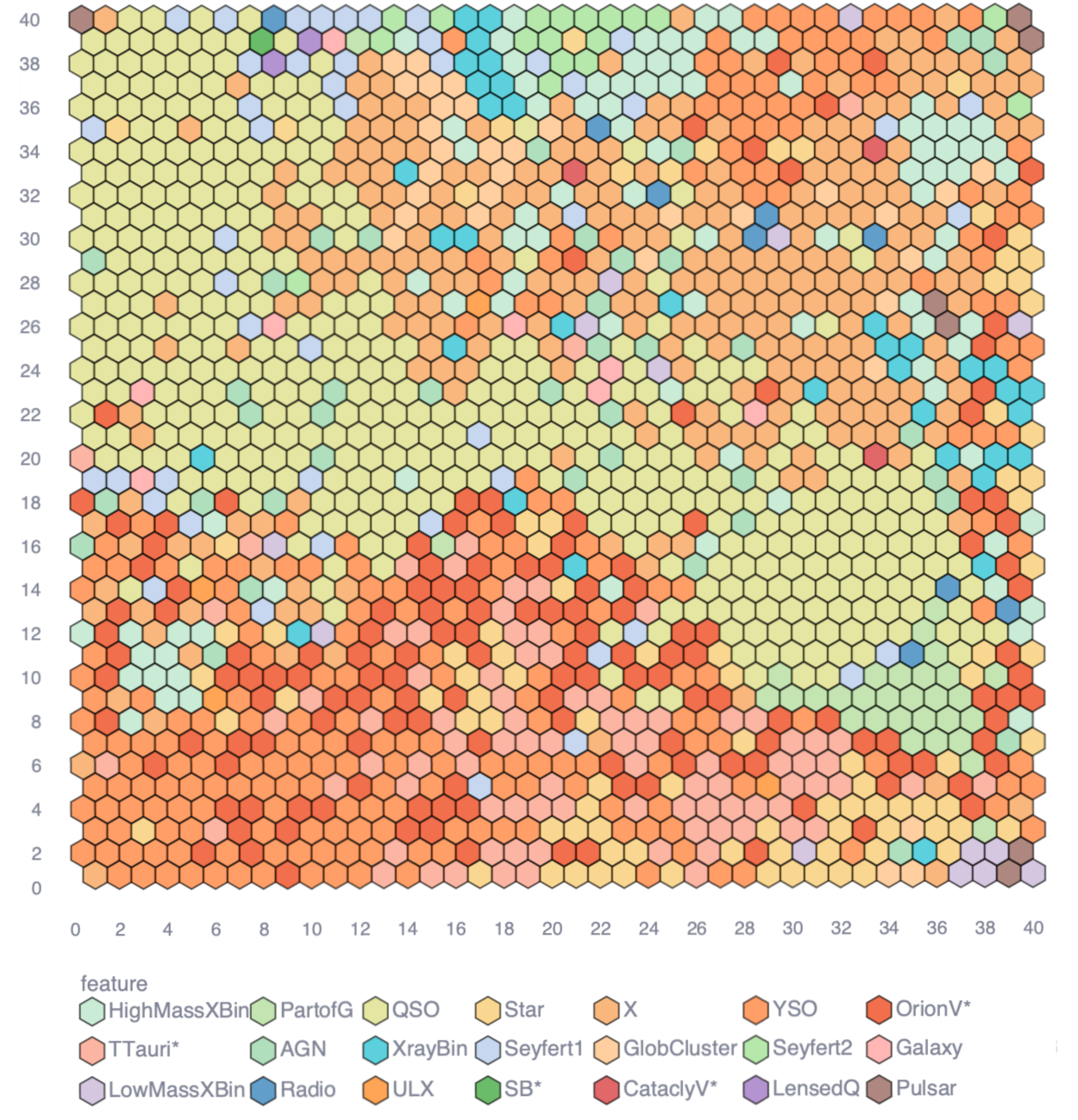}
        \includegraphics[width=0.48\linewidth]{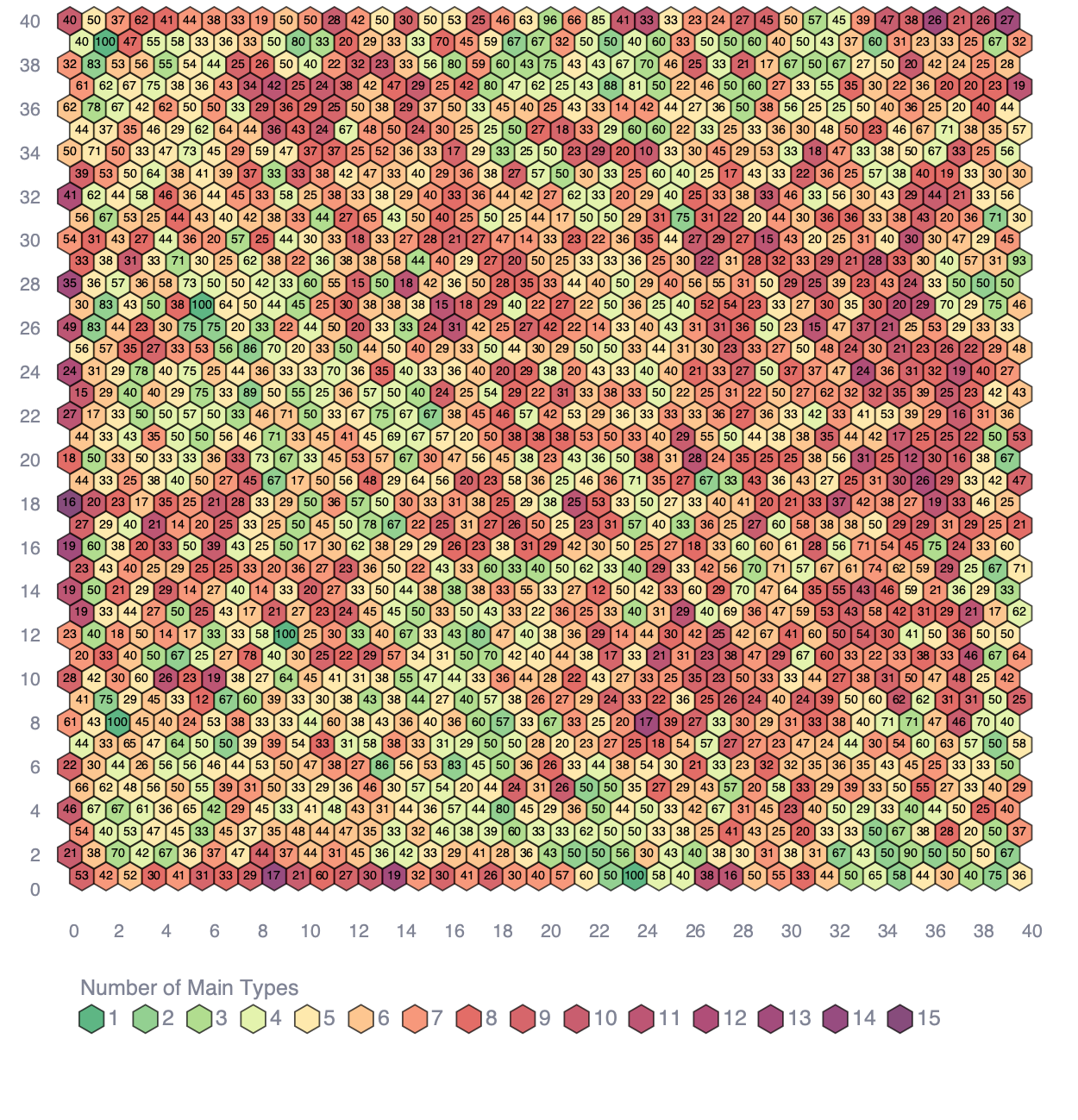}
        \includegraphics[width=0.48\linewidth]{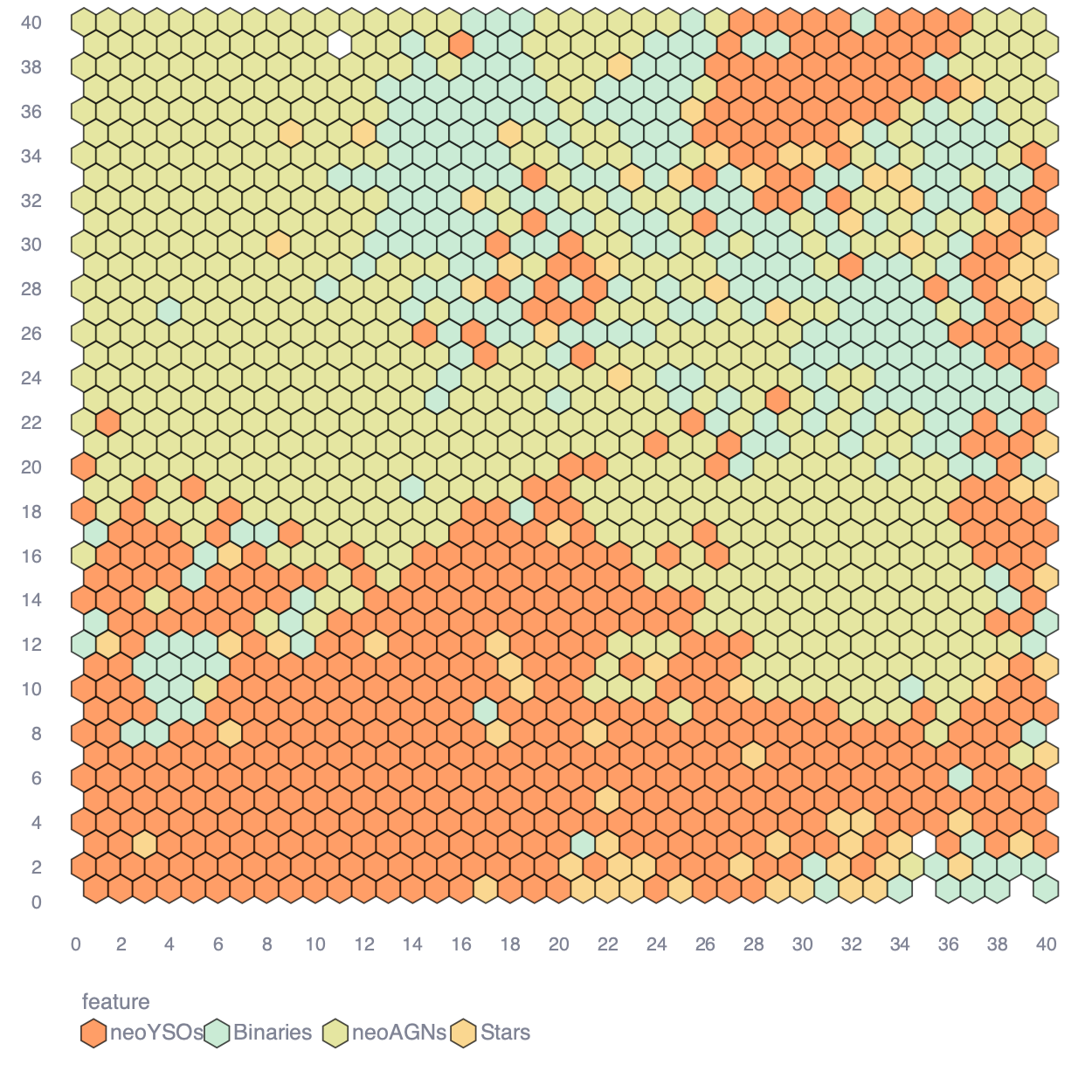}
        \includegraphics[width=0.48\linewidth]{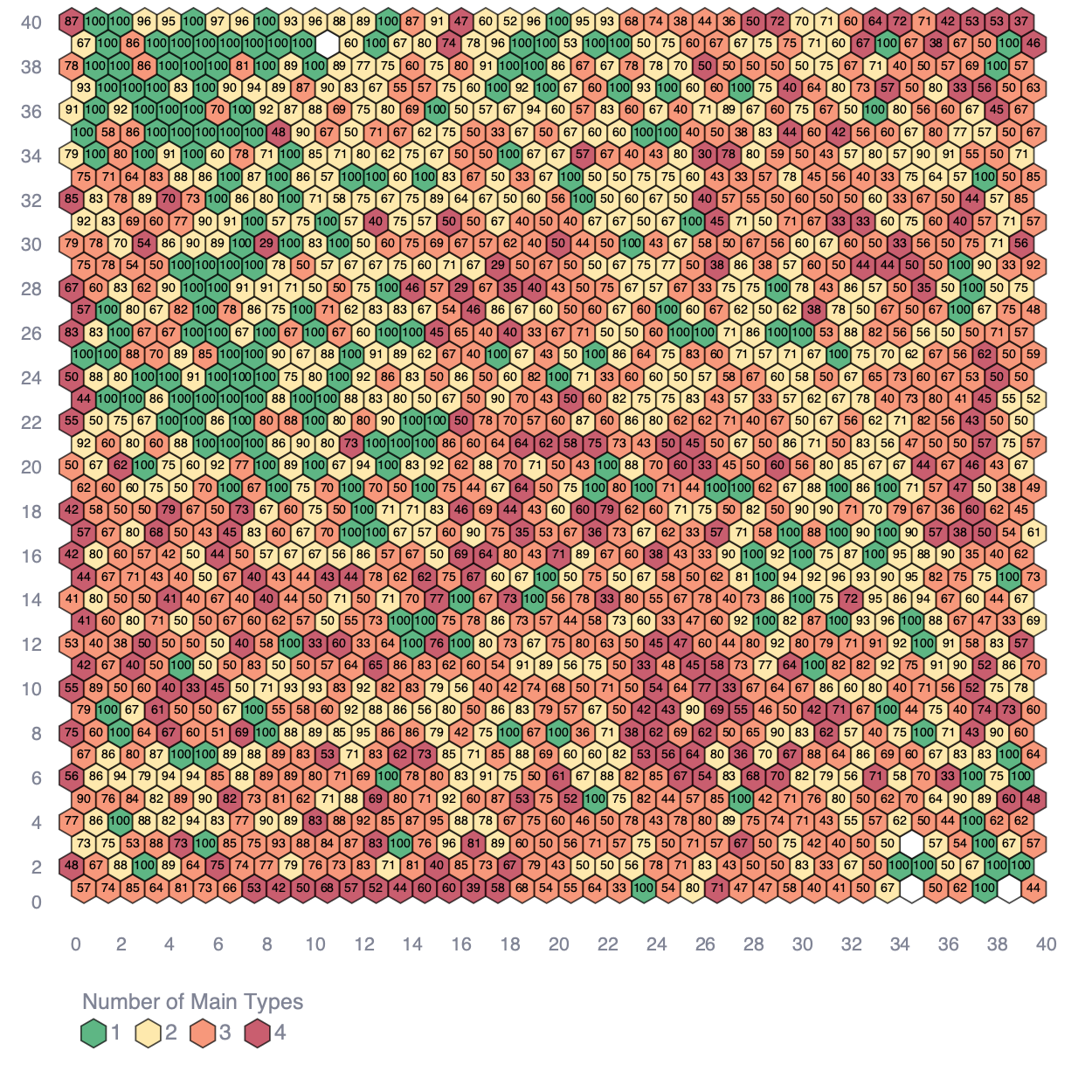}
        \caption{Left: Distribution of the most common labels
        on a per-neuron base in the trained SOM map for original SIMBAD \texttt{main\_type} 
        (upper), and aggregated labels (lower plot, see Section~\ref{subsec:simbad}). 
        Right: Class heterogeneity and dominance ratio per neuron for original labels (upper)
        and aggregated labels (lower). Each hexagon is colored according to the 
        number of different source classes it contains, while the number indicates the percentage of detections 
        belonging to the dominant class.}
        { \label{fig:main_type_cluster}}
    \end{center} 
\end{figure}

Figure~\ref{fig:main_type_cluster} shows how the source types in the dataset are 
distributed across the SOM map, according to the twenty-one most common original
\texttt{main\_type} classification (upper plots), and the aggregated labels defined in 
Section~\ref{subsec:simbad} (lower plots). In the plots on the left, each cell of the 
SOM two-dimensional map is colored according to the most common class 
associated with that specific cell by the SOM, while the plots on the right
of Figure~\ref{fig:main_type_cluster} display the number of different classes associated
with detections in each cell of the SOM map (color-coding) and the percentage of the most
represented class (number in each cell). The maps for the aggregated labels
(lower plots in Figure~\ref{fig:main_type_cluster}) do not include labels whose original 
labels do not provide information of the nature of the source but only about observational
\textit{features} and/or position relative to other sources (\texttt{X} and \texttt{PartOfG}). 

Figure~\ref{fig:main_type_cluster} shows that the spatial distribution of cells
of the SOM map dominated by the class of detections, both for original \texttt{main\_type} labels from SIMBAD (upper plots) and the 
aggregated labels based on the original \texttt{main\_type} values (lower panels),
reveals distinctive patterns, described below:

\begin{itemize}
    \item Cells dominated by detections classified as~\texttt{QSO}~(\textit{light yellow} cells), \texttt{AGN} (\textit{green} cells) and the less common~\texttt{Seyfert1},~\texttt{Seyfert2}
    according to the SIMBAD \texttt{main\_type} in the upper plots in Figure~\ref{fig:main_type_cluster} 
    and as~\texttt{neoAGNs} (\textit{green} cells) in the aggregated labels map 
    (lower left in Figure~\ref{fig:main_type_cluster}) tend to occupy similar areas of the SOM map. 
    These cells are mostly located in an elongated region
    stretching across the center of the SOM map, that extends
    from the left–middle toward the right–middle corners, 
    with additional presence toward the top-left corner. These regions are associated with 
    relatively small average distance between SOM cells (see U-Matrix, left plot in 
    Figure~\ref{fig:som_analysis}), and are characterized by relatively large hardness ratio 
    \texttt{hard\_hm} between the Medium and Hard energy bands (upper plot in 
    Figure~\ref{fig:features_values_maps_1}) but softer spectra as measured by \texttt{hard\_ms} and 
    \texttt{hard\_hs}. Interestingly, the area dominated by~\texttt{neoAGNs} features low variability probability and amplitudes 
    in the Broad energy band as measured by \texttt{var$\_$prob$\_$b} and \texttt{var$\_$ratio$\_$b} 
    (second row from the top in Figure~\ref{fig:features_values_maps_2} and top row in 
    Figure~\ref{fig:features_values_maps_3}, 
    respectively), but it also matches regions of large variability probability and amplitudes in the Soft and 
    Hard bands (lower two rows in Figure~\ref{fig:features_values_maps_2} and Figure~\ref{fig:features_values_maps_3}, 
    respectively). This diversity of spectral behaviors is likely caused by the wide range of different emission mechanisms 
    that the sources labeled as~\texttt{neoAGNs} exhibit depending
    on the specific component of the AGN where the X-rays are 
    produced. For example, 
    the two main contributors to the total X-ray flux of AGNs are the IC scattering, responsible for the emission from the X-ray Corona, and the thermal black-body 
    emission from the accretion disk. The large variance of the \textit{feature} values associated with cells
    dominated by AGNs are also possibly influenced by other
    factors like the inclination of the AGN relative to the observer, 
    the mass of the supermassive black hole that powers the AGNs, 
    host galaxy characteristics (including the morphology and distance), or the activity level. 

    \item Cells dominated by detections labeled as~\texttt{YSO} and~\texttt{neoYSOs}
    (\textit{dark orange} in both maps of the original and aggregated labels) 
    are mostly observed in the lower-left region of the 
    original label maps (upper left plot in Figure~\ref{fig:main_type_cluster}) and in the lower area of the aggregated labels SOM map (lower left plot). Additional areas populated by~\texttt{neoYSOs} can be found along the right side of the SOM map and in a small region along the upper 
    boundary of the map. These regions are characterized by wide range of median values of Hardness
    Ratios (upper three plots in Figure~\ref{fig:features_values_maps_1}), with harder 
    detections located in the lower left corner and along the right and upper borders
    of the map, and softer ones in the right section of the large area along the lower border of 
    the map. The median values of the index of the power-law model (\texttt{powlaw\_gamma}) and the 
    temperature (\texttt{bb\_kt}) of the best-fit black-body model of the X-ray spectrum for \texttt{neoYSOs}-dominated cells on the SOM map are comprised, respectively, 
    in the $[0,3]$ and $[0.5,1.5]$ kT intervals, which are consistent with the average values of these
    parameters for the whole sample of detections used to train the SOM.
    On the other hand, the YSO-dominated areas of the SOM, both according to the original labels 
    and in the aggregate map, are characterized by large values of the variability probabilities 
    in all bands 
    \texttt{var\_prob\_(b,s,h)} (lower three plots in Figure~\ref{fig:features_values_maps_2}) and the 
    variability amplitude ratios \texttt{var\_ratio\_(b,s,h)} (Figure~\ref{fig:features_values_maps_3}). 
    In general, the YSO-dominated areas are adjacent to QSOs-dominated regions of the SOM map
    but they feature different spectral and variability behaviors, suggesting that the SOM mapping 
    captures the fundamental difference between the physical mechanisms responsible for the X-ray radiation 
    emitted by these two macro-classes of sources. 
    
    \item The aggregate class \texttt{Binaries} (\textit{light green} in the aggregate labels 
    SOM map~(lower left plot in Figure~\ref{fig:main_type_cluster}) is composed of all SIMBAD 
    \texttt{main\_type} classes whose phenomenology can be associated with binary stellar systems 
    (i.e., \texttt{HMXB},~\texttt{XRayBin},~\texttt{LMXB},
    ~\texttt{CataclyV*},~\texttt{SB*} and \texttt{BYDrav*}). The majority of
    \texttt{Binaries}-dominated cells in the aggregated SOM map are located in the upper-right quadrant. This
    region is characterized by relatively large median hardness ratios values 
    (upper plots in Figure~\ref{fig:features_values_maps_1}), low values of the power-law index
    and large values of the black-body best-fit model temperature (low plot in 
    Figure~\ref{fig:features_values_maps_1} and upper plot in Figure~\ref{fig:features_values_maps_2}), 
    and relatively low variability probabilities and ratios in the Broad and Hard bands, but large median 
    values in the Soft band for \texttt{var\_prob\_s} and \texttt{var\_ratio\_s}, as evident in the
    third from top plot in Figure~\ref{fig:features_values_maps_2} and second
    from top plot in Figure~\ref{fig:features_values_maps_3}, respectively. It is worthwhile to point
    out that cells whose majority of labeled detections are \texttt{Binaries} occupy the same area populated by 
    SIMBAD \texttt{main\_type} generic label \texttt{X}, indicating that they share the same profile in the 
    \textit{feature} space mapped by the SOM and suggesting that at least a subset of these unclassified
    X-ray detections might be binary stellar systems.
    
    \item Stars (\textit{orange}) dominate cells mostly located in the lower-right
    side of the SOM map for the original labels (upper left plot of Figure~\ref{fig:main_type_cluster}), 
    characterized by small values of hardness ratios and large \texttt{powlaw\_gamma} 
    (Figure~\ref{fig:features_values_maps_1}), small median black-body temperatures and variability 
    properties (Figure~\ref{fig:features_values_maps_2} and~\ref{fig:features_values_maps_3}, 
    respectively). Some cells where the majority of detections is labeled as~\texttt{Stars} can also 
    be found along the right margin of the SOM projection, in a region with a mix of dominant 
    labels, and scattered across the upper region mostly occupied by \texttt{AGN} and 
    \texttt{Binaries}-dominated cells. In the SOM map for the aggregated labels 
    (lower left plot of Figure~\ref{fig:main_type_cluster}), the number of \texttt{Stars}-dominated 
    cells shrink significantly because several cells located in the lower right section 
    of the original labels 
    SOM map where the dominant label \texttt{Stars} accounted for less than 50$\%$ of the total number of  
    detections in the cell (upper right plot of Figure~\ref{fig:main_type_cluster}) are now dominated by 
    aggregated \texttt{neoYSOs} class (the aggregated class contains  
    the original \texttt{YSO} labels plus detections classified as \texttt{OrionV*} and \texttt{TTauri*}).
\end{itemize}

\begin{figure}[t]
    \begin{center}
    \includegraphics[width=0.48\linewidth]{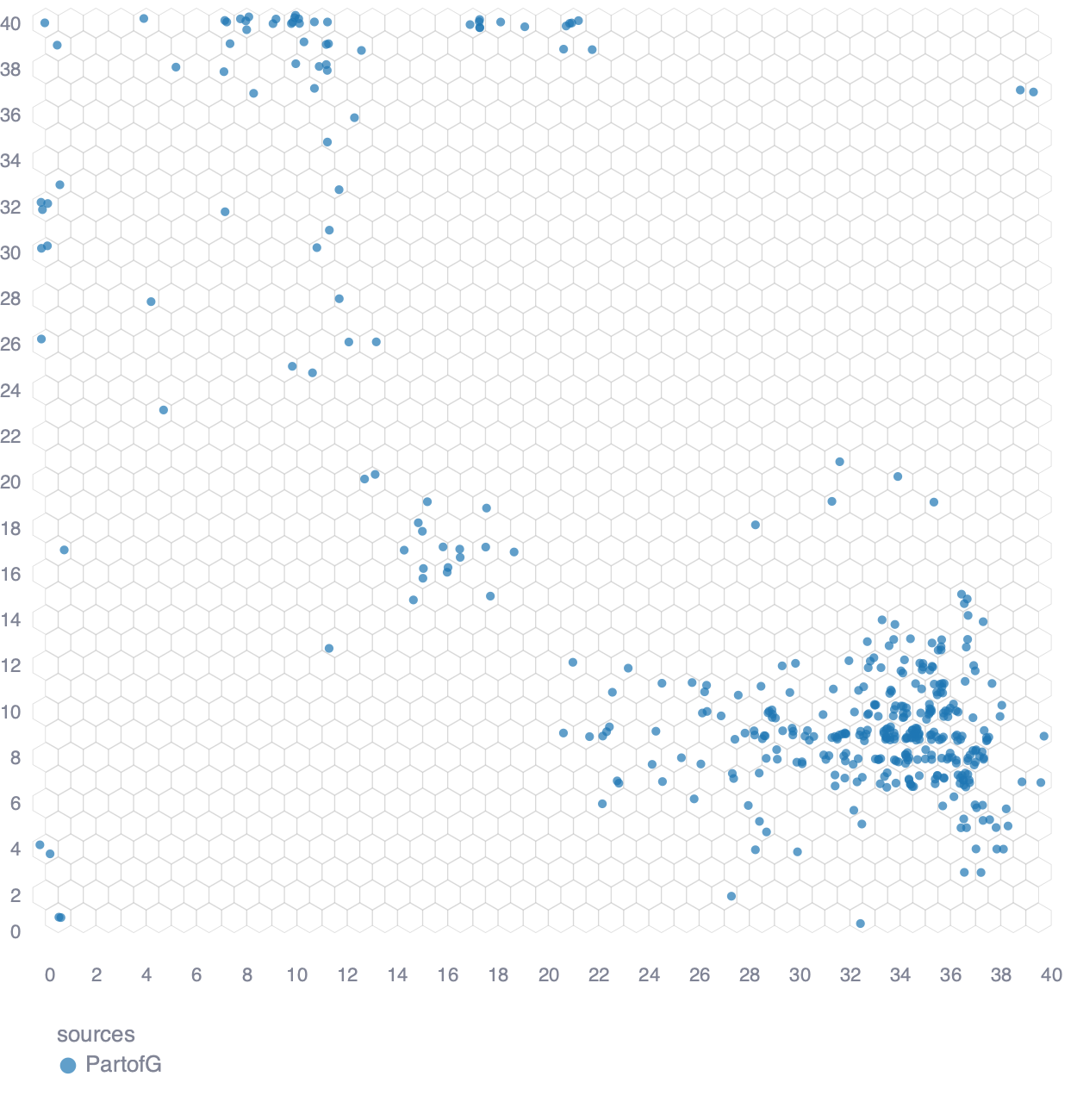}
    \caption{Positions of the single detections classified as 'Part of 
    Galaxy' (\texttt{PartOfG}) from the SIMBAD source classification, in the 
    SOM map trained on the ACIS-SV-CSC dataset.}
    {\label{fig:partofg_scatter}}
    \end{center} 
\end{figure}

An example of how the SOM mapping of the parameter space distribution of CSC 
detections can help our understanding of X-ray sources whose nature is uncertain or unknown, 
we can investigate the location of the detections associated with CSC sources 
classified as \texttt{PartOfG}. This label belongs to the \textit{“8. BLENDS, ERRORS, NOT WELL DEFINED 
OBJECTS”} category of SIMBAD, as defined in the~\citet{SIMBADotypes} classification scheme, and is used for 
sources that are spatially associated with galaxies but otherwise not clearly matched to sources of known nature. 
We can infer the nature of the detections associated with these ambiguous labels by examining where 
each \texttt{PartOfG} detection lands on the SOM map. Figure~\ref{fig:partofg_scatter} shows 
the distribution of the 427 detections in the ACIS-SV-CSC dataset labeled 
as \texttt{PartOfG}, for a total of 
19 single master sources, 5 of which have only a single detection. The \texttt{PartOfG} detections cluster 
in the lower-right region of the SOM map, approximately centered on the neuron ($x\!=\!36$, $y\!=\!9$), which correspond, 
in the SOM map of the dominant aggregated labels (lower left plot of Figure~\ref{fig:main_type_cluster}) 
to a transitional area containing cells dominated by detections classified as \texttt{neoAGNs} and 
\texttt{neoYSOs}. A few additional detections are scattered across the whole SOM map, with two small groups, 
one located in the center of the SOM map in an area again mostly populated by 
detections labeled as \texttt{AGN} and \texttt{YSO}, and the other in the upper left corner respectively, 
where neurons are mostly dominated by AGNs.
The positions of the \texttt{PartOfG} detections on the SOM map suggest that the typical values of the 
X-ray \textit{features} of these sources resemble those usually associated with AGNs or young stellar objects. 
The inspection of the sky positions of the three CSC sources in the \texttt{PartOfG} sample with the largest 
number of single detections (\texttt{2CXO J123048.6+122332} with 124 detections, \texttt{2CXO J123049.2+122328} with 121 and 
\texttt{2CXO J123049.3+122328} with 79 accounting, together, for $\sim\!76\%$ of all \texttt{PartOfG} labels) reveals 
that these X-ray sources are associated with three spatially resolved knots of the relativistic 
jet emitted by the nearby radio-galaxy M87~\citep{marshall2002}. Their X-ray emission is 
usually attributed to the synchrotron mechanism, one of the most common non-thermal emission mechanisms
responsible for the X-ray emission of supernova remnants and radio galaxies (i.e.,  systems where particles
can be accelerated to relativistic velocities). While it is reasonable to assume that the nature of the 
remaining detections labeled as \texttt{PartOfG} may be similar to that of the sources that have been 
investigated in detail, it will only be confirmed through follow-up observations in the 
X-rays and other wavelengths.

This analysis underscores how a trained SOM can be used to investigate the 
distribution of similar classified sources on the projected SOM map to determine the 
potential common emission mechanisms responsible for their similarities. We
also highlighted the presence of clear spatial patterns in the distribution of labeled 
detections on the SOM projected map, which can be used to refine our knowledge of the 
nature of uncertain or unlabeled sources. 
A statistically robust method that takes advantage of the trained SOM and of the subset of labeled 
detections in the ACIS-SV-CSC dataset as training set to classify the unlabeled
X-ray sources according to the aggregated SIMBAD labels is presented in the next section.

\subsection{Classification}
\label{subsec:classification}

We perform detection-level classification for detections with missing or uncertain labels by leveraging the 
topology of the trained SOM. In order to do so, we retrained the SOM 
on the same ACIS-SV-CSC dataset using only the training and unlabeled subsets, holding out the 
validation and test subsets; this yields a quantization error (QE) of 0.16 and a topographic error 
(TE) of 0.03. By taking advantage of the 
trained SOM map, we build \textit{SOMClassifier}, a local-neighborhood method that assigns labels by pooling posterior 
support around the BMU of each detection on the hexagonal grid, thereby respecting the map topology and 
preserving neighborhood consistency. We partition the detections contained in the ACIS-SV-CSC dataset into 
training/validation/test (70\%/15\%/15\%) with a source-grouped scheme, so all detections from the same source fall 
into exactly one split (no cross-split leakage).

The \textit{SOMClassifier} assigns a class label to a given detection by evaluating the local neighborhood around its BMU. Rather than relying solely on the BMU, the classifier aggregates information from nearby neurons, adjusting their influence based on how similar they are to the BMU. To ensure high-confidence predictions, the model only outputs a label if the surrounding neighborhood satisfies statistical criteria; otherwise, it abstains (i.e., it does not return a classification for that detection). We consider a classification \textit{valid} when these conditions, detailed below as support, purity, and margin thresholds, are fully met. The step-by-step procedure is described next.

The first step involves using training detections to accumulate a per-cell, per-class histogram hist, 
where \( \mathrm{hist}[i,j,k] \) is the number of training detections of class \(k\) whose BMU is cell \( (i,j) \). 
We also compute per-class totals \( \mathrm{class\_counts}[k] = \sum_{i,j} \mathrm{hist}[i,j,k] \).

We mitigate class imbalance with a global per-class weight applied to the training histograms, with \(\gamma\) 
controlling the strength of reweighting (a free hyperparameter to be tuned). Let \(c\) be the mean of the per-class 
counts (over \(k\)) and \(\varepsilon=10^{-9}\).
The tempered weight is
\[
w_k=\left(\frac{c}{\text{class\_counts}_k+\varepsilon}\right)^{\gamma}.
\]

For the BMU \((i,j)\) of each detection, we consider its hex-grid neighbors within graph radius \(R\). Each neighbor \((u,v)\) is weighted by the similarity between its prototype \(\mathbf{w}_{uv}\) and the 
BMU prototype \(\mathbf{w}_{ij}\), measured with a radial-basis kernel whose width is controlled by \(\sigma_f\). In this 
way, prototype similarity determines how strongly each neuron contributes: neighbors whose prototypes are closer 
to \(\mathbf{w}_{ij}\) contribute more. The resulting weighted sums yield per-cell, per-class counts, which are then 
converted into Laplace-smoothed posteriors.

In order to assign a class to a detection with BMU \((i,j)\), we consider a predefined set of hex-grid radii and select 
the smallest \(R\) such that the pooled support \(S_R(i,j)\ge\text{MIN\_SUPPORT}\); if no such radius exists, the 
classifier abstains. Given \(R\), let \(\mathbf{p}_R(i,j)\) denote the Laplace-smoothed posterior over classes. 
We assign
\[
\hat{y}(i,j)=\arg\max_k p_{R,k}(i,j)
\]
with confidence \(\max_k p_{R,k}(i,j)\) if both thresholds below are satisfied; otherwise, we abstain:
\begin{itemize}
  \item \textbf{Purity}: \(\max_k p_{R,k}(i,j) \ge \text{MIN\_PURITY}\)
  \item \textbf{Margin}: \(p_{(1)}(i,j)-p_{(2)}(i,j) \ge \text{MARGIN}\), where \(p_{(1)}\ge p_{(2)}\) are the top two components of \(\mathbf{p}_R(i,j)\)
\end{itemize}

\subsubsection{Hyperparameters and tuning}
\label{subsubsec:hyper}

The \textit{SOMClassifier} has a set of hyperparameters that regulate neighborhood pooling on the 
hex grid, Laplace smoothing, class reweighting, and abstention thresholds. We tune them on the validation split 
to maximize macro-$F_1$, 
then fix the final setting before any test evaluation. After this tuning process, we adopt the following setting: we use 
the radius set $R\in\{1,2,3,4\}$, Laplace parameter $\alpha=0.5$, class-reweighting exponent $\gamma=0.7$, support 
threshold \texttt{MIN\_SUPPORT} $=1$, purity threshold \texttt{MIN\_PURITY} $=0.20$, margin threshold \texttt{MARGIN} $=0.3$.
We also set $\sigma_f=0.05$ by measuring the \textit{feature}-space step between adjacent neurons on the 
trained SOM: for each neuron, we computed the Euclidean distances between its weight vector and those of 
its six immediate hex-grid neighbors across the map. The median of these distances is $0.104$, and we 
set $\sigma_f$ to the value closest to this median that maximized macro-$F_1$.

Our classification considers as target labels the four aggregated classes described in
Section~\ref{subsec:simbad}:~\texttt{neoYSOs} (6,008 detections),~\texttt{Stars} (1,723 detections), \texttt{neoAGNs} (5,957 detections)~and \texttt{Binaries} (3,250 detections). 
The exact mapping from SIMBAD types to our four targets follows the procedure described in the SIMBAD cross-match section 
(Section~\ref{subsec:simbad}). The generic labels from SIMBAD \texttt{X} and \texttt{PartOfG} are excluded because they are too vague to support reliable detection-level classification.

\subsubsection{Validation and testing}
\label{subsubsec:validationtesting}

Hyperparameters are fixed by maximizing macro-$F_1$ under a coverage constraint ($\ge 50\%$) on the 
validation split. Table~\ref{tab:metrics} reports macro- and micro-$F_1$, per-class $F_1$ scores, and the 
number of valid (non-abstained) detections over the total $N$ for validation and test.

\begin{deluxetable}{lccccccc}
\tablecaption{Metrics for the selected configuration (valid = non-abstained).}
\label{tab:metrics}
\tablehead{
\colhead{Split} &
\colhead{Macro-$F_1$} &
\colhead{Micro-$F_1$} &
\colhead{Valid / $N$} &
\colhead{\texttt{neoYSOs} $F_1$} &
\colhead{\texttt{neoAGNs} $F_1$} &
\colhead{\texttt{Binaries} $F_1$} &
\colhead{\texttt{Stars} $F_1$}
}
\startdata
Val  & 0.665 & 0.756 & 1273 / 2500 & 0.796 & 0.857 & 0.428 & 0.577 \\
Test & 0.654 & 0.737 & 1446 / 2700 & 0.760 & 0.830 & 0.551 & 0.474 \\
\enddata
\end{deluxetable}

Depending on the application, the hyperparameters can be relaxed or tightened to trade precision for coverage. Overall, \texttt{neoYSOs} and \texttt{neoAGNs} achieve the highest $F_1$ scores. They are also the best represented 
classes in the dataset, yielding stronger neighborhood support on the map and cleaner posteriors. \texttt{Stars} also perform reasonably well, while \texttt{Binaries} remain the most challenging class.

Table~\ref{tab:cm} reports the confusion matrix for the valid (non-abstained) predictions on the \textit{test set} (rows=true, columns=predicted). For the aggregated \texttt{Stars} class, only $\sim\!48\%$ of the detections are correctly classified, while the majority of the misclassifications are \texttt{Stars}$\rightarrow$\texttt{neoYSOs} (36 of 59 errors; $61\%$), which can be caused by either the~\texttt{neoYSOs} label containing several stellar subclasses whose properties are similar to those of the detections included in the \texttt{Stars} aggregated class, at least in the limited \textit{feature} 
space explored by the SOM in this paper, or by detections labeled as \texttt{Stars} that are misclassified as~\texttt{neoYSOs}. For \texttt{Binaries}, the most significant
confusion is \texttt{Binaries}$\rightarrow$\texttt{neoAGNs} (73 of 128 errors; $57\%$), which is likely caused by the similar structure of these two classes of 
sources, where the same physical components, i.e., an accretion disk and a corona, emit in the X-ray due to similar physical mechanisms (i.e., black-body emission and IC scattering, respectively). The similarities in the X-ray observational properties of these two aggregated classes are evident in the projected SOM map of the dominant aggregated labels~(lower left plot in Figure~\ref{fig:main_type_cluster}), where cells dominated by the class~\texttt{Binaries} are located near \texttt{neoAGNs}-rich cells.

\begin{deluxetable}{lrrrr}
\tablecaption{Confusion matrix on the \textit{test set} (valid = non-abstained; rows = true, columns = predicted).}
\label{tab:cm}
\tablehead{
\colhead{} &
\colhead{\textbf{P:\texttt{neoYSOs}}} &
\colhead{\textbf{P:\texttt{Stars}}} &
\colhead{\textbf{P:\texttt{neoAGNs}}} &
\colhead{\textbf{P:\texttt{Binaries}}}
}
\startdata
\textbf{T:\texttt{neoYSOs}}   & 351 &  45 &  35 &  23 \\
\textbf{T:\texttt{Stars}}     &  36 &  55 &  22 &   1 \\
\textbf{T:\texttt{neoAGNs}}   &  39 &   7 & 538 &  45 \\
\textbf{T:\texttt{Binaries}}  &  44 &  11 &  73 & 121 \\
\enddata
\end{deluxetable}

The ACIS-SV-CSC dataset contains multiple detections associated with the same astrophysical source, but our 
classification may produce different classifications for detections of the same source.
For this reason, it is interesting to investigate the per-source aggregated 
classifications of detections. Table~\ref{tab:cm_sources} reports a 
per-source confusion matrix for sources in the \textit{test set}: off-diagonal entries count the number of \emph{sources} of true class $C_i$ 
for which \emph{at least one} valid detection is predicted as $C_j$ ($j\neq i$), and we report 
the corresponding fraction with respect to the total number of sources in row $C_i$. Diagonal 
entries instead report the fraction of sources whose valid detections are 
\emph{all} assigned to the true class of the sources. Since a single source may 
have detections assigned to multiple classes, row sums are not normalized to unity. 

\begin{deluxetable}{lrrrr}
\tablecaption{Per-source confusion summary on the \textit{test set} (valid = non-abstained). Off-diagonal entries report the number (and fraction) of sources of true class $C_i$ for which at least one detection is predicted as $C_j$ ($j \neq i$). Diagonal entries report sources for which \emph{all} detections are predicted as the true class.}
\label{tab:cm_sources}
\tablehead{
\colhead{} &
\colhead{\textbf{P:\texttt{neoYSOs}}} &
\colhead{\textbf{P:\texttt{Stars}}} &
\colhead{\textbf{P:\texttt{neoAGNs}}} &
\colhead{\textbf{P:\texttt{Binaries}}}
}
\startdata
\textbf{T:\texttt{neoYSOs}} (235)   & 164 (69.8\%) & 26 (11.1\%) & 29 (12.3\%) & 21 (8.9\%) \\
\textbf{T:\texttt{Stars}} (68)      &  27 (39.7\%) & 30 (44.1\%) & 10 (14.7\%) &  1 (1.5\%) \\
\textbf{T:\texttt{neoAGNs}} (355)   &  25 (7.0\%)  &  7 (2.0\%)  & 294 (82.8\%) & 33 (9.3\%) \\
\textbf{T:\texttt{Binaries}} (64)   &  19 (29.7\%) &  9 (14.1\%) & 27 (42.2\%) & 16 (25.0\%) \\
\enddata
\end{deluxetable}

In general, among the 216 multi-detection sources in the test set (i.e., excluding the 506 
single-detection sources), 
agreement remains high: 152 sources (70.4\%) are unanimous, while 64 sources (29.6\%) show variability 
across detections. The most likely detection-level misclassification associated 
with~\texttt{neoYSOs} sources is~\texttt{neoAGNs}, while $\sim\!40\%$
of the sources classified as~\texttt{Stars} have at least one detection classified 
as~\texttt{neoYSOs}. Detections belonging to sources in the~\texttt{neoAGNs} class
are very often assigned to the correct label, with $\sim\!82\%$ of the confirmed~\texttt{neoAGNs} sources 
having no mis-classified detections. On the other 
hand,~\texttt{Binaries} with multiple detections are often mis-classified: 
$\sim\!42\%$ of sources whose real class is~\texttt{Binaries} have at least
one detection classified as~\texttt{neoAGNs}, with $\sim\!30\%$ and $15\%$ of the 
confirmed~\texttt{Binaries} sources have at least
one detection classified as either~\texttt{neoYSOs}
or~\texttt{Stars}, respectively. This pattern suggests that the X-ray observational 
properties of binaries, observed at different epochs, can vary sufficiently to cause 
significant changes to their positions on the SOM map and, in turn, their per-detection 
classification.

\subsubsection{Classification of unlabeled detections}

We applied \textit{SOMClassifier} to 11,712 ACIS-SV-CSC detections without SIMBAD labels. The classifier 
assigned valid labels to 6,225 detections (coverage $\sim\!53\%$) and abstained on 5,487.

\begin{figure}[t]
\begin{center}
    \includegraphics[]{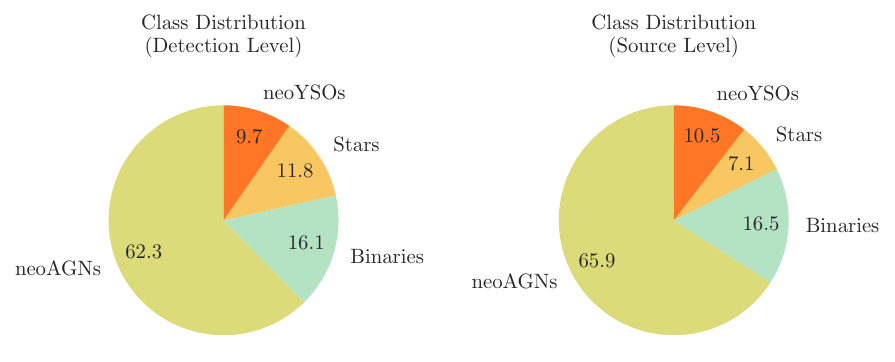}%
    \caption{Class distributions for the unlabeled sample classified with \textit{SOMClassifier}. Left: detection-level distribution over the four target classes. Right: source-level distribution after aggregating per-detection predictions. Values shown inside the charts are percentages.}
    \label{fig:unlab_class_dist}
\end{center} 
\end{figure}

Among the valid classifications retained, the class distribution is the following: \texttt{neoAGNs}
3,881 ($62.35\%$),~\texttt{Binaries} 1,005 ($16.14\%$),~\texttt{Stars} 737 ($11.84\%$), 
and~\texttt{neoYSOs} 602 ($9.67\%$). Posterior confidence (per-class median) is highest for 
\texttt{neoAGNs} ($0.74$), and remains solid for \texttt{Stars} ($0.67$), \texttt{neoYSOs}
($0.64$), and \texttt{Binaries} ($0.63$). The overall median purity across valid predictions is $0.7$, 
and the effective neighborhood support is solid (median $15.7$). The detection-level class distribution is summarized 
in the left panel of Figure~\ref{fig:unlab_class_dist}.

\begin{figure*}[t]
\centering
\includegraphics[width=\linewidth,height=0.91\textheight,keepaspectratio]{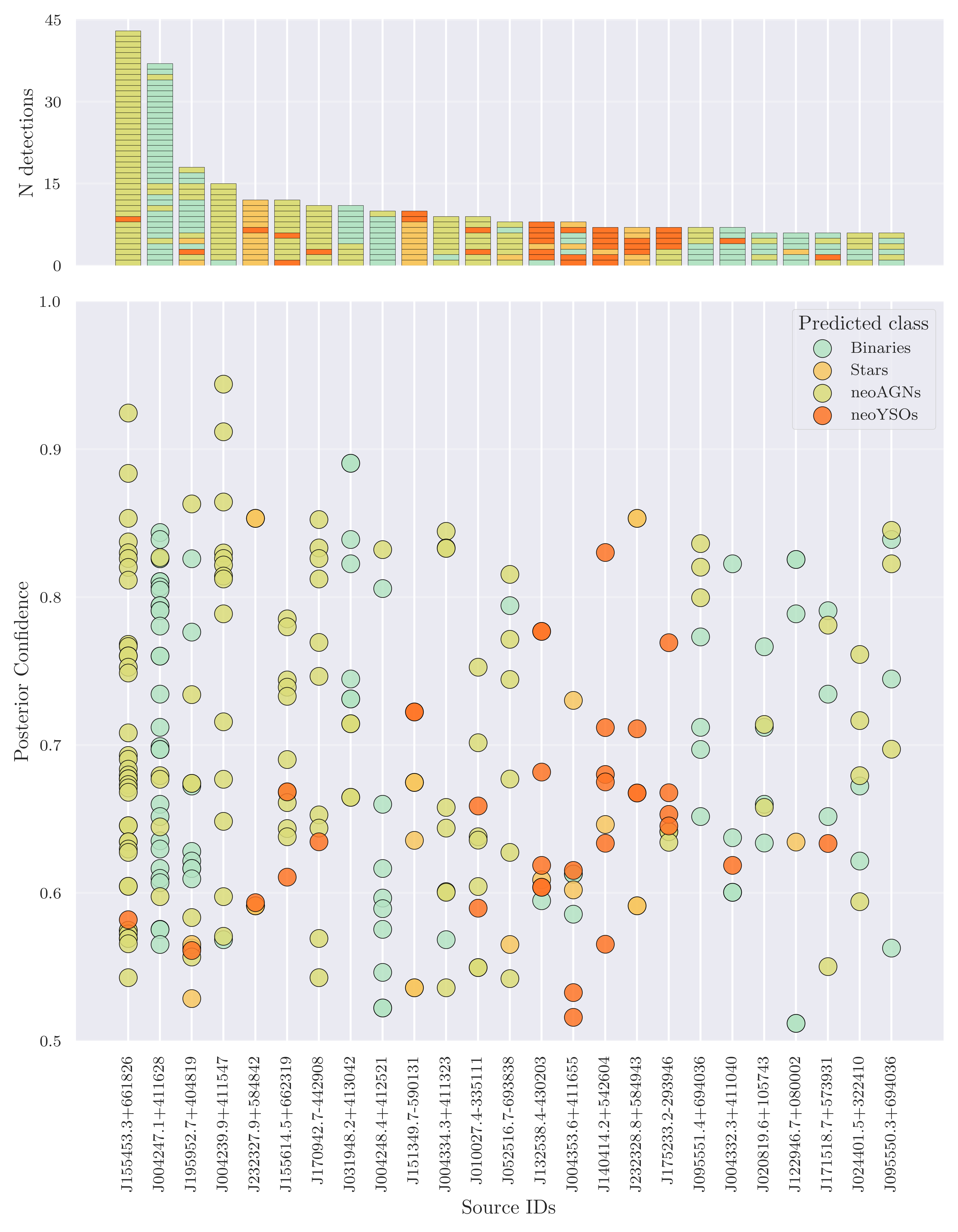}
\caption{\textit{SOMClassifier} predictions for the most frequently detected sources with non-unanimous per-detection classifications. The main panel displays 
the posterior confidence for 
each individual detection, also colored by predicted class. The top panel shows the number of (non-abstained) 
detections per source. The ``2CXO'' prefix has been omitted from the source names for readability.}
\label{fig:unlab_source_consistency}
\end{figure*}

\textit{SOMClassifier} can also be used to determine source-level classifications for sources with multiple
detections. We aggregated per-detection predictions at the source level (7,849 unique sources) by majority vote with a 
confidence-based tie break. This yields 4,577 sources with at least one classification. The resulting per-source class 
distribution is:~\texttt{neoAGNs} 3,016 ($65.89\%$),~\texttt{Binaries} 754 
($16.47\%$),~\texttt{neoYSOs} 480 ($10.49\%$),~\texttt{Stars} 327 ($7.14\%$). 
Relative to the detection-level, consolidation shifts weight from~\texttt{Stars} 
to~\texttt{neoAGNs}/\texttt{neoYSOs}. Among the 762 multi-detection sources, agreement is high 
($77.17\%$ of source classifications are unanimous, i.e., all the detections of the 
same source are assigned the same class), and the mean agreement across sources is $90.50\%$. The right panel 
of Figure~\ref{fig:unlab_class_dist} shows the source-level distribution of classifications. 
Figure~\ref{fig:unlab_source_consistency} shows how
\textit{SOMClassifier} predictions vary across multiple detections of the
same source, for sources with non-unanimous per-detection classifications. The main panel shows the posterior confidence of each individual detection, 
colored by its predicted class, for the sources with the most classified detections (sorted by decreasing 
number of detections). The top panel displays the number of detections per source.

Among the classification results, we identified 2,100 detections ($17.93\%$) whose BMU lies on the outer boundary of the SOM. 
For these detections located in marginal cells of the SOM map, the hex-grid neighborhood is truncated 
and asymmetric, resulting in a decreased effective support 
and a potentially biased RBF-weighted aggregation. As a result, their posteriors tend to be 
less reliable than interior neurons.
The full table of classification results, including per-detection and per-source assignments, posterior confidences, and abstention flags, is publicly available on Zenodo {\citep{scaudo2026_zenodo}}\footnote{\href{https://doi.org/10.5281/zenodo.20932854}{doi:10.5281/zenodo.20932854}}. To facilitate the use of these data, we calculated and included a \texttt{heterogeneity\_score} in the per-source table. This classified-only score provides a simple way to identify sources with the most heterogeneous detection-level classifications, which may be particularly interesting for examining the diversity of classification outcomes across their detections.

To assess the robustness of these source-level classifications, we compare them with the source classifications reported 
by~\citet{2024MNRAS.528.4852P}. Although that study adopts a Gaussian Mixture Model (GMM), a different unsupervised machine-learning approach, and was 
developed with different goals in mind, the comparison remains informative because both analyses characterize the topological 
properties of the distribution of the same set of CSC spectral and variability \textit{features} and employ the same SIMBAD-derived 
labels as a training set for their classification method.

For this purpose, we applied the same aggregation scheme introduced in Section~\ref{subsec:simbad} to their SIMBAD-derived 
labels. Because their target classification does not include an equivalent to our~\texttt{Stars} category, we excluded it from the comparison and restricted the analysis to the three shared classes:~\texttt{neoAGNs},~\texttt{Binaries}, 
and~\texttt{neoYSOs}.

\begin{figure}[t]
\begin{center}
    \includegraphics[width=0.5\textwidth]{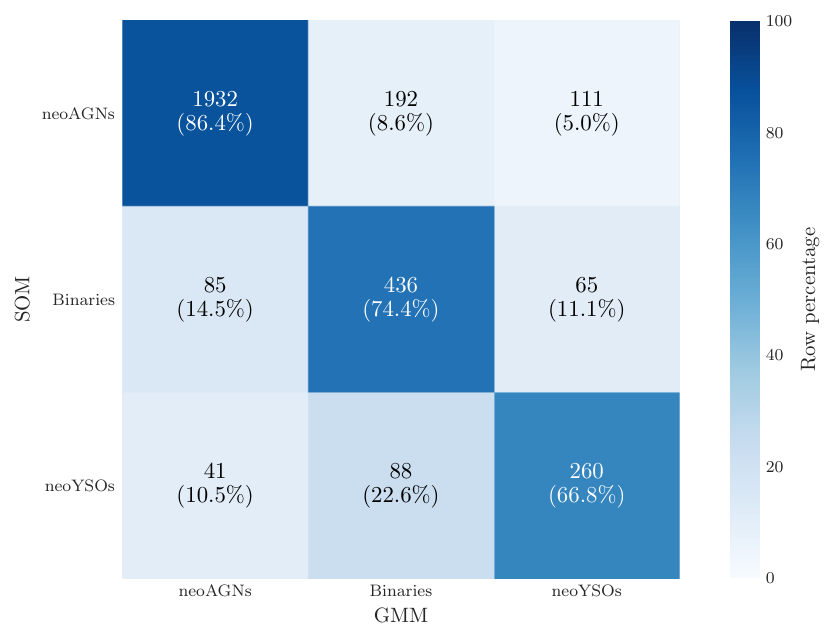}%
    \caption{Row-normalized confusion matrix comparing our source-level classifications with those of~\citet{2024MNRAS.528.4852P}. Their SIMBAD-derived labels were remapped using the aggregation scheme of Section~\ref{subsec:simbad}.}
    \label{fig:confusion_matrix_vs_GMM}
\end{center} 
\end{figure}

The confusion matrix presented in Figure~\ref{fig:confusion_matrix_vs_GMM} compares the classifications obtained with 
the two approaches and shows a good overall agreement, with 86.4\% of~\texttt{neoAGNs}
also classified as~\texttt{neoAGNs} by the GMM method, 74.4\% agreement for~\texttt{Binaries}, and 
66.8\% for~\texttt{neoYSOs}. The most significant cross-contamination is observed between~\texttt{Binaries} 
and~\texttt{neoYSOs}, similarly to what is observed for the classification
based on the SOM map alone (see Table~\ref{tab:cm}), further suggesting that these two macro-classes
share similar traits that make their separation sometimes difficult. A consensus classification
based on this work and the~\citet{2024MNRAS.528.4852P} results is outside the scope of this paper, but
would be beneficial to determine high-likelihood X-ray sources for each of the merged categories 
based on SIMBAD labels, and \textit{bona fide} sources whose X-ray properties consistently
change across different detections as candidates for follow-up observations.

\section{Comparison with other models}
\begin{figure}[t]
\begin{center}
\includegraphics[width=\linewidth]{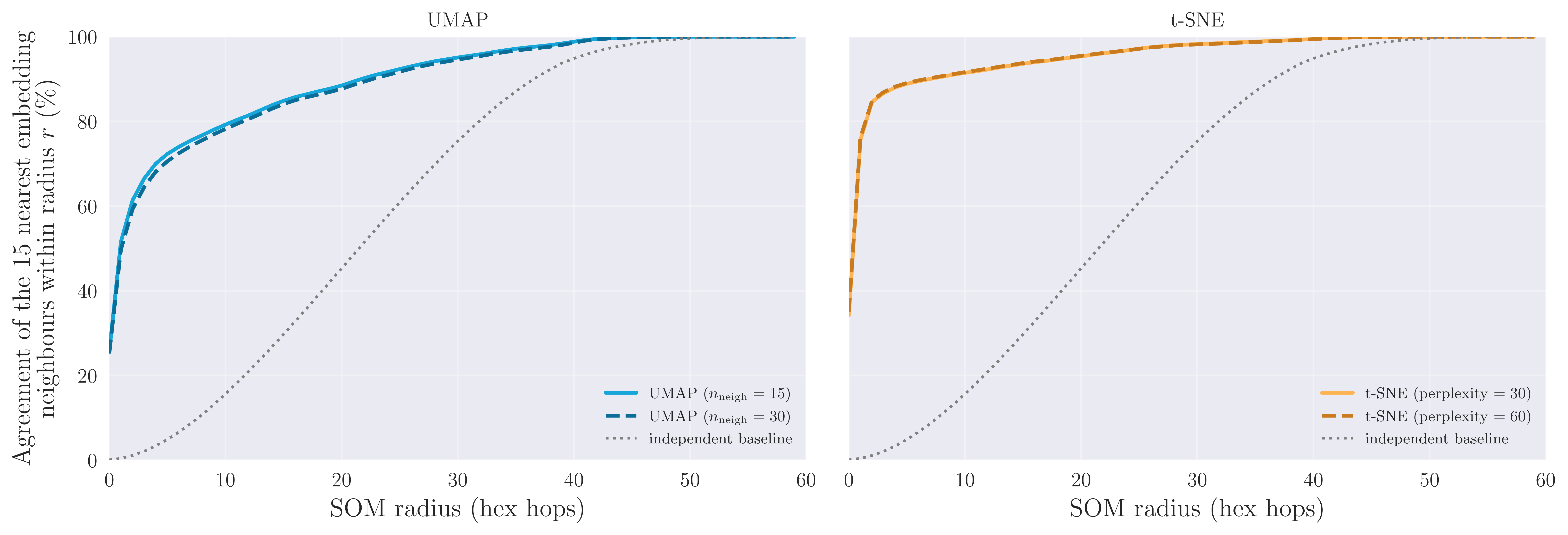}
\caption{Cumulative fraction of the 15 nearest neighbors per point in the
UMAP (\textit{left}) and t-SNE (\textit{right}) embeddings that fall within a
radius $r$ (hexagonal hops) of the point's BMU on the SOM grid, compared with
the independent (chance) baseline probability. Curves are shown for 
two sets of hyper-parameters for each model. The two curves for the t-SNE model overlap almost exactly.}
{\label{fig:step5_radius_local_agreement}}
\end{center} 
\end{figure}

We validate the low-dimensionality projection produced by the SOM model applied on our set of CSC 2.1.1 detections by training other models that produce two-dimensional embeddings of high-dimensional datasets, and comparing the results. We applied two widely used techniques for non-linear dimensionality reduction, UMAP~\citep{2018arXiv180203426M} and t-SNE~\citep{JMLR:v9:vandermaaten08a}, to 
the same set of ACIS-SV-CSC detections and compared the structure of the resulting embeddings with projection produced by the SOM. Figure~\ref{fig:step5_radius_local_agreement} displays a diagnostic of the agreement between SOM and UMAP (left) and t-SNE (right) models. In order to quantitatively assess the similarities of the two-dimensional embeddings produced by the three different models, we use a metric defined as follows: for each detection in the SOM map, we determined its position in the UMAP and t-SNE embeddings, and its 15 nearest neighbors in these embeddings, and measured the fraction of those detections that fall within a radius $r$ (in hexagonal hops) from its BMU on the SOM grid. 

\begin{figure}[t]
\begin{center}
\includegraphics[width=\linewidth]{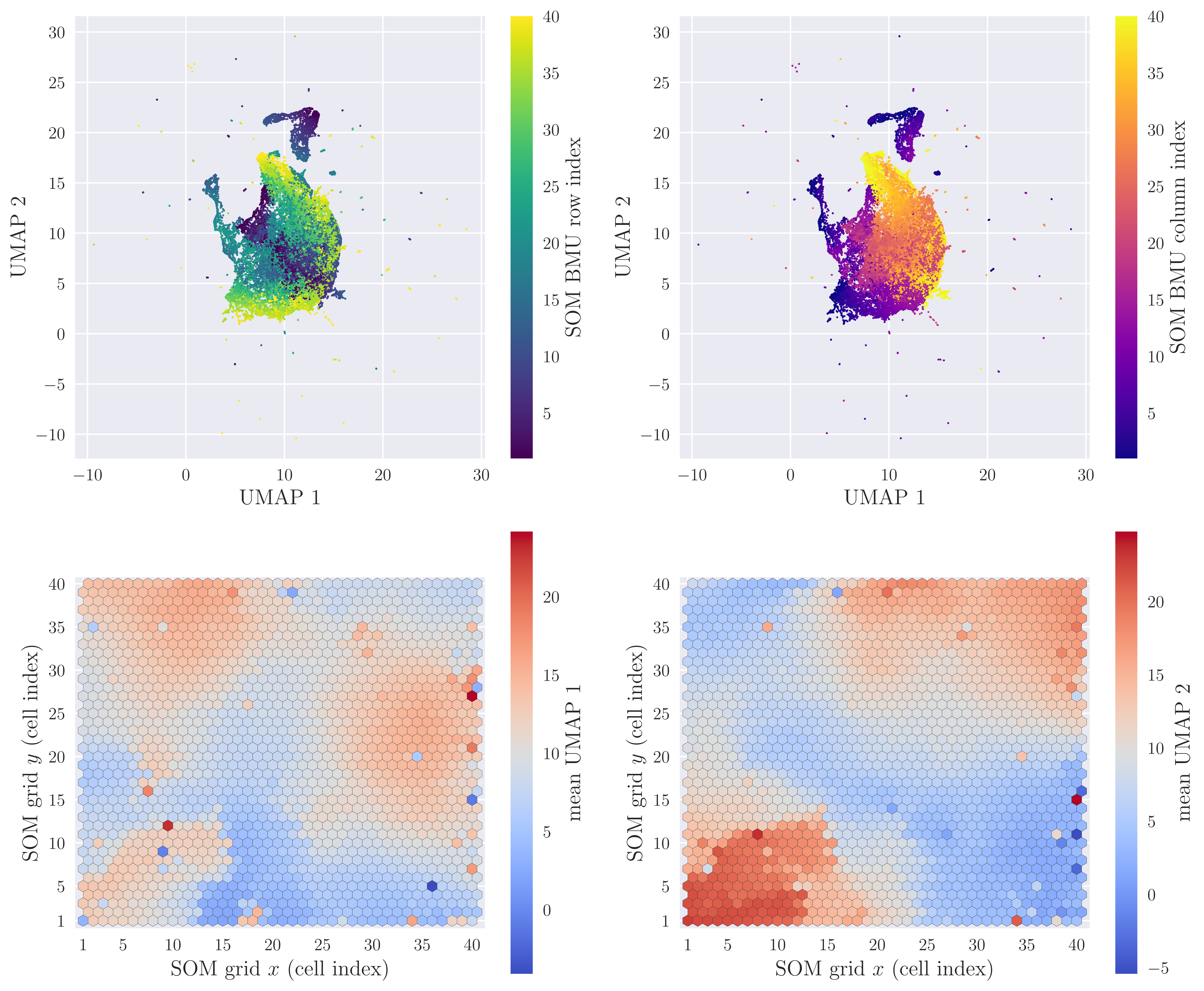}
\caption{Upper plots: UMAP embedding of the ACIS-SV-CSC detections colored by the SOM grid
row (\textit{left}) and column (\textit{right}) indeces of their BMU.
\textit{Lower plots:} mean UMAP coordinates of the detections assigned to each SOM
cell along the axis 1 (left) and axis 2 (right) of the UMAP map.}
{\label{fig:step5_umap_nn15_som_relationships}}
\end{center}
\end{figure}

\begin{figure}[t]
\begin{center}
\includegraphics[width=\linewidth]{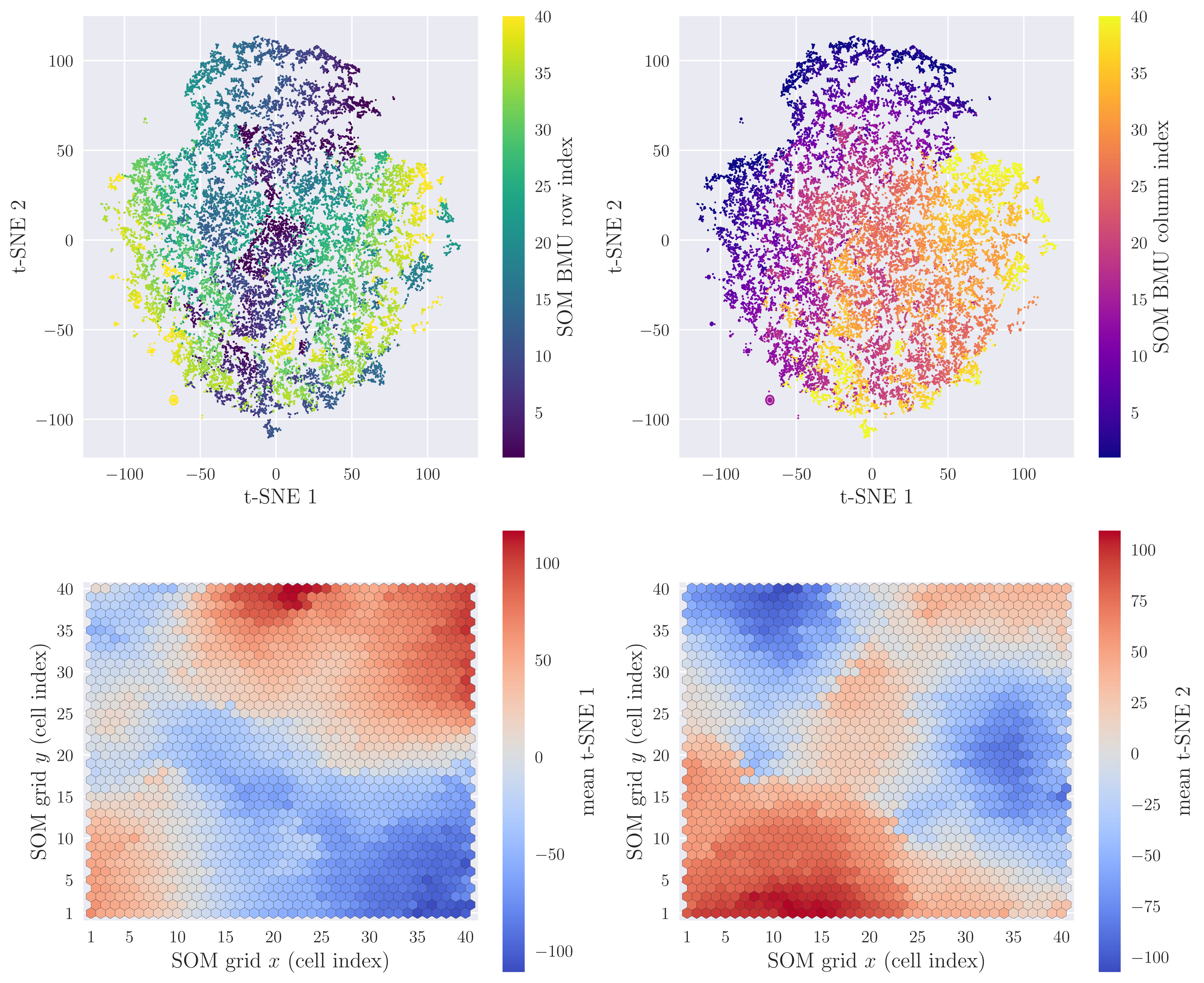}
\caption{As in Figure~\ref{fig:step5_umap_nn15_som_relationships} for t-SNE embedding of the ACIS-SV-CSC detections.}
{\label{fig:step5_tsne_perp30_som_relationships}}
\end{center} 
\end{figure}

At $r=1$ (same or adjacent cell), the agreement is $51.6\%$ for UMAP and $75.6\%$ for t-SNE, against the probability that two detections drawn at random are assigned to the same or to adjacent BMUs, which is only $\sim\!0.4\%$. The curves
are insensitive to the UMAP neighborhood size (15 vs.\ 30) and to the t-SNE perplexity (30 vs.\ 60), indicating that the agreement is not an artifact of the embedding hyperparameters. Figure~\ref{fig:step5_umap_nn15_som_relationships}
and~\ref{fig:step5_tsne_perp30_som_relationships} provides a visualization of the level of correlation between the two-dimensional maps produced by UMAP and t-SNE, respectively, and SOM from a different perspective. In the upper plots of both Figures each point in the embedding produced by the UMAP or t-SNE models is colored according to the row (left) and column (right) indices of its BMU, as a proxy of its position in the SOM map. The lower plots of Figure~\ref{fig:step5_umap_nn15_som_relationships} and 
Figure~\ref{fig:step5_tsne_perp30_som_relationships} show the SOM map where each cell is colored according to the average UMAP and t-SNE coordinates along the x (left) and y (right) axes, respectively, of all the detections associated with the cell. All plots display large-scale spatial patterns characterized by smooth gradients, clearly showing that the embeddings obtained by the two different methods are correlated and that nearby points in the UMAP and t-SNE spaces are mapped to nearby cells on the SOM grid, and, conversely, the mean embedding coordinates of the detections assigned to each SOM cell vary smoothly across the maps produced by the UMAP and t-SNE models. 

We also verified that this agreement does not depend on the specific SOM configuration by retraining the SOM with a different number of neurons ($20\times20$ and $60\times60$ hexagonal grids) or a different topology ($40\times40$ rectangular grid), while keeping all other hyper-parameters unchanged. The fraction of UMAP neighbors falling within $r=1$ decreases as the number of neurons increases ($69.8\%$, $51.6\%$, and $43.1\%$ for the $20\times20$, $40\times40$, and $60\times60$ grids, respectively), together with the chance baseline ($1.61\%$, $0.44\%$, and $0.20\%$), since at higher SOM resolution the neighborhood of each neuron represents a smaller fraction of the \textit{feature} space, while the rectangular grid yields an agreement ($55.0\%$) similar to that of the hexagonal grid of the same size.

Overall, the structure of the map produced by our SOM model shows strong \textit{local} coherence with the independent nonlinear embeddings produced by both the UMAP and t-SNE methods, confirming the SOM as an interpretable representation of the local structure of the ACIS-SV-CSC dataset in its \textit{feature} space.

\section{Conclusion}
\label{sec:conclusion}
SOMs provide a locally topology-preserving projection that is well suited to revealing structure in high-dimensional astronomical datasets. We applied this technique to the CSC 2.1.1, mapping observation-level X-ray detections onto a two-dimensional grid based on their spectral and variability properties. We extracted the ACIS-SV-CSC dataset from the full catalog, obtaining 36,904 high statistical significance ACIS-only detections belonging to 17,927 unique sources, and trained a 40$\times$40 hexagonal SOM on 11 distance-independent \textit{features}. A positional $1\arcsec$ cross-match with the SIMBAD database provided 
astrophysical labels for 25,192 of these detections ($\sim\!68\%$), which we consolidated into four aggregated target classes at a comparable hierarchical level:~\texttt{neoYSOs},~\texttt{neoAGNs},~\texttt{Binaries}, and~\texttt{Stars}. 
The \textit{feature}-projection maps of the trained SOM reveal coherent gradients in spectral shape and variability behavior 
across the two-dimensional grid, reflecting the diversity and continuous transition of physical emission 
mechanisms present in the population of X-ray sources in the CSC. We investigated the nature of the detections
and interpret the observed patterns followed by the \textit{features} on the SOM map by overlaying SIMBAD-derived labels, assumed to be representative of the real nature of the subtending sources, that are available for a subset of ACIS-SV-CSC detections. The analysis of the most represented labels per cell and as a function of the positions of the cell 
on the SOM map confirms that, while physically distinct populations (identified either according to the original inhomogeneous SIMBAD \texttt{main\_type} labels or 
using the four statistically
similar and of comparable granularity aggregated classes~\texttt{neoAGNs}, \texttt{neoYSOs}, \texttt{Binaries},
and~\texttt{Stars}) tend to occupy 
different regions of the SOM grid, different classes of sources with similar emission mechanisms can occupy
the same regions on the SOM map and cannot be reliably disentangled using exclusively the X-ray 
spectral and variability \textit{features} employed in this paper. We discuss these findings in the context of 
the different observational strategies and diverse scientific objectives that contribute to the heterogeneous nature of the
sample of X-ray sources represented in the CSC.

In order to leverage these topological structures highlighted by the SOM mapping of the 
original \textit{feature} space distribution of labeled ACIS-SV-CSC sources for classification, we introduced 
\textit{SOMClassifier}, a local-neighborhood method that assigns labels by pooling posterior support around the best-matching 
unit of each detection on the hex grid. On the test set, \textit{SOMClassifier} achieved the highest per-class 
scores for~\texttt{neoAGNs} and~\texttt{neoYSOs}, reflecting their stronger training-set representation and more distinctive 
spectral and variability signatures, while the performance of the classification of sources in the~\texttt{Binaries} 
and~\texttt{Stars} classes proved to be overall worse, likely due to the intrinsic heterogeneity of the \textit{feature} distribution of these two classes. Applied to the 11,712 detections in the ACIS-SV-CSC dataset lacking 
SIMBAD labels, \textit{SOMClassifier} assigned valid (non-abstained) classifications to $\sim\!53\%$ of them. The aggregation of these predictions at the source level by majority vote yielded unique classifications for 4,577 distinct sources; among the 762 multi-detection sources, 77.17\% received unanimous per-detection labels, 
demonstrating high consistency.

The trained SOM model, the \textit{SOMClassifier} implementation, and interactive visualization and classification workflows are 
publicly available through a web application~\footnote{\href{https://somcsc.streamlit.app/}{somcsc.streamlit.app}}. Future work includes 
the investigation of general astronomical questions, like the characterization of the 
time-dependent variability of the X-ray observable across multi-detection sources spanning wide regions across the 2D map and the study of the drivers behind their changeable behavior. 
We will also characterize the position on the SOM map of interesting, rare, and/or transient X-ray sources, selected based 
on their peculiar spectral and time-dependent characteristics, and use this information to extract potential unknown 
candidate X-ray sources of similar traits for follow-up investigations.

\begin{acknowledgments}
\noindent A.S. is supported by Italian Research Center on High Performance Computing Big Data and Quantum Computing (ICSC), project funded by European Union - NextGenerationEU - and National Recovery and Resilience Plan (NRRP) - Mission 4 Component 2 within the activities of Spoke 2 (Fundamental Research and Space Economy), (CN 00000013 - CUP C53C22000350006). 
R.D'A. is supported by NASA contract NAS8-03060 (\textit{Chandra} X-ray Center). R.M.G. acknowledges support from AstroAI.
\end{acknowledgments}

\appendix

\section{Web application}
\label{sec:webapp}
To support the exploration and understanding of the trained Self-Organizing Map (SOM), we created an interactive web interface using the \texttt{Streamlit} platform, accessible at \href{https://somcsc.streamlit.app/}{somcsc.streamlit.app}. This tool enables users to engage with various 2D maps produced by the SOM. These visualizations allow researchers to deepen into different characteristics of the data. The plots presented in this Appendix are shown for reference only and are not directly associated with the results discussed in the main body of the paper; they are included solely to illustrate the functionality of the web app. Each \textit{feature} projection tool within the interface provides two display options: a \textit{scatter view}, where individual detections are plotted as points colored according to their \textit{feature} values, and a \textit{hexbin view}, which emphasizes the most common element within each neuron. The subsections below describe the range of visualization options available in the web interface.

\subsection{The U-Matrix map}
The U-Matrix, shown in Figure~\ref{fig:som_analysis_row1} (left), illustrates the distances between adjacent neurons in the SOM using a color gradient from red to green. This visualization is especially effective for revealing the cluster organization of the data. Areas in green signify greater separation between neurons, marking cluster boundaries, whereas red areas indicate closer similarity among neurons, often within the same cluster.
The U-Matrix is calculated by measuring the distances between the weight vectors of neighboring neurons in the trained SOM grid. For each neuron, the average distance to its immediate neighbors is computed and assigned as the U-Matrix value of that neuron.
\subsection{The Activation Response map}
\begin{figure*}[htbp]
    \includegraphics[width=0.48\textwidth]{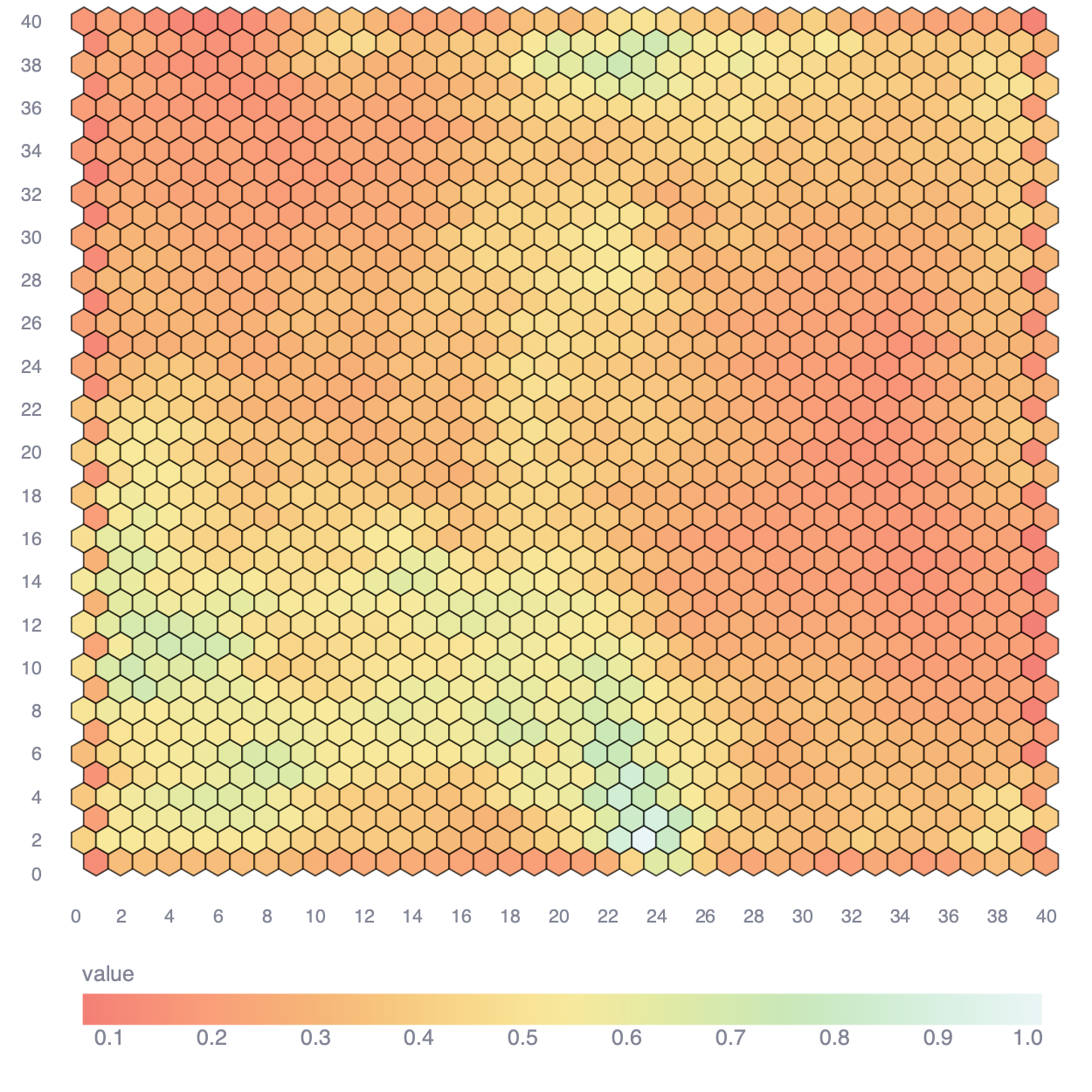}
    \includegraphics[width=0.48\textwidth]{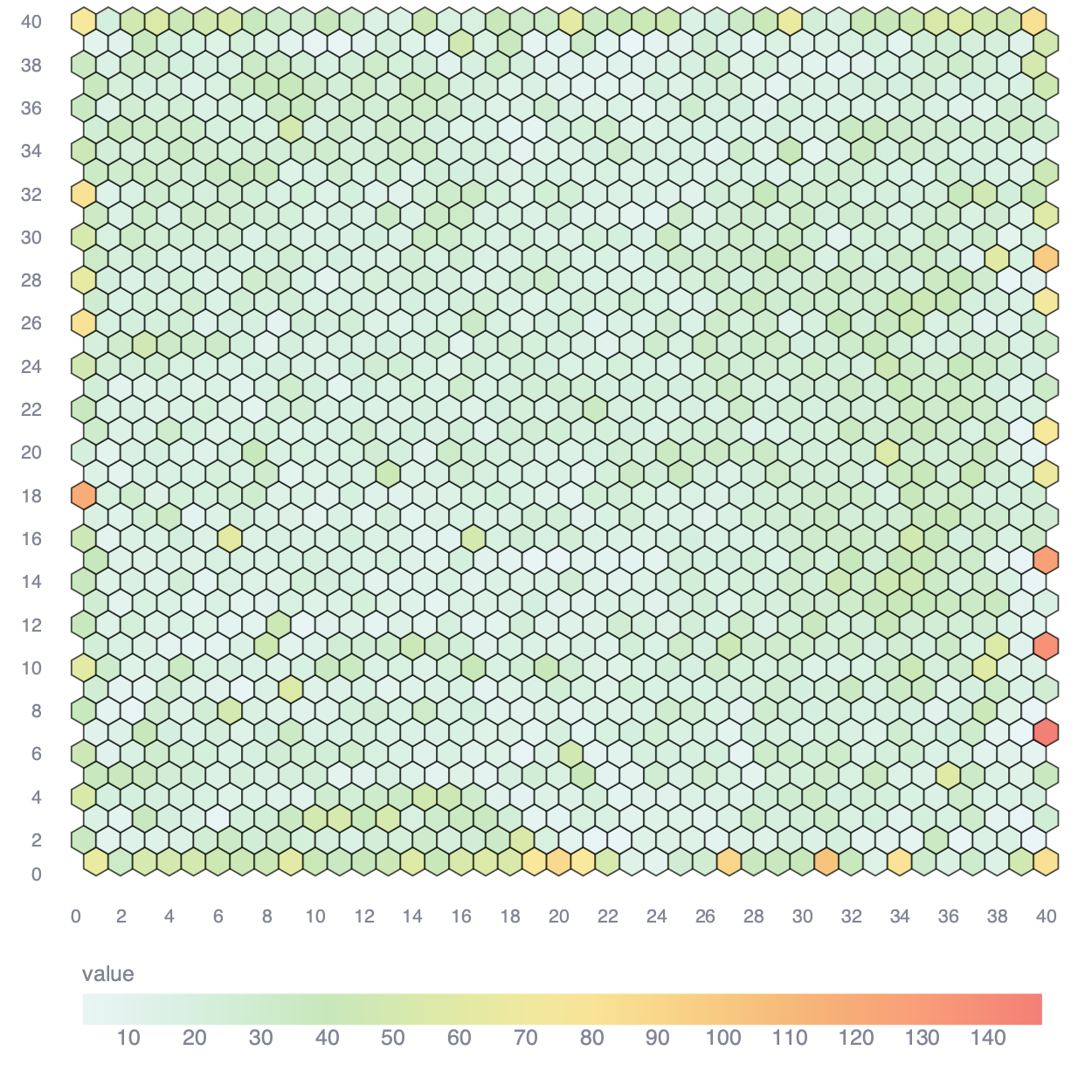}
    \caption{Visualization of SOM: U-Matrix (left) and Activation Response (right)}
    \label{fig:som_analysis_row1}
\end{figure*}

Illustrated in Figure~\ref{fig:som_analysis_row1} (right), this visualization tracks how often each neuron in the SOM was chosen as the Best-Matching Unit (BMU) during training for the input data. Neurons with higher frequency values suggest they are repeatedly selected as BMUs, pointing to regions of the map that correspond to dense concentrations in the dataset. This tool helps identify the areas of the map that most reflect the data.

\subsection{Training Feature Space map}
\begin{figure*}[t]
    \includegraphics[width=0.48\textwidth]{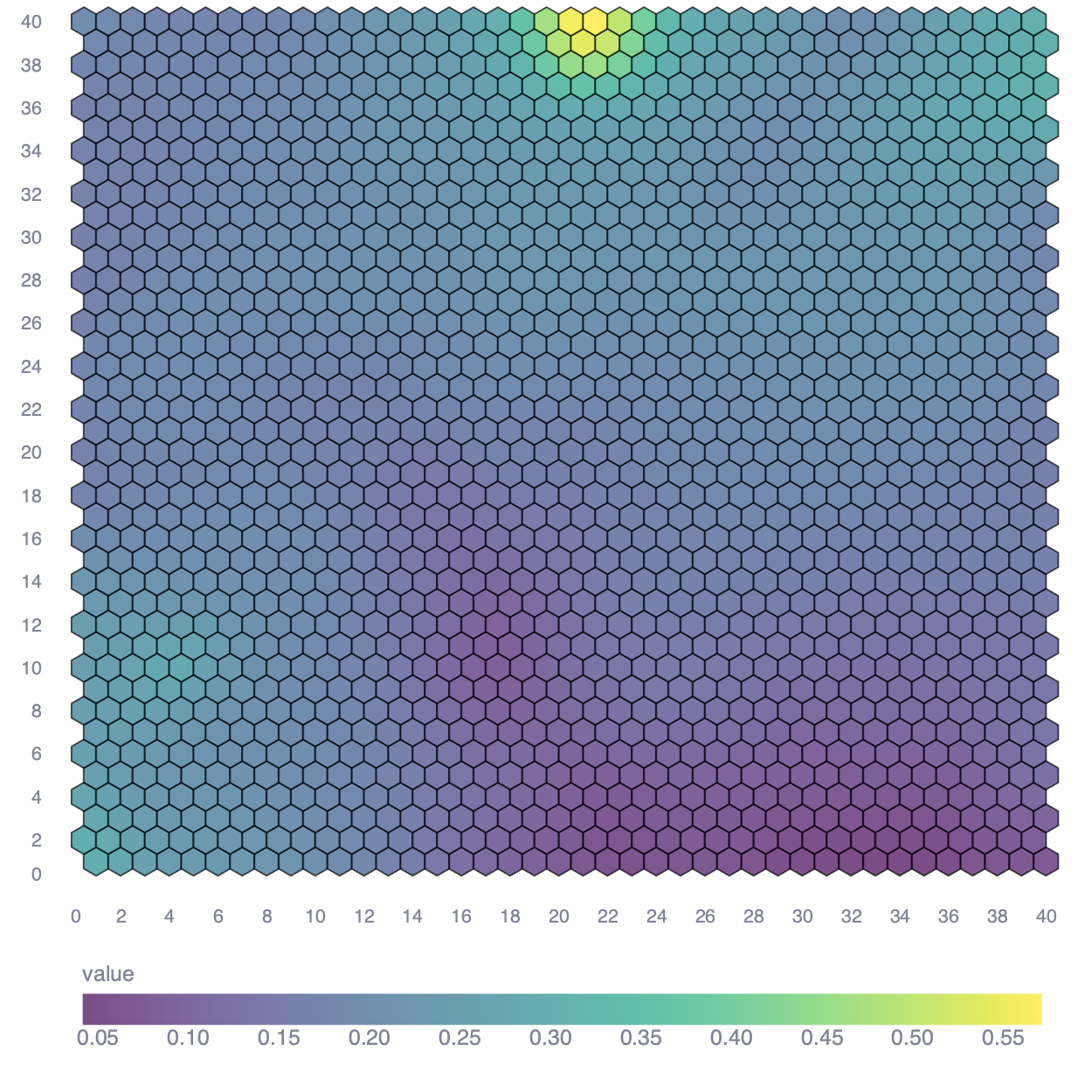}
    \includegraphics[width=0.48\textwidth]{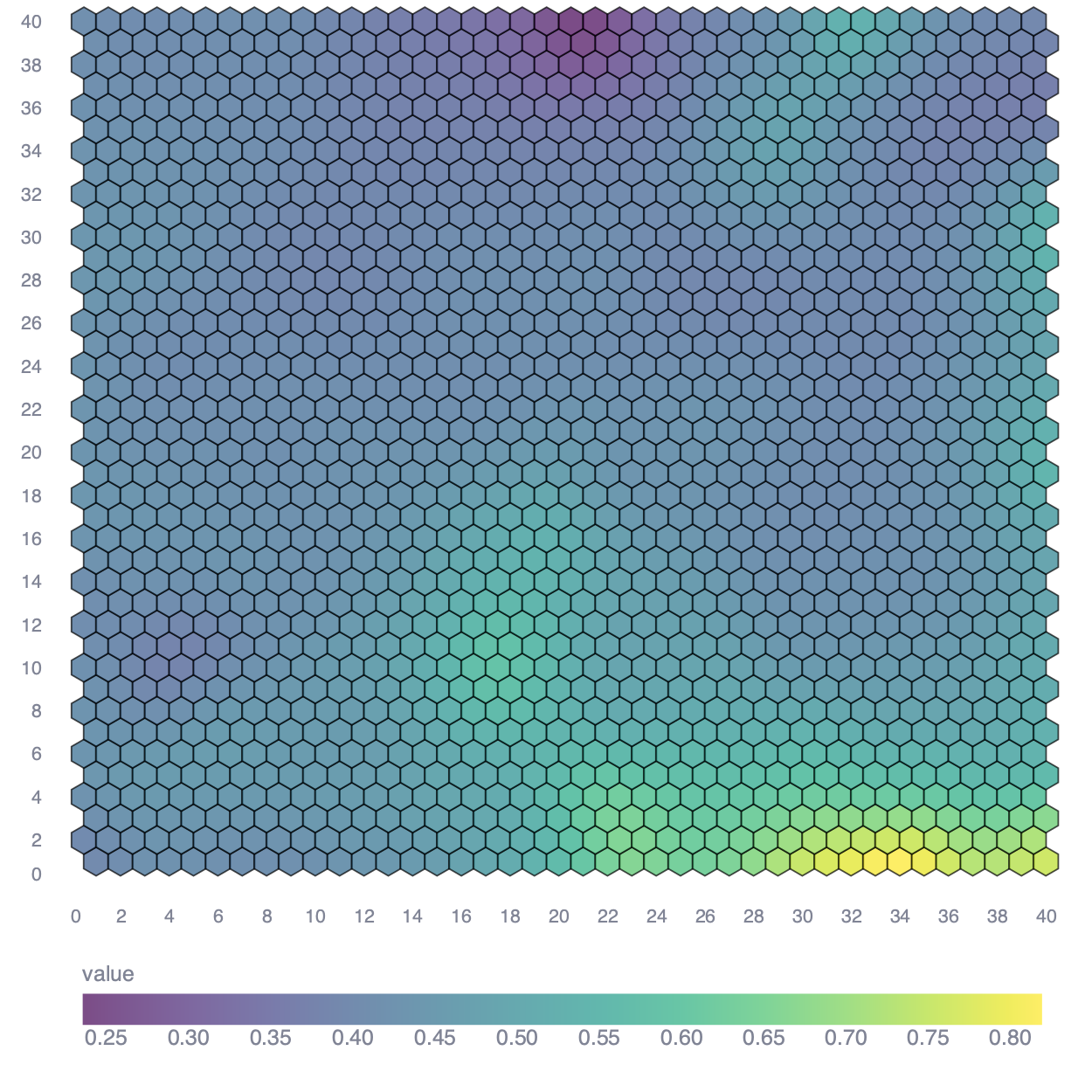}
    \caption{Visualization of SOM weights: Feature Visualization of Black Body temperature (left) and Power-law Index (right)}
    \label{fig:feature_vis}
\end{figure*}

This tool displays the spread of weight values across the SOM for the \textit{features} employed in training. Each neuron reflects a set of weight combinations tied to the input \textit{features}, allowing users to see how these attributes shape the clustering of the data. This visualization also provides insight into which \textit{features} play the most significant roles in defining the SOM layout. Figure~\ref{fig:feature_vis} presents two examples, the weights of black-body temperature and power-law index, illustrating their complementary distributions across the SOM, consistent with their expected physical relationships.

\subsection{Source Dispersion map} 
This tool visualizes how detections for chosen sources are distributed across the SOM map. Users can pick specific sources to examine, sorted by the \textit{Dispersion Index}, which calculates the average pairwise distance between detections, providing insights into the extent of the detection spread of a source. The idea behind this tool is that the extent of the detection spread of a source may reflect its time-varying behavior: widely dispersed detections suggest that the emission properties of the source shift notably over time, potentially landing in distant SOM regions, while a tight spread indicates consistent behavior across observations. Users can also define a minimum and maximum detection count for sources to be included.

\subsection{Source Name Visualization map}
This tool lets users investigate specific sources from the Chandra Source Catalog (CSC) by plotting their names on the SOM, with each source marked as a colored dot. This helps users see how sources are distributed across the map.

For instance, Figure~\ref{fig:source_name_main_type_visualization} (left) highlights three CSC sources with the most detections:
\begin{itemize}
    \item 2CXO J123048.6+122332 (orange-red) – Part of a Galaxy, 124 detections
    \item 2CXO J123049.2+122328 (orange) – Part of a Galaxy, 121 detections
    \item 2CXO J004238.6+411603 (blue) – X-ray Binary, 114 detections
\end{itemize}

The orange sources, both tied to galaxy parts (PartOfG), cluster in a similar SOM region, suggesting shared \textit{feature} traits. Conversely, the blue X-ray binary (XRayBin) occupies a separate area, indicating distinct \textit{feature} characteristics. This separation underscores the SOM ability to differentiate astrophysical sources with unique physical emission mechanisms, aiding users in linking sources to clusters and understanding their relationships.

\subsection{Main Type Visualization map}
This feature displays the \texttt{main\_type} classification assigned to sources, as matched with the SIMBAD database. It enables users to investigate the grouping patterns of various astronomical object types within the SOM and reveal how they relate to one another. In addition to the native SIMBAD \texttt{main\_type}, the map can also project the \emph{aggregated} main-type labels used for classification (Section~\ref{subsec:simbad}).

\begin{figure*}[t]
    \includegraphics[width=0.48\textwidth]{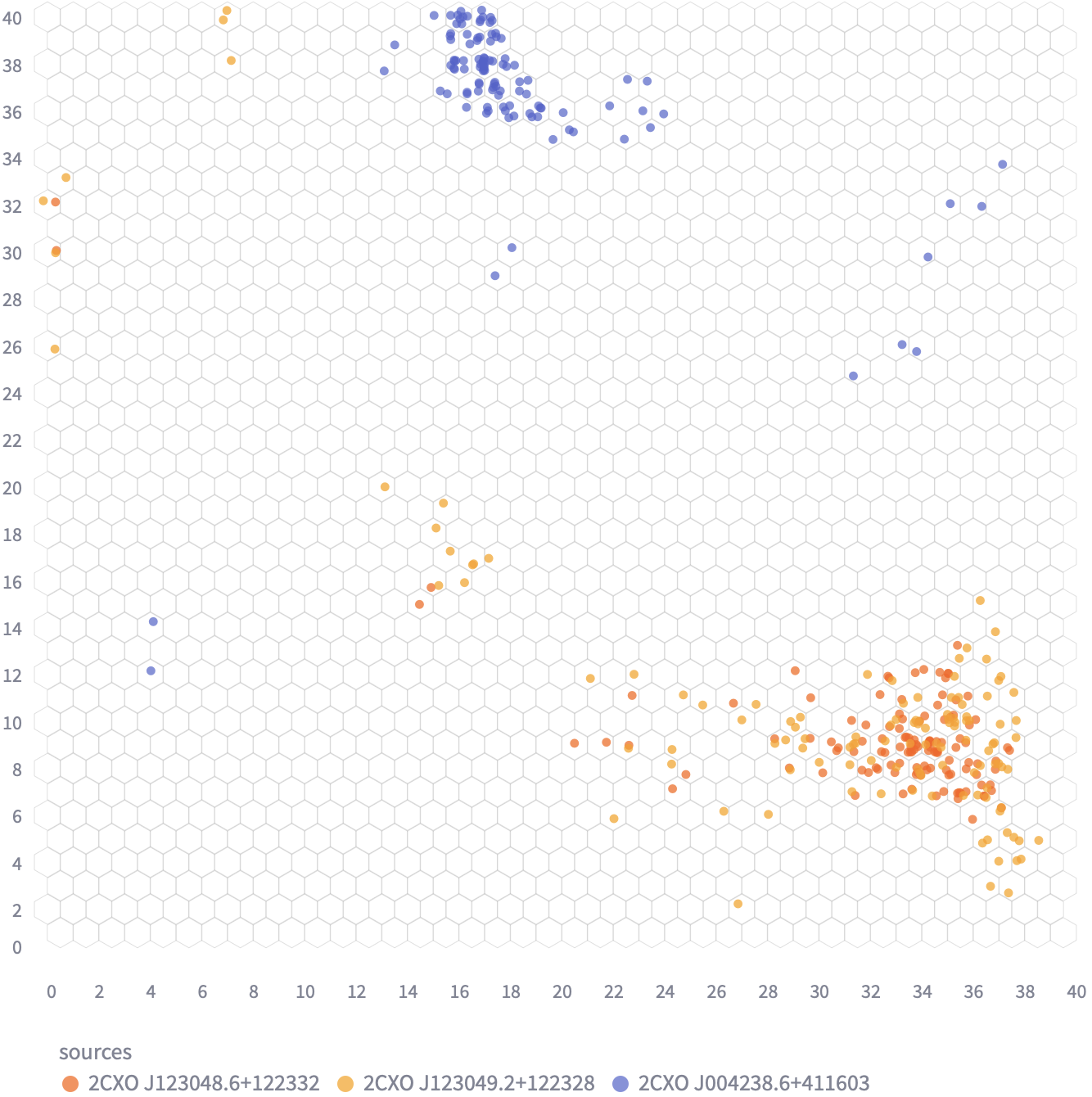}
    \includegraphics[width=0.48\textwidth]{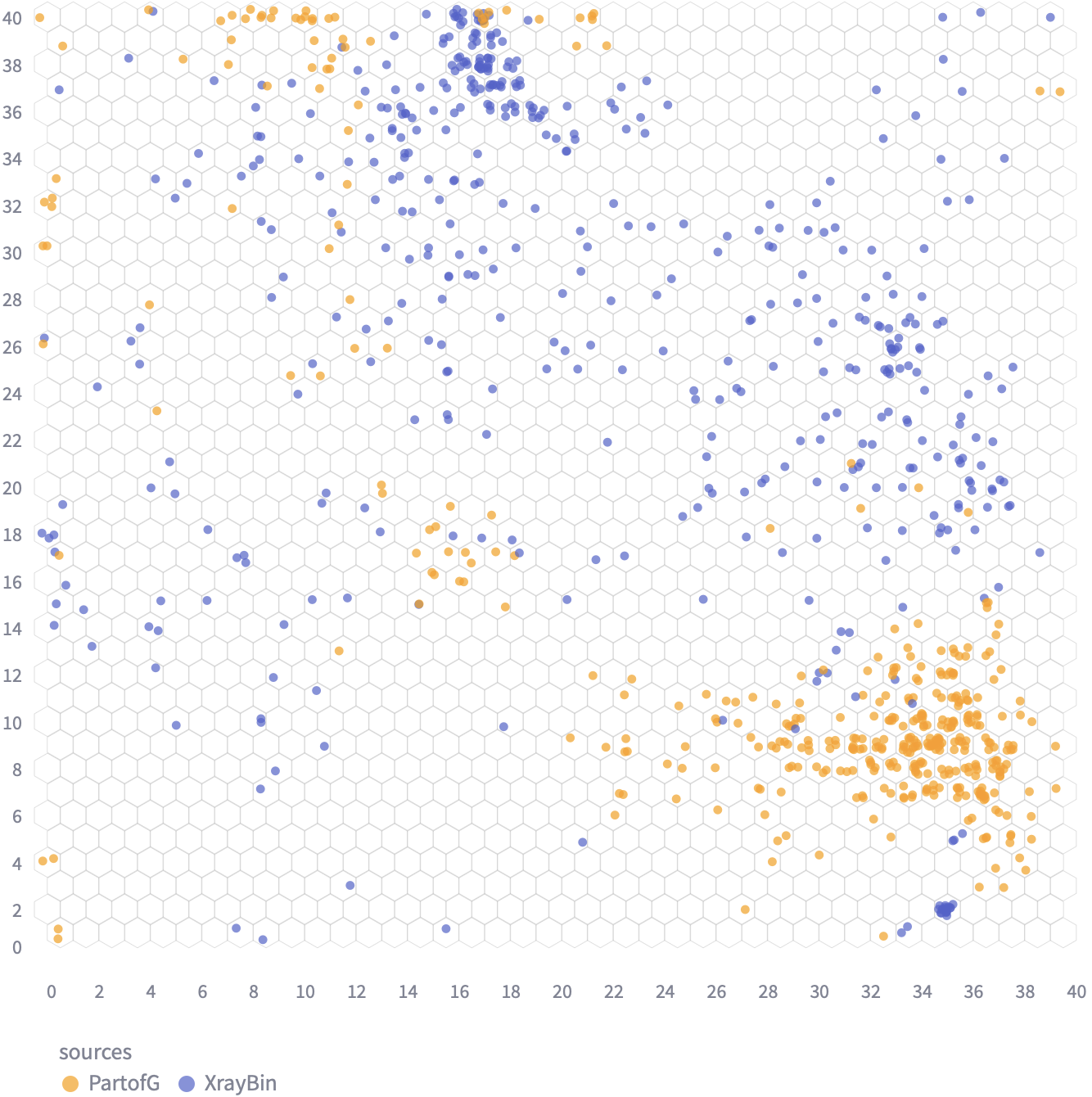}
    \caption{Left: Three distinct sources from the Chandra Source Catalog, each marked by a unique color. Right: Distribution of two primary source types across the SOM, 'Part of a Galaxy' (orange) and 'X-ray Binary' (blue), highlighting their spatial arrangement.  }
    \label{fig:source_name_main_type_visualization}
\end{figure*}

In Figure~\ref{fig:source_name_main_type_visualization} (right), two labels are shown:
\begin{itemize}
    \item Part of a Galaxy (PartOfG) (orange).
    \item X-ray Binary (XRayBin) (blue).
\end{itemize}

The figure reveals that galaxy components tend to group in certain areas, especially in the bottom-right, while X-ray binaries spread more broadly, with a cluster in the top-left. This indicates differing \textit{feature} profiles between the two labels.

\subsection{Features Visualization map}
This tool allows users to map any \textit{feature} from the CSC onto the SOM. Beyond source names or types, it supports in-depth analysis of dataset \textit{features}. Users can also import external datasets, as long as they include the SOM training \textit{features}. Additionally, when multiple numeric \textit{features} align with the same BMU, users can choose a scaling approach to adjust how these values appear on the map.

\subsection{Classification}
This tool provides the \textit{SOMClassifier} used in the paper and lets users train and apply it. When the hyperparameter set is chosen, the app automatically trains a dedicated SOM for classification (separate from the visualization SOM) and then runs \textit{SOMClassifier} on that map. Users can evaluate on the validation and test splits or classify unlabeled detections.

During training, BMUs are attached to the labeled training detections to build per-cell class histograms; counts are then pooled over hex-grid neighborhoods with an RBF kernel in \textit{feature} space. Pooled counts are converted to Laplace-smoothed posteriors. For each detection, the smallest radius that meets the support threshold is chosen, and a label is emitted only if both purity and margin exceed the user-set thresholds, otherwise the classifier abstains.

After training, the app reports, separately for validation and test, macro-$F_1$, micro-$F_1$, coverage, per-class $F_1$, valid/$N$, and the confusion matrix.

For unlabeled detections, the app summarizes assigned and abstained predictions, coverage, class distribution, confidence-by-class, and support/purity statistics. The full prediction table can be downloaded from the dedicated section.

\end{document}